\documentclass[prb,twocolumn,preprintnumbers,amsmath,amssymb,showpacs,superscriptaddress,floatfix,dvips]{revtex4-2}
\usepackage{graphicx}% Include figure files
\usepackage{dcolumn}% Align table columns on decimal point
\usepackage{bm}% bold math
\begin{document}

%\preprint{APS/123-QED}
%
\title{Pseudogap and strange metal states in the square-lattice Hubbard model: A comprehensive study}
\author{Arata Tanaka}%
\affiliation{Quantum Matter Program, Graduate School of Advanced Science and Engineering, Hiroshima University, Higashi-hiroshima 739-8530, Japan}
\date{\today}% It is always \today, today,
\begin{abstract}
To clarify the origin of the pseudogap and strange metal states as well as their mutual relationship in cuprate superconductors, a comprehensive study on the spectral function, Fermi surface, resistivity and dynamical spin susceptivity of the Hubbard model on the square lattice has been conducted by means of the ladder dual-fermion approximation with an electron self-energy correction similar in spirit to Moriyaesque $\lambda$ correction.
It is found that the appearance of these two states requires that the characteristic hole concentration below which the Mott--Heisenberg and Slater mechanisms of electron localization occurs $p_{\rm MS}$ nearly coincides with the hole concentration where the Van Hove singularity (VHS) point, i.e., the renormalized quasiparticle energy $\tilde{\varepsilon}^*_{\bm{X}}$ at the X point $\bm{k}=(\pi,0)$, is in the vicinity of the Fermi level. When this condition is met, $\tilde{\varepsilon}^*_{\bm{X}}$ is pinned at which the nesting condition of the antiferromagnetic (AFM) fluctuations is fulfilled almost everywhere on the Fermi surface in a wide range of the hole concentration in a metallic state, i.e., the strange metal state. 
 The spin fluctuations of the strange metal state are nearly quantum critical and the dynamical spin susceptivity is well described by overdamped spin wave having the $\omega/T$ scaling with the relaxation rate at the Planckian limit, $\hbar\Gamma \approx 2k_{\rm B}T$. Because of these distinctive features of the strange metal state, the $\bm{k}$ dependence of scattering rate of electrons is small and electrons behave as the marginal Fermi liquid, where the imaginary part of their self-energy has Im\,$\Sigma_{\bm{k}}(\omega)=Tf(\omega/T)$ scaling, resulting in $T$-linear resistivity. In contrast, the pseudogap state is magnetically in the renormalized classical regime and the pseudogap is formed near the X point where the nesting condition of the short-range AFM order is fulfilled. Although the main origin of these two states is the AFM fluctuation, the end point of the pseudogap phase $p^*$ is placed near $p_{\rm MS}$ and the strange metal state is present not just around the quantum critical point (QCP) of the AFM fluctuation $p_{\rm QCP}$ at low temperatures as expected in the Hertz--Millis--Moriya theory but extends toward $p^*$ due to the pinning of the VHS point.
These results are consistent with the resistivity in La$_{2-x}$Sr$_x$CuO$_4$ (LSCO), where the linear coefficient is maximum at $p^*$, the first-order-like sudden drop in the intensity of the angle resolved photoemission spectra at the Brillouin zone boundary at $p^*$ found in $({\rm Bi},{\rm Pb})_2$Sr$_2$CaCu$_2$O$_{8+d}$ and the extended quantum critical behavior of spin fluctuations at low temperatures in the inelastic neutron scattering experiments of LSCO in the strange metal state.
\end{abstract}
\maketitle
\section{Introduction}
Although four decades have passed since the first cuprate superconductor was discovered by Bednorz and M\"{u}ller in 1986 \cite{JBednorz1986}, cuprate superconductors remain one of the most actively researched topics in the field of strongly correlated electron systems \cite{PALee2007}. Not only as the archetypal doped Mott insulator \cite{MImada1998}, the origins of various phases appeared in cuprates and their mutual relationship are still controversial and even now subjects of stimulating discussion. In particular, the origin of the pseudogap state has long been one of the most intriguing enigmas in cuprate superconductors from their discovery. 

The presence of the pseudogap in cuprates is first found in nuclear magnetic resonance measurements as a reduction in the intensity of the low-frequency spin excitations below a temperature $T^*$, which is higher than the superconducting transition temperature $T_{\rm c}$ \cite{WWWarren1989,HAlloul1989}. Since then the phenomena have been observed in other experimental methods \cite{BKeimer2015,OCyrChoiniere2018} including the angle resolved photoemission spectroscopy (ARPES), where parts of the Fermi surface are lost due to the pseudogap formation in the spectral function at the Fermi level \cite{MHashimoto2014}. It is still unknown the pseudogap state arises from some kind of long-range order or how it can be related to the strange metal state \cite{PWPhillips2022}, another mystery in cuprate superconductors rival with the pseudogap state. A variety of theories has been put forward to explain the origin of the pseudogap state: spontaneous symmetry breaking \cite{SAKivelson1998,SChakravarty2001,Kaminski2002,MESimon2002}, topological orders \cite{MSScheurer2018a,MSScheurer2018b,WWu2018,SSachdev2019,WWu2020}, emergent composite-particles \cite{YYamaji2011}, slave fermion approach \cite{ZLong2023} and a new quantum statics \cite{JZaanen2011}. The pseudogap state also has been discussed in relation to the Mott physics: the structure of the poles and zeros of the electron Green's function \cite{TDStanescu2009,SSakai2010,SSakai2023}, the momentum-selective Mott transition \cite{MFerrero2009,PWerner2009,EGull2009,EGull2010} and the Widom line \cite{GSordi2012,CWalsh2013}.

On the other hand, since the parent compound of cuprates without carrier doping La$_2$CuO$_4$ is well described by a Mott insulator with an antiferromagnetic (AFM) ground state \cite{SChakravarty1988,PHasenfratz1991,RColdea2001}, it would be plausible to consider the pseudogap state in cuprates is caused by a short-range AFM order disturbed or modulated by doped carriers. To clarify the effects of the AFM fluctuation in doped Mott insulators, theories capable to describe the strong electron correlation are required and various numerical methods have been developed, e.g., exact diagonalization, quantum Monte Carlo (QMC) \cite{MQin2020,HXu2022,SSorella2023,WWu2017,FSimkovicIV2024}, density matrix renormalization group (DMRG) and its tensor extensions \cite{Stoudenmire2012,AWietek2021,QLi2023}. The dynamical mean-field theory (DMFT) \cite{AGeorges1996} is also one of the choice. However, to deal with the AFM fluctuations accurately, the effects of spatial correlations must be taken into account, which are considered only the mean-field level in DMFT. To include the short-range spatial correlation, cluster extensions of DMFT have been developed \cite{MHHettler1998,AILichtenstein2000,GKotliar2001,TMaier2005}, such as the dynamical cluster approximation (DCA) \cite{MHHettler1998} and the cellular DMFT (CDMFT) \cite{GKotliar2001}. Indeed, the pseudogap appeared in the spectral function due to the short-range AFM correlation in the calculations with these approach \cite{CHuscroft2001,AMacridin2006,BKyung2006,OGunnarsson2015,MMeixner2024,MKlett2022}. The properties of the pseudogap were also studied with the diagrammatic Monte Carlo \cite{WWu2017,FSimkovicIV2024} and tensor network \cite{AWietek2021}. Another way to include the effects of spatial correlations is perturbative extensions of DMFT \cite{HKusunose2006,AToschi2007,ANRubtsov2008,SBrener2008,ANRubtsov2009,GRohringer2018} such as the dynamical vertex approximation (D$\Gamma$A) \cite{AToschi2007} and the dual fermion approximation (DFA) \cite{ANRubtsov2008,SBrener2008,ANRubtsov2009}. The effects of long-range spatial correlations can be handled with the ladder or parquet diagrammatic expansion with these methods. The importance of the long-range AFM correlation in the metal-insulator transition in the Hubbard model at the half-filling has been discussed in the previous D$\Gamma$A \cite{TSchaefer2015} and DFA \cite{HHafermann2009a,JOtsuki2014,ATanaka2019} studies with the particle-hole ladder diagram. The superconducting state in cuprates has been also discussed those including both the particle-hole and particle-particle ladder diagrams \cite{MKitatani2019,MKitatani2023,GVAstretsov2020}. 

In addition to the origin of the pseudogap state itself, there has been a long standing debate about the nature of $p^*$, the hole concentration at which the pseudogap phase ends \cite{NEHussey2018,JAyres2020}. If the pseudogap state arises from some kind of long-range order, $p^*$ can be the quantum critical point (QCP) of it and the superconducting state would be caused by the critical fluctuation of this order as expected in the Hertz--Millis--Moriya theory \cite{JAHertz1976,AJMillis1993,TMoriya1985,HvLoehneysen2007}. Although the pseudogap and strange metal phases across $p^*$ in low temperatures are ``hidden'' by the presence of the superconducting phase at low temperatures, experiments with the strong magnetic field, where the superconducting state is suppressed, have brought about opportunity to obtain information of them \cite{CProust2019,RDaou2009,BMichon2018,BMichon2019,CGirod2021,CCollignon2017}. Rather abrupt drop of $T^*$ has been found just above $p^*$ in La$_{1.6-x}$Nd$_{0.4}$Sr$_x$CuO$_4$ (Nd-LSCO) and La$_{1.8-x}$Eu$_{0.3}$Sr$_x$CuO$_4$ (Eu-LSCO) \cite{BMichon2018,BMichon2019}. Similar abrupt change in the intensity of the photoemission spectra at the Brillouin zone boundary at $p^*$ has been also observed in a recent ARPES study on $({\rm Bi},{\rm Pb})_2$Sr$_2$CaCu$_2$O$_{8+d}$ (Bi2212) \cite{SDChen2019}. In contrast, the temperature dependence of the electronic specific heats have found to have a critical behavior $C_e/T \sim \ln(1/T)$ near $p^*$ in La$_{2-x}$Sr$_x$CuO$_4$ (LSCO), Nd-LSCO, Eu-LSCO and Bi$_{2+y}$Sr$_{2-x-y}$La$_x$CuO$_{6+\delta}$ (Bi2201) \cite{BMichon2019,CGirod2021}. The Fermi surface transformation at $p^*$ in Nd-LSCO has been inferred from the resistivity and Hall coefficient, which is consistent with previous LSCO and YBa$_2$Cu$_3$O$_y$ (YBCO) studies \cite{RDaou2009,CCollignon2017} and recent angle resolve magnetoresistance experiments on Nd-LSCO \cite{YFang2022}.

The strange metal (SM) state in cuprates can be characterized by their $T$-linear resistivity, which is inconsistent with Landau's Fermi liquid theory. Some of cuprates exhibit $T$-linear resistivity extends over a wide range of temperature: from $T_{\rm c}$ to 600~K (YBCO), 700~K (Bi2201) and more than 1100~K (LSCO) \cite{MGurvitch1987,SMartin1990}. Furthermore, if the superconducting state is suppressed by the strong magnetic field, the $T$-linear resistivity persists down to $T=0$~K just above the electron concentration of the AFM QCP in electron-doped cuprates and near $p^*$ in hole-doped cuprates \cite{RACooper2009,RDaou2009,BMichon2018,PGiraldoGallo2018,ALegros2019}. It has been argued that the transport scattering rate $1/\tau_{\rm tr}$ of electrons in the SM phase is near the Planckian time scale $\tau_{\rm P}=\hbar/k_{\rm B}T$, which is the shortest time scale of inelastic relaxation \cite{JZaanen2004,JZaanen2019}, $\tau_{\rm tr}\sim \tau_{\rm P}$ and the transport scattering rate of overdoped cuprates in the SM regime, indeed, have been estimated at the limit of the Planckian dissipation as $1/\tau_{\rm tr}=\alpha k_{\rm B}T/\hbar$ with $\alpha\sim 1$ \cite{ALegros2019}. However, the existence of the Planckian bound of metals and how it can be defined or measured in experiments are still controversial problems \cite{NRPoniatowski2021,SAhn2022,SAHartnoll2022}. In addition to cuprates, strange metals with the $T$-linear resistivity have been found in many strongly correlated systems including heavy fermion compounds, iron pnictides, organic molecular solids and magic angle twisted bi-layer graphene \cite{JANBruin2013,SAHartnoll2022,MTaupin2022,PWPhillips2022}.

To explain behavior of the SM state of cuprates, the marginal Fermi liquid (MFL) hypothesis has been proposed, where electrons are assumed to be scattered by critical bosonic excitations over a wide range of momentum \cite{CMVarma1989}. This hypothesis is consistent with experimental results of the $T$-linear resistivity, $C_e/T \sim \ln(1/T)$ behavior of the electronic specific heat and the $\omega/T$ scaling of the optical conductivity of the SM state \cite{CMVarma1989,CMVarma2020,BMichon2023}.
The $T$-linear dependence of the resistivity in the two-dimensional (2D) Hubbard model at high temperatures has been found in studies with the determinantal Monte Carlo calculations \cite{EWHuang2019} and the cold atom experiments \cite{PTBrown2019}. The $T$-linear electron scattering rate has been found to occur in the 2D Hubbard model with DCA in a non-Fermi liquid  (NFL) phase placed between the pseudogap and the Fermi liquid phases \cite{WWu2022}.  

Recently, the Sachdev-Ye-Kitaev (SYK) approach has been proposed to describe many-body systems without quasiparticle excitations by introducing randomness in the interaction and the SM state has been studied from various aspects \cite{AAPetal2019,PCha2020,PTDumitrescu2022,AAPatel2023,CLi2024,DChowdhury2022,SSachdev2023}. It has been pointed out that there are similarities between $p^*$ in cuprates and the Kondo destruction QCP in heavy-fermion SM materials, where the phase transition accompanied with transformation of the Fermi surface occurs \cite{SKirchner2020,HHu2024}.    
In fact, the $T$-linear resistivity at a Kondo destruction transition has been found in a large-$N$ slave boson theory including randomness in the interaction similar to the SYK mode \cite{EEAldape2022}. A study of a Kondo destruction QCP with the periodic Anderson model treated by two-site CDFMT has also found the dynamical staggered-spin susceptibility with $\omega/T$ scaling and the SM behavior of the optical conductivity and resistivity \cite{AGleis2024,AGleis2025}. Unlike the SYK approach, this method does not required interaction disorder.

In an attempt to explain various phases in cuprates, the so-called Van Hove scenario has been proposed, where the instability toward MFL \cite{PCPattnaik1992,DMNewns1995} or NFL \cite{IEDzyaloshinskii1996} in the presence of the Van Hove singularity (VHS) point in the vicinity of the Fermi level in a 2D fermion system further leads to a magnetic or superconducting ground state. In fact, it has been found that both the AFM and $d$-wave superconducting states are stable even with a small on-site Coulomb interaction by means of a renormalization group (RG) analysis \cite{CJHalboth2000,DZanchi2000,CHonerkamp2001,PAIgoshev2011} and the parquet approximation \cite{ATZheleznyak1997,NFurukawa1998,VYIrkhin2001,VYIrkhin2002,AAKatanin2003}. The flattening of the bands and pinning of the VHS point at the Fermi level in relation to the concept of the Fermi condensation \cite{VAKhodel1990,GEVolovik1991,PNozieres1992,MVZverev1999} have been also discussed \cite{VYIrkhin2002} and recently demonstrated in the Hubbard model on the triangular lattice by means of the ladder DFA \cite{DYuin2014}.

In this paper the pseudogap and strange metal states and their mutual relationship in the Hubbard model on the square lattice are discussed in connection with the AFM fluctuation and the VHS point by means of DFA including the particle-hole ladder diagram (LDFA) with an electron self-energy correction similar in spirit to Moriyaesque $\lambda$ correction \cite{TMoriya1973,TMoriya1985}. To do so, precise information of the long-range AFM correlation near QCP and subtle change in the Fermi surface are required and for the purpose, the Lanczos exact-diagonalization technique was employed \cite{ATanaka2019}, which is an efficient and accurate means to obtain the local four-point vertex function of the effective impurity Anderson model of LDFA and essential in this study in addition to the self-energy correction.

The plan of the paper is the following: in Sec.~\ref{Overview}, the $p$-$T$ phase diagrams of the four representative parameter sets discussed in this paper and an overview of the results obtained in this study are presented. The LDFA is outlined and the self-energy correction employed in this paper is explained in Sec.~\ref{Method}. The LDFA results of the DOS, spectral function, Fermi surface, self-energy, electrical resistivity, dynamical spin susceptivity and pair susceptivity are presented in Sec.~\ref{Results}. Discussions are given in Sec.~\ref{Discussion} and the paper is closed with a brief summary in Sec.~\ref{Conclusion}.

\section{$p$-$T$ phase diagram\label{Overview}}
\begin{figure*}
\includegraphics[width=17cm]{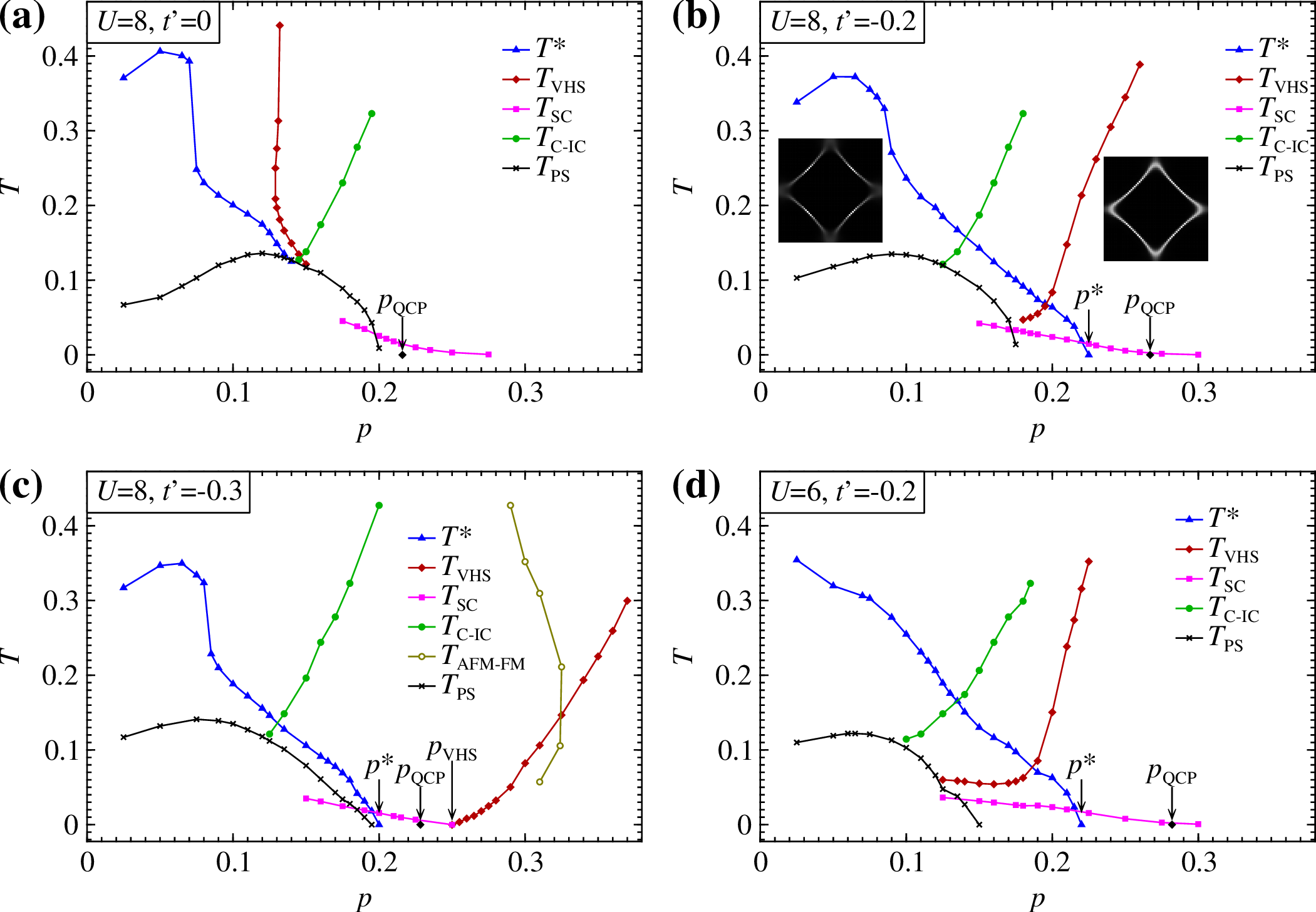}
  \caption{\label{pTphase}$p$-$T$ phase diagrams for $U=8$ with $t'=0$ (a), $t'=-0.2$ (b) and $t'=-0.3$ (c) and for $U=6$ with $t'=-0.2$ (d). In each panel the temperatures of pseudogap formation $T^{*}$, the VHS point at the Fermi level $T_{\rm VHS}$, transition to the $d$-wave superconductor $T_{\rm SC}$, commensurate to incommensurate crossover of AFM fluctuations $T_{\rm C-IC}$ and metal-insulator phase separation $T_{\rm PS}$ are presented. The positions of the end point of the pseudogap phase $p^*$ and the AFM QCP $p_{\rm QCP}$ are indicated by the arrows. In (c), the crossover temperature between the AFM and ferromagnetic fluctuations $T_{\rm AFM-FM}$ is also depicted and the hole concentration where $T_{\rm VHS}=0$ $p_{\rm VHS}$ is indicated by the arrow. In (b), the Fermi surfaces typical in the PG phase (at $p=0.2$ and $T=1/28$) and the SM phase (at $p=0.235$ and $T=1/28$) within the first Brillouin zone are also indicated (for details see Fig.~\ref{FSh02} in Sec.~\ref{Fsurf}).}
\end{figure*}
Before discussing any detail of this study, in Fig.~\ref{pTphase}, we present phase diagrams with hole concentration $p$ and temperature $T$ obtained with four representative parameter sets, which will be referred throughout in this paper, and overview the salient results in this study. In this paper the origin of the pseudogap and strange metal states and their mutual relationship are discussed in relation to the AFM spin fluctuation, the Mott--Heisenberg and Slater mechanisms of electron localization and the Van Hove singularity point. In this regards, the characteristic temperatures each of which indicates change in particular aspects of the electronic states as a function of $p$ are shown in each panel.

The Hamiltonian of the Hubbard model on the square lattice is expressed as
\begin{align}
  {\cal H}=\sum_{i,j,\sigma}t_{i,j}c^{\dagger}_{i\sigma}c_{j\sigma}+U\sum_i \hat{n}_{i\uparrow}\hat{n}_{i\downarrow},
\end{align}
where $c^{\dagger}_{i\sigma}$ ($c_{i\sigma}$) is the creation (annihilation) operator of an electron at site $i$ with the spin $\sigma~(=\uparrow,\downarrow)$, $\hat{n}_{i\sigma}\equiv c^{\dagger}_{i\sigma}c_{i\sigma}$, $t_{i,j}$ denotes the hopping integral between sites $i$ and $j$, and $U$ represents the on-site Coulomb interaction. In this study, the first $t_{i,j}=-t$ and second $t_{i,j}=-t'$ nearest neighbor hopping integrals are considered and thus the one-body energy of an electron with the wave vector $\bm{k}$ is written as $\varepsilon_{\bm{k}}=-2t(\cos k_x +\cos k_y )-4t'\cos k_x\cos k_y$. In the following, we set $t=1$ and use it as the unit of energy and temperature.

The Van Hove singularity (VHS) temperature $T_{\rm VHS}$ is the temperature at which the renormalized quasiparticle energy $\tilde{\varepsilon}^*_{\bm{k}}$ at the X point $\bm{X}=(\pi,0)$ is placed at the Fermi level $\tilde{\varepsilon}^*_{\bm{X}}=0$. Here,  $\tilde{\varepsilon}^*_{\bm{k}}$ is defined as $\tilde{\varepsilon}^*_{\bm{k}}=z_{\bm{k}}\tilde{\varepsilon}_{\bm{k}}$ and 
\begin{align}
\tilde{\varepsilon}_{\bm{k}}=\varepsilon_{\bm{k}}+{\rm Re}\,\Sigma_{\bm{k}}(\tilde{\varepsilon}^*_{\bm{k}})-\mu,
\end{align}
where $\Sigma_{\bm{k}}(\omega)$ denotes the electron self-energy, $\mu$ represents the chemical potential and $z_{\bm{k}}$ is the quasiparticle residue
\begin{align}
  z_{\bm{k}}= \left[1-\frac{\partial}{\partial\omega}{\rm Re}\,\Sigma_{\bm{k}}(\omega)\Big|_{\omega=\tilde{\varepsilon}^*_{\bm{k}}}\right]^{-1}.\label{zk}
\end{align}
The stability of the pseudogap and strange metal states is strongly influenced by the relative positions of $T_{\rm VHS}$ (the red lines in Fig.~\ref{pTphase}) and the characteristic hole concentration $p_{\rm MS}$ below which tendency toward the electron localization of the Mott--Heisenberg and Slater mechanisms occurs: the metal-insulator phase separation with $U\gtrsim 3$, i.e., tendency toward charge disproportionation owing to the Mott electron localization, and the pseudogap formation by the AFM fluctuations akin to the gap formation at the magnetic Brillouin zone boundary of the Slater mechanism. On the other hand, with $p\gtrsim p_{\rm MS}$ electrons are prone to be mobile at the cost of the increase of effective mass, i.e., reduction of $z_{\bm{k}}$, caused by the strong scattering owing to the AFM fluctuations. The detailed discussions on $p_{\rm MS}$ are in Sec.~\ref{Chemi} and the values for $U=6$ and $U=8$ are $p_{\rm MS}\approx 0.18$ and $p_{\rm MS}\approx 0.2$, respectively. The energy position of the VHS can be controlled by $t'$ and three typical cases where the $T_{\rm VHS}$ line is placed far below ($t'=0$), around ($t'=-0.2$) and far above ($t'=-0.3$) $p_{\rm MS}$ with fixed value of $U=8$ are shown in Figs.~\ref{pTphase}(a), (b) and (c), respectively.
To show the effects of $U$, the results of $U=6$ and $t'=-0.2$ is also depicted in Fig.~\ref{pTphase}(d).

Note that the appearance of the metal-insulator phase separation at low temperatures is considered to be an artifact caused by the insufficient treatment of the spatial charge correlation in the present study, where only the ladder diagram of the particle-hole channel of the dual fermion is considered, and in reality, the phases at $T=0$ are expected to be the spin-density wave (SDW), stripe and $d$-wave superconductor ($d$-SC) in studies with unbiased large-scale QMC calculations \cite{MQin2020,HXu2022,SSorella2023}. Instead, the results obtained in this study well reproduce those in the experiments with temperatures above these low-temperature phases or a strong magnetic field, where the superconducting state is suppressed.

As will be discussed in Sec.~\ref{Spectra}, in the pseudogap phase a part of the Fermi surface is lost because of the formation of the pseudogap due to the AFM fluctuations on the quasiparticle peak at the Fermi level in the spectral function $A_{\bm{k}}(\omega)$. Here, $A_{\bm{k}}(\omega)$ is defined as 
\begin{align}
  A_{\bm{k}}(\omega)\equiv -\frac{1}{\pi}{\rm Im}\,G_{\bm{k}}(\omega),
\end{align}
where $G_{\bm{k}}(\omega)$ denotes the electron Green's function.
Similar to the long-range AFM order, the pseudogap is formed at the Fermi surface where the nesting condition of the AFM fluctuations is matched to the wave vector $\bm{Q}$ at which  the spin susceptivity $\chi^{\rm sp}(\bm{q},\omega=0)$ takes its maximum as presented in Sec.~\ref{Fsurf}. The pseudogap formation temperature $T_{\bm{k}}^*$ at a $\bm{k}$ point on the Fermi surface can be characterized by the temperature where $A_{\bm{k}}(\omega =0)$ takes the maximum value as a function of $T$, below which the pseudogap is formed on the quasiparticle peak as the AFM correlation length grows and the intensity of the quasiparticle peak at the Fermi level is reduced with decreasing $T$. Hence, the upper boundary of the pseudogap phase $T^*$ can be defined as the maximum $T_{\bm{k}}^*$ among $\bm{k}$ where the pseudogap is formed on the Fermi surface. In each panel, $T^*$ is show by the blue line and the end point of the pseudogap phase $p^*$ is also indicated by the arrow in Figs.~\ref{pTphase}(b)-(d). Note that since LDFA abides the Mermin--Wagner theorem \cite{NDMermin1966,JOtsuki2014}, no long-range AFM order is expected at any finite temperature and $T^*$ is not a transition temperature but only a crossover temperature.

Upon hole doping from the half filling ($p=0$), the Fermi level abruptly shifts to the lower Hubbard band top located near $\bm{k}=(\pi/2,\pi/2)$ and the pseudogap is formed almost whole square-shaped Fermi surface owing to nearly perfect AFM nesting with the nesting vector $\bm{Q}=(\pi,\pi)$ having particularly high $T^*$ below $p\approx 0.08$ except for $U=6$ and $t'=-0.2$, where such a change is not clearly seen. With further doping, the pseudogap is first lost around the $\Gamma$--M symmetry line, i.e., the nodal direction, and four portions of the Fermi surface is recovered, where the nesting condition is not satisfied and the quasiparticle peak appears at the Fermi level in each of $A_{\bm{k}}(\omega)$. This so-called Fermi arcs spread toward the $\Gamma$--X symmetry line, i.e., the antinodal direction with increasing $p$. As the mismatch of the nesting increases with increasing $p$, a crossover from the commensurate $\bm{Q}=(\pi,\,\pi)$ to incommensurate $\bm{Q}=(\pi,\,\pm\delta)$, $(\pm\delta,\,\pi)$ AFM fluctuations takes place at $T_{\rm C-IC}$ depicted by the green lines with decreasing $T$. The four magnetic peaks at the incommensurate $\bm{Q}$'s correspond to the parallel spin stripe order observed in the neutron scattering experiments in LSCO, La$_{2-x}$Ba$_x$CuO$_4$ and Nd-LSCO \cite{SWakimoto1999,QMa2021}. 

The metal-insulator phase separation occurs at the temperatures below $T_{\rm PS}$ shown by the black lines within $p\lesssim p_{\rm MS}$ and the $T_{\rm C-IC}$ lines are ended at the $T_{\rm PS}$ lines near the critical point of the phase separation, indicating instability toward a charge disproportionation. The intersection occurs at $p_{\rm C-IC}\approx 0.125$ for $U=8$ with $t'=-0.2$ and $-0.3$. Indeed, $p\approx 0.125$ coincides with the hole concentration where the temperature below which short-range charge-density-wave (CDW) correlations appear $T_{\rm CDW}$ takes its maximum in LSCO \cite{OCyrChoiniere2018}.  

The pseudogap state is stabilized when the VHS point is located in the vicinity of the Fermi level with $p\lesssim p_{\rm MS}$, since the AFM fluctuations are enhanced by the flatten band nearby the X point and the correlation length of the short-range AFM order grows at low temperatures as detailed in Sec.~\ref{epsX}. In particular, when the $T_{\rm VHS}$ line is located near $p_{\rm MS}$, the stabilization of the pseudogap state is prominent as seen in Figs.~\ref{pTphase}(b) and (d). While the pseudogap phase is ended at $p=0.14$ just above $T_{\rm PS}$ for $U=8$ and $t'=0$ in Fig.~\ref{pTphase}(a), the end point of the pseudogap is extended up to $p^*=0.225$ for $U=8$ with $t'=-0.2$ as in Fig.~\ref{pTphase}(b). In addition, the region of the phase separation is reduced with $t'=-0.2$ and the AFM quantum critical point $p_{\rm QCP}$ is shifted from $p_{\rm QCP}=0.215$ of $t'=0$ to $p_{\rm QCP}=0.265$. However, when the $T_{\rm VHS}$ line is located at much higher doping side of $p_{\rm MS}$, the stabilization of pseudogap state is less prominent as shown in Fig.~\ref{pTphase}(c): $T^*$ is lower than that of $t'=-0.2$ as well as $p^*=0.2$ and $p_{\rm QCP}=0.228$. In contrast, if $U$ is reduced from $U=8$ to $U=6$ with $t'=-0.2$, the exchange interaction of spins $J\sim t^2/U$ is increased and thus the effects of the AFM fluctuations are expected to be more enhanced. Indeed, both $T^*$ and $p_{\rm QCP}$ are increased and the region of the phase separation is reduced as shown in Fig.~\ref{pTphase}(d).

As will be discussed in Sec.~\ref{epsX}, the pseudogap state ($T\lesssim T^*$) is magnetically in the renormalized classical regime and the strange metal state ($T\gtrsim T^*$) is in the quantum critical regime with $p<p^*$. Hence, $T^*$ corresponds to the crossover temperature from the quantum critical to renormalized classical regime with $p<p^*$. However, when the VHS point is located near the Fermi level, $T^*$ is rather steeply reduced near $p^*$ with increasing $p$ and located nearby $p_{\rm MS}\sim 0.2$ instead of $p_{\rm QCP}$ as expected in the Hertz--Millis--Moriya theory. This is because electrons are tend to mobile with $p\gtrsim p_{\rm MS}$ and they can be scattered to the flat band near the X point slightly off the Fermi surface. This causes further flattening of the band near the X point due to the mass enhancement of electrons, resulting in the pinning of the VHS point $\tilde{\varepsilon}^*_{\bm{X}}$ in the vicinity of the Fermi level. This extends the range of $p$ where nearly quantum critical metallic state with almost perfect nesting of the AFM fluctuations is stable, i.e., the strange metal state. The pinning of the VHS point nearby the Fermi level is the origin of the strange metal state and its peculiar behavior: the MFL properties (Sec.~\ref{SelfE}), $T$-linear resistivity (Sec.~\ref{Resistivity}) and extended quantum critical behavior at low temperatures in the dynamical spin susceptibility (Sec.~\ref{chi}). 

It is also found that the first-order-like abrupt disappearance of the pseudogap occurs in the spectral function $A_{\bm{k}}(\omega)$ near the X point with $\tilde{\varepsilon}^*_{\bm{k}}$ off the Fermi level as $p$ approaches $p^*$ in the PG phase, as detailed in Appendix~\ref{pGap}. This is consistent with the abrupt change in the intensity at the intersection of the Brillouin zone boundary and the hole-like Fermi surface of the photoemission spectra at $p^*$ in the ARPES experiments on Bi2212 \cite{SDChen2019}.

As presented in Sec.~\ref{chi}, another essential ingredient for understanding the properties of the SM phase is the behavior of the dynamical spin susceptibility at $p_{\rm QCP}$. This behavior is well described by an overdamped spin wave that exhibits $\omega/T$ scaling with a relaxation rate at the Planckian limit, $\hbar\Gamma \approx 2k_{\rm B}T$, consistent with inelastic neutron scattering experiments \cite{GAeppli1998,MZhu2023}. Furthermore, the dynamical exponent $z=1$ is incompatible with the Hertz--Millis--Moriya theory, which expects $z=2$ for 2D antiferromagnets. How these peculiar properties of the dynamical spin susceptibility related to the $T$-linear dependence of ${\rm Im}\,\Sigma_{\bm{k}}(\omega=0)$ and the $T$-proportionality of the resistivity in the SM state is discussed in Sec.~\ref{Discussion}.

The superconducting transition temperature $T_{\rm SC}$ estimated from the pair susceptibility including effects of the charge fluctuations within the ladder approximation of the particle-particle channel are show by the magenta lines. Clearly the range of $p$ where the superconducting phase is present is extended beyond $p^*$ and ended at $p_{\rm QCP}$ for $U=6$ and $U=8$ with $t'=-0.2$ and at $p_{\rm VHS}$ for $U=8$ with $t'=-0.3$. These extended range of $p$ is exactly where the pinning of the VHS point occurs and the AFM fluctuations are enhanced (Sec.~\ref{pair}).

\section{method\label{Method}}
\subsection{Ladder dual-fermion approach\label{LDFA}}
In DMFT, while the on-site temporal correlation effects are accurately considered, the spatial correlations are taken into account only the mean-field level \cite{AGeorges1996}. 
The dual fermion approximation (DFA) is a perturbative extension of DMFT to include the spatial correlations \cite{ANRubtsov2008,SBrener2008,ANRubtsov2009,HHafermann2009a,JOtsuki2014,GRohringer2018}. 
Instead of directly performing the perturbative calculations, in DFA, a fermionic auxiliary field, which is called the dual fermion, is introduced by a Hubbard--Stratonovich transformation. The original action can be mapped onto that of the dual fermion by integrating out the real electron field. In this way, one can separate the problem of solving the effective impurity Anderson model (IAM) to obtain the local approximation and the perturbative corrections to the spatial correlations in terms of the dual fermions, where as the effective interaction, the reducible four-point vertex function of IAM, which is the lowest order, is included. The neglect of the higher-order vertex functions causes internal inconsistencies \cite{JGukelberger2017,EGCPvanLoon2018,TRibic2018} and this requires correction to the electron self-energy as will be explained in Sec.~\ref{Selfenergy}.
To consider effects of long-range spin fluctuations, in this work, the ladder diagram of the particle-hole channel of the dual fermion \cite{HHafermann2009a} is taken into account and both the self-energy of the dual fermion and the hybridization function of IAM are self-consistently solved. Since there is one-to-one relation between the real electron and dual fermion fields, the Green's functions and response functions of electrons can be obtained from those of dual fermions. 

To solve the IAM problem, the Lanczos exact-diagonalization technique to calculate the local four-point vertex function is applied in this study \cite{ATanaka2019}. This technique is advantageous to access low temperatures and obtain accurate Green's functions on the real axis with the maximum entropy method. The same calculation methods to the previous study on the half-filled Hubbard model \cite{ATanaka2019} is used. The number of discretized momentum points $N$ in the first Brillouin zone of the square lattice used in the calculations is $N=64\times 64$. The number of the bath energy levels for the effective IAM model assumed at high temperatures is $N_{\rm b}=7$ and one of them is placed at the Fermi level. For low temperatures, one additional level is inserted above the Fermi level ($N_{\rm b}=8$). 

To obtain the dynamical spin susceptibility on the real frequency axis $\chi^{\rm sp}(\bm{q},\omega)$, that on the Matsubara axis 
\begin{align}
&\chi^{\rm sp}(\bm{q},i\omega_n)\nonumber\\
  &~~~~\equiv\int_0^{\beta}d\tau\,e^{i\omega_n\tau}\sum_j e^{i\bm{q}\cdot\bm{R}_j}\left\langle S_z(\bm{R}_j,\tau)S_z(\bm{0},0)\right\rangle,
\end{align} 
where $i\omega_n= 2\pi n/\beta$ is the bosonic Matsubara frequency, $\bm{R}_j$ denotes the position of the $j$-th site and $S_z(\bm{R}_j,\tau)\equiv\frac{1}{2}(\hat{n}_{j\uparrow}(\tau)-\hat{n}_{j\downarrow}(\tau))$, is first calculated within the ladder diagram of the particle-hole channel of LDFA \cite{SBrener2008} and then analytic continuation to the real axis is performed numerically by the maximum entropy method.

\subsection{Self-energy correction\label{Selfenergy}}
In the LDFA employed in this study, only the reducible four-point vertex function of IAM is considered as the effective interaction in the dual-fermion action. However, within this approximation, there is a discrepancy between the electron occupation number par site with spin $\sigma$ $n_\sigma$ obtained from the electron Green's function and the asymptotic form of the electron self-energy.
To avoid this problem, in this section, a simple method to make a correction to the electron self-energy is introduced.

In the Hubbard model the asymptotic form of the self-energy of electron can be written as
\begin{align}
&\Sigma_{\bm{k},\sigma}(i\omega_n)\nonumber\\
  &~~~~=Un_{-\sigma}+\frac{U^2n_{-\sigma}(1-n_{-\sigma})}{i\omega_n}+{\cal O}(1/(i\omega_n)^2)\label{sigma_asym}
\end{align}
 and this can be also related to the electron Green's function as
\begin{align}
  n_{\sigma}=\frac{2T}{N}\sum_{\bm{k},n>0}{\rm Re}[G_{\bm{k},\sigma}(i\omega_n)]+\frac{1}{2}.\label{n_G}
\end{align}
While $n_{\sigma}$ obtained with Eq.~(\ref{n_G}) is consistent with that in the asymptotic form in Eq.~(\ref{sigma_asym}) within DMFT, in DFA these equations are not consistent with each other except for the presence of the electron-hole symmetry, e.g., half filling with $t'=0$ in the present model. The self-energy of electron in DFA consists of that of DMFT and its perturbative correction:
\begin{align}
  \Sigma^{\rm DFA}_{\bm{k},\sigma}(i\omega_n)=\Sigma^{\rm DMFT}_{\sigma}(i\omega_n)+\Delta\Sigma^{\rm DFA}_{\bm{k},\sigma}(i\omega_n).\label{sigma_DFA}
\end{align}
Since asymptotic relation in Eq.~(\ref{sigma_asym}) is satisfied within DMFT, the inconsistency is caused by the DFA correction. Although the electron number within DMFT and that of DFA are generally different, $\Delta\Sigma^{\rm DFA}_{\bm{k},\sigma}$ contains only correction to the higher order terms more than $1/(i\omega_n)$ in Eq.~(\ref{sigma_asym}) and no correction to the leading term $Un_{-\sigma}$ occurs.
This is because only the reducible four-point vertex function of IAM is included and the higher-order vertex functions are neglected in the dual fermion action in the present study \cite{JGukelberger2017,EGCPvanLoon2018,TRibic2018}.

To avoid this discrepancy, here, it is assumed that the discrepancy is mainly caused by the lack of correction to the leading term $Un_{-\sigma}$ in Eq.~(\ref{sigma_asym}) within the approximation made in the dual fermion action, and this can be mended by adding a correction term to Eq.~(\ref{sigma_DFA}) as
\begin{align}
\Sigma_{\bm{k},\sigma}(i\omega_n)=\Sigma^{\rm DMFT}_{\sigma}(i\omega_n)+\Delta\Sigma^{\rm DFA}_{\bm{k},\sigma}(i\omega_n)+U\delta n_{-\sigma}, \label{sigma_DFA2} 
\end{align}
where $\delta n_{\sigma}$ is the difference between the electron number within DMFT $n^{\rm DMFT}_{\sigma}$ and that of DFA with this correction $n_{\sigma}$:
\begin{align}
   &\delta n_{\sigma}=n_{\sigma}-n^{\rm DMFT}_{\sigma}\nonumber\\
  &=\frac{2T}{N}\sum_{\bm{k},n>0}{\rm Re}[G_{\bm{k},\sigma}(i\omega_n)]-2T\sum_{n>0}{\rm Re}[G^{\rm DMFT}_{\sigma}(i\omega_n)].\label{delta_n}
\end{align}
$n_{\sigma}$ is chosen to satisfy Eqs.~(\ref{sigma_DFA2}) and (\ref{delta_n}) simultaneously.
Note that this correction of $n_{\sigma}$ is made after the conventional DFA solution is obtained and it is assumed that the values of $\Sigma^{\rm DMFT}_{\sigma}$ and $\Delta\Sigma^{\rm DFA}_{\bm{k},\sigma}$ of the conventional DFA are not affected by this correction.

The procedure for the self-energy correction proposed here is as follows:
\begin{enumerate}
  \item[{i)}] Obtain the DFA solution without correction and calculate $\delta n^{(0)}_{\sigma}$ using Eq.~(\ref{delta_n}) as the initial value.
  \item[{ii)}] Using $\delta n^{(l)}_{\sigma}$ obtained previous step $l$, calculate $\Sigma^{(l)}_{\bm{k},\sigma}$ by Eq.~(\ref{sigma_DFA2}) and compute Green's function
\begin{align}
    G^{(l)}_{\bm{k},\sigma}(i\omega_n)=\frac{1}{i\omega_n-\varepsilon_{\bm{k}}+\mu-\Sigma^{(l)}_{\bm{k},\sigma}(i\omega_n)}.
\end{align}
  \item[{iii)}] Calculate a new $\delta n^{\rm new}_{\sigma}$ with $G^{(l)}_{\bm{k},\sigma}$ by Eq.~(\ref{delta_n}) and obtain 
\begin{align}  
  \delta n^{(l+1)}_{\sigma}=(1-\alpha)\delta n^{(l)}_{\sigma}+\alpha\delta n^{\rm new}_{\sigma},
\end{align}
where $\alpha$ is a constant ($1\ge\alpha>0$).
 \item[{iv)}] Repeat (ii)-(iii) untile $n_{\sigma}$ converges.
\end{enumerate}

Note that similar inconsistency occurs in susceptibilities $\chi^{(r)}(\bm{q},\omega)$ in the ladder D$\Gamma$A due to the violation of several sum rules and in such case, correction to asymptotic behavior of susceptibilities can be made by introducing adjustable parameters $\lambda_r$, i.e., the so called Moriyaesque $\lambda$ correction: $[\chi_{\lambda}^{(r)}(\bm{q},\omega)]^{-1}=[\chi^{(r)}(\bm{q},\omega)]^{-1}+\lambda_r$ \cite{GRohringer2018,TMoriya1973,TMoriya1985}.

\begin{figure}
\includegraphics[width=8cm]{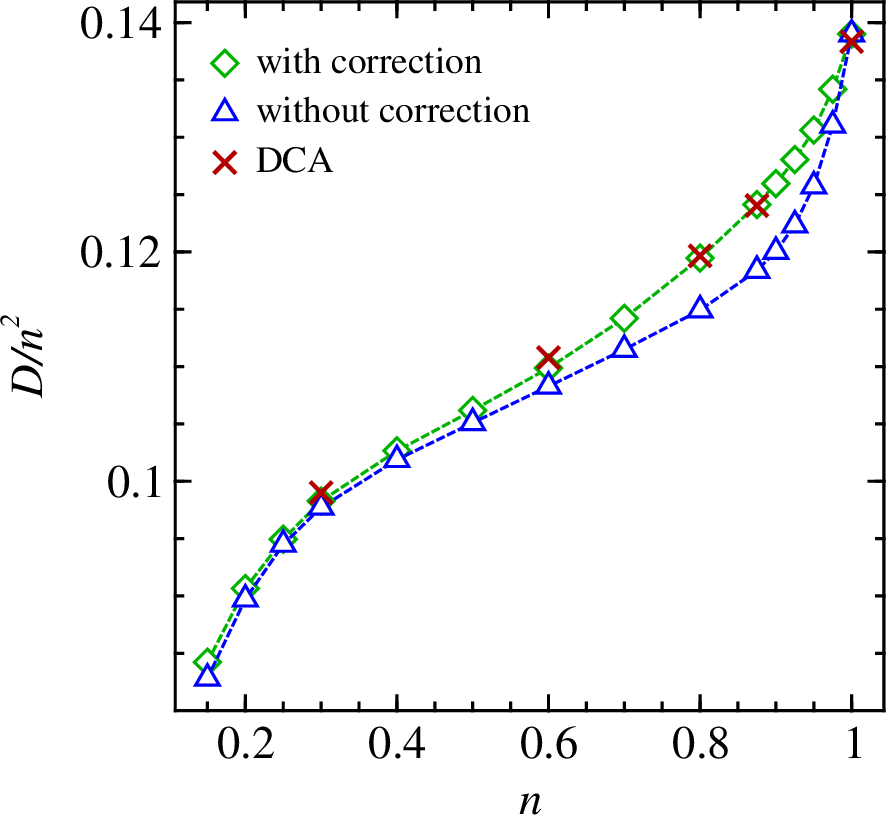}
\caption{\label{Dn}$D/n^2$ versus $n$ plots. The results of LDFA with (diamonds) and without (triangles) the self-energy correction for $U=4$, $t'=0$ and $\beta=4$ are shown. For comparison the DCA data reproduced from Ref.~\onlinecite{JPFLeBlanc2015} are also depicted as crosses.}
\end{figure}
To show to what extent this self-energy correction affects the results, in Fig.~\ref{Dn} $D/n^2$ as a function of the electron density $n$ with and without the correction are compared with the DCA results in Ref.~\onlinecite{JPFLeBlanc2015}, where the double occupancy $D\equiv\langle \hat{n}_{i\uparrow}\hat{n}_{i\downarrow}\rangle$ is calculated using the Migdal-Galitskii formula \cite{Galitskii1958}. While good agreement between the results of DCA and LDFA with the self-energy correction can be found, substantial deviations can be seen for the LDFA results without the correction. 

\section{Results\label{Results}}
\subsection{Behavior of the chemical potential\label{Chemi}}
\begin{figure}
\includegraphics[width=8cm]{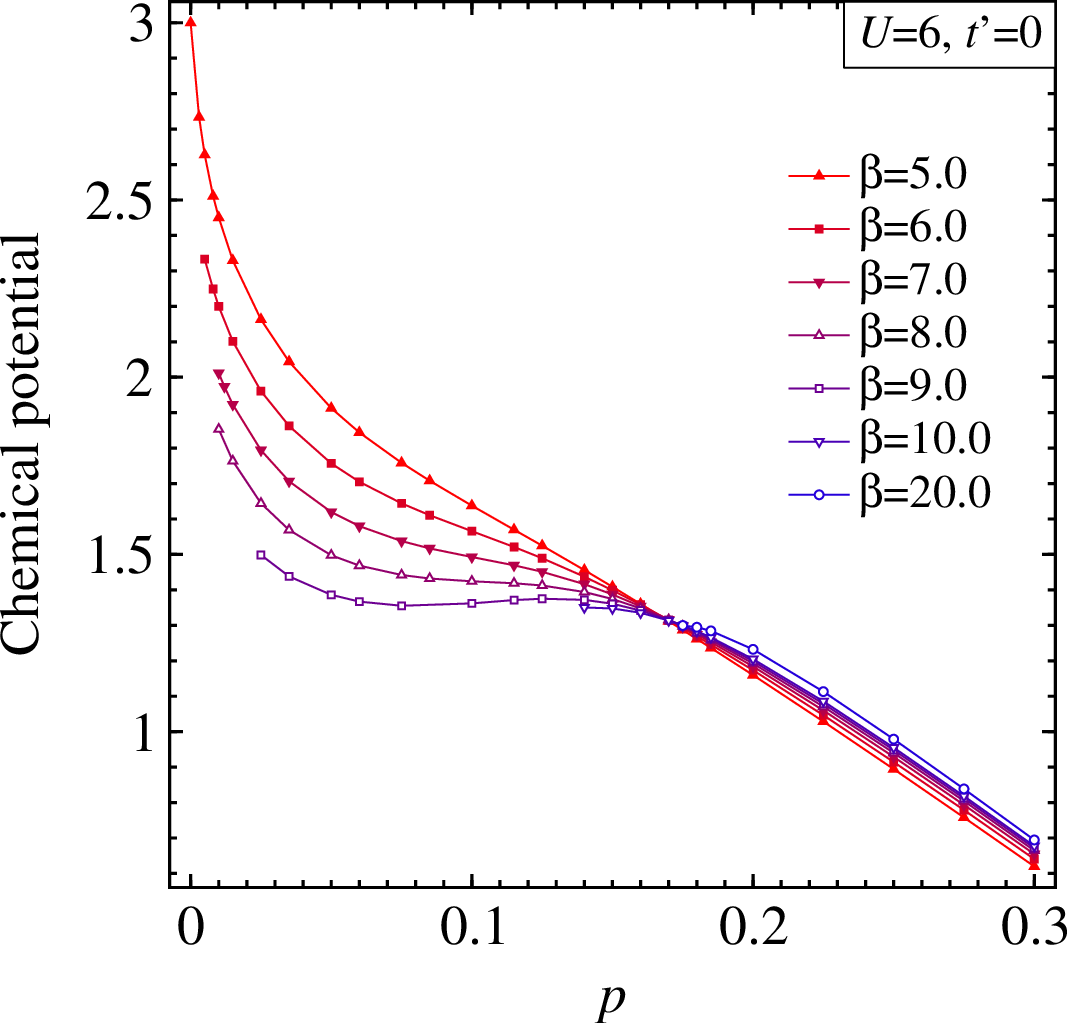}
\caption{\label{muU60}Chemical potential $\mu$ versus hole concentration $p$ curves for various temperatures for $U=6$ with $t'=0$.}
\end{figure}
In Fig.~\ref{muU60}, the chemical potential $\mu$ as a function of the hole concentration $p\equiv 1-n$ for $U=6$ and $t'=0$ is shown for various temperatures. As temperature decreases the value of $\mu$ between $p=0$ and $p=0.176$ rapidly reduced, indicating the formation of the Mott insulating gap at half-filling $p=0$, where $\mu$ is abruptly changed at low temperatures: it moves to the lower Hubbard-band top upon hole doping or the upper Hubbard-band bottom upon electron doping in the vicinity of half-filling. In contrast, for $p>0.176$, the $T$ dependence of $\mu$ is rather limited and only slightly increases with decreasing $T$, i.e., the behavior consistent with ordinal Fermi liquid. The hole concentration at which $\mu$ is unaffected by the change of $T$, therefore, can be regarded as the characteristic hole concentration below which tendency toward the Mott--Heisenberg and Slater mechanisms of electron localization occurs. Here, we denote the hole concentration $p$ at which $(\partial \mu/\partial T)_p=0$ satisfy as $p_{\rm MS}$. However, a further complication arises below the critical temperature $T_{\rm CPMI}=1/7.393$, where the positive $(\partial \mu/\partial p)_T$ region of $p$ appears as seen in the curves for $\beta=8$ and 9 in the figure and a metal-insulator phase separation occurs.

\begin{figure}
\includegraphics[width=8cm]{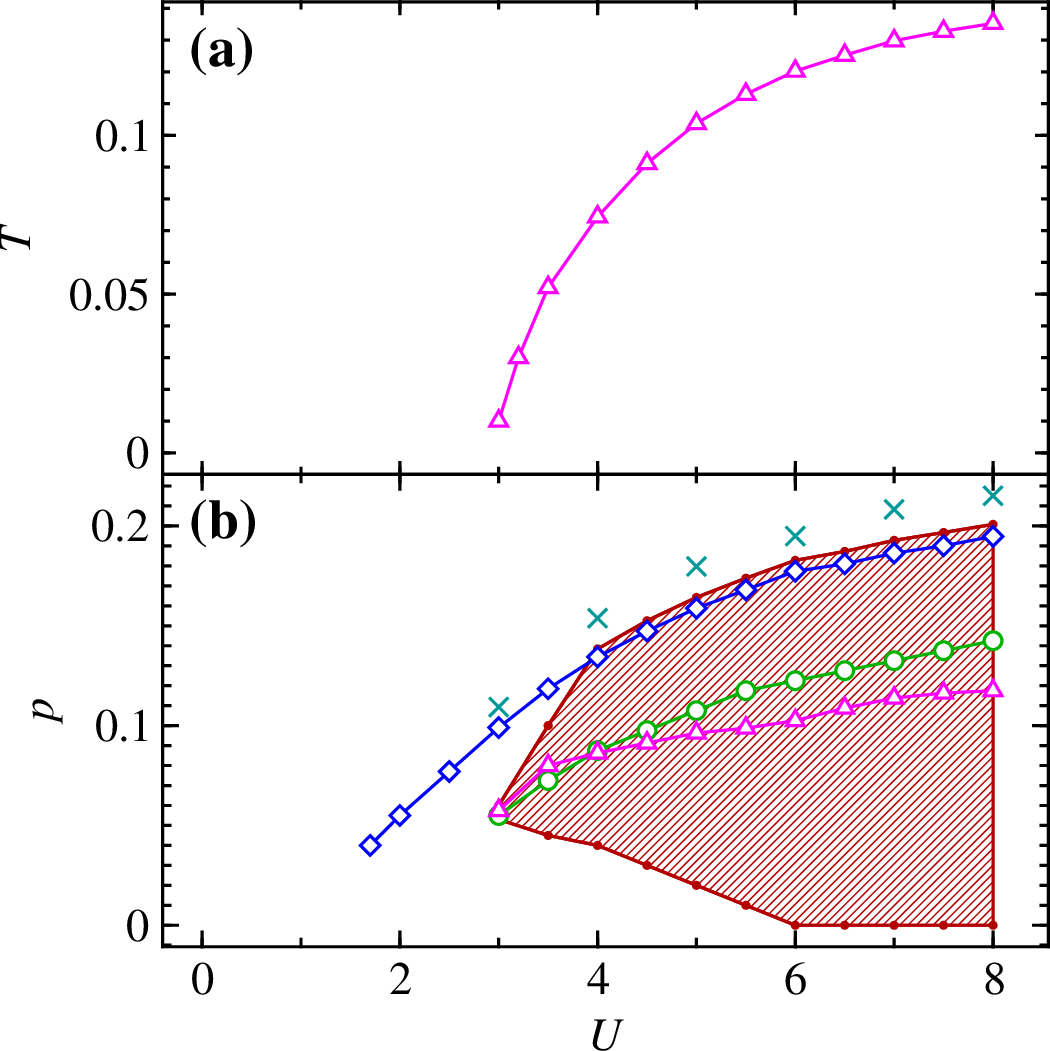}
  \caption{\label{Mott}(a) $T_{\rm CPMI}$ (triangles) as a function of $U$; (b) $U$ dependence of $p_{\rm MS}$ (diamonds), $p_{\rm CPMI}$ (triangles) and $p_{\rm C-IC}$ (circles) and the metal-insulator phase separation region (hatched red area) on $U$--$p$ plane together with the AFM  quantum critical points $p_{\rm QCP}$ for $U=3$, 4, 5, 6, 7 and 8 (crosses).}
\end{figure}
To discuss mutual relation among the characteristic hole concentration of tendency toward electron localization $p_{\rm MS}$, the metal-insulator phase separation and the AFM fluctuation, $U$ dependence of those related quantities are depicted in Fig.~\ref{Mott}. In Fig.~\ref{Mott}(b), $U$ dependence of $p_{\rm MS}$ at $T=0$ and the critical point of the metal-insulator phase separation $p_{\rm CPMI}$ and the phase separation region at $T=0$ on $U$--$p$ plane are presented. Here, the upper boundary of the phase separation region with respect to $p$ at $T=0$  $p_{\rm PS}$ is approximated by the hole concentration at which the linear extrapolation of $(\partial \mu/\partial p)_T$ is zero. In addition, the AFM quantum critical point $p_{\rm QCP}$ and the commensurate-incommensurate crossover point of AFM fluctuations just above $T_{\rm PS}$ $p_{\rm C-IC}$ are shown. In Fig.~\ref{Mott}(a), $T_{\rm CPMI}$ as a function of $U$ is drawn.

The upper boundary of phase separation region $p_{\rm PS}$ nearly coincides with the $p_{\rm MS}$ line with $U\ge 4$ and $p_{\rm QCP}$ are also placed near the $p_{\rm MS}$ line as shown in Fig.~\ref{Mott}(b). The boundary between the AFM and paramagnetic phases at $T=0$, i.e., $p_{\rm QCP}$, is consistent with the previous auxiliary field QMC (AFQMC) study of the Hubbard model \cite{HXu2022}. As already discussed in Sec. \ref{Overview}, the end point of the PG state $p^*$ and $p_{\rm PS}$ are also located near $p_{\rm MS}$ with $t'=-0.2$ and $-0.3$. In addition, $p_{\rm CPMI}$ and $p_{\rm C-IC}$ lines are close together (see also Fig.~\ref{pTphase}(a)) and coincide at the end point with respect to $U$ of the phase separation region $U\approx 3$, where $T_{\rm CPMI}$ decreases continuously to zero with decreasing $U$, indicating the presence of the second-order transition point at $T=0$ as in Fig.~\ref{Mott}(a). As mentioned in Sec. \ref{Overview}, the metal-insulator phase separation is considered to an artifact of insufficient treatment of the spatial charge correlation in this study and one can regard it as an indication of instability toward a charge disproportionation. In fact, SDW states accompanied by CDW is found with large $U \ge 4$ in the AFQMC study \cite{HXu2022}.

Throughout this paper, $p_{\rm MS}$ obtained with $t'=0$ is used as the characteristic hole concentration around which tendency toward partial electron localization is end. This is one of the two essential factors to understand the behaviors of the pseudogap and strange metal states. The other is the relative position of the VHS point to the Fermi level, which is controlled by the value of $t'$. The value of $p_{\rm MS}$ for $U=6$ and $U=8$ with $t'=0$ are  $p_{\rm MS}\approx 0.18$ and $p_{\rm MS}\approx 0.2$, respectively.

\subsection{Doping and temperature dependence of DOS\label{DOS}}
\begin{figure*}
\includegraphics[width=5.5cm]{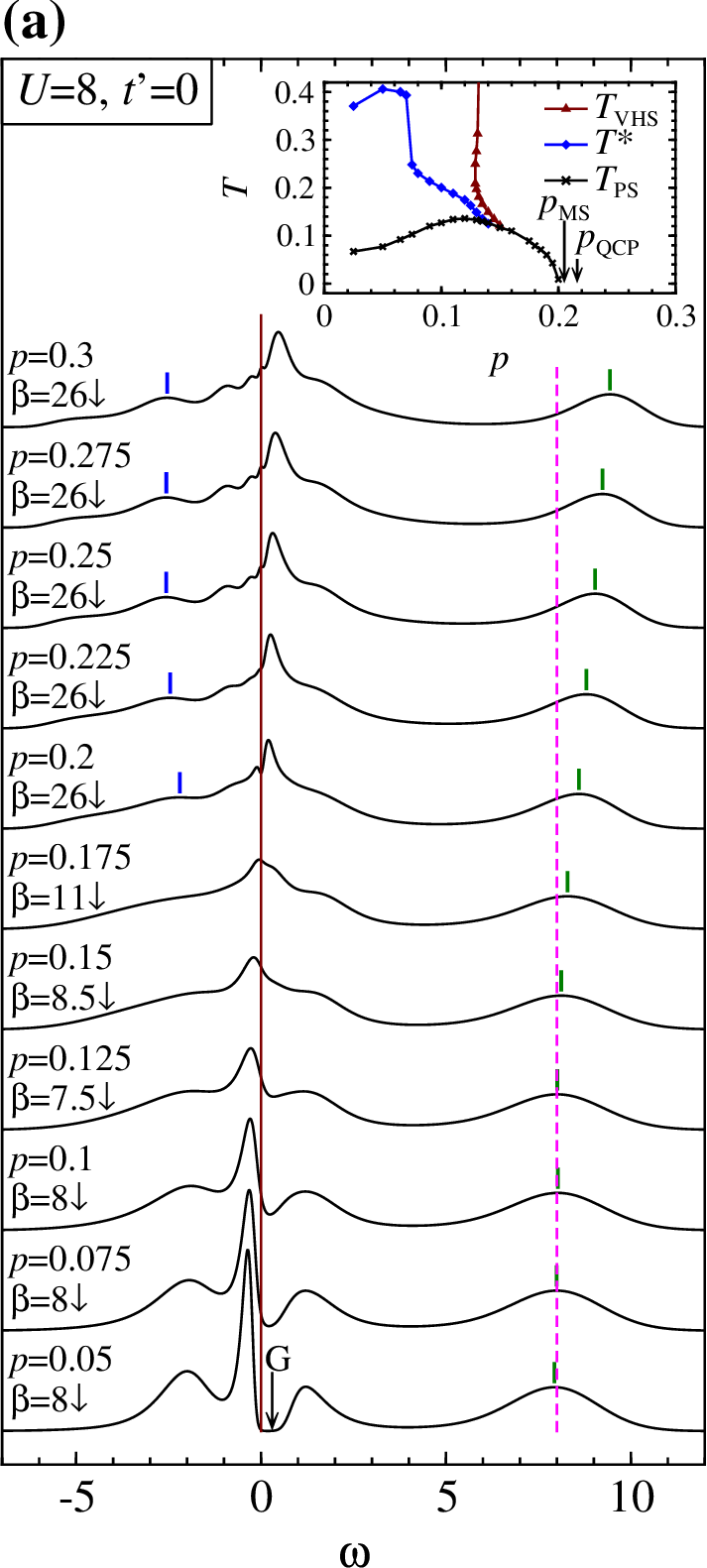}
\includegraphics[width=5.5cm]{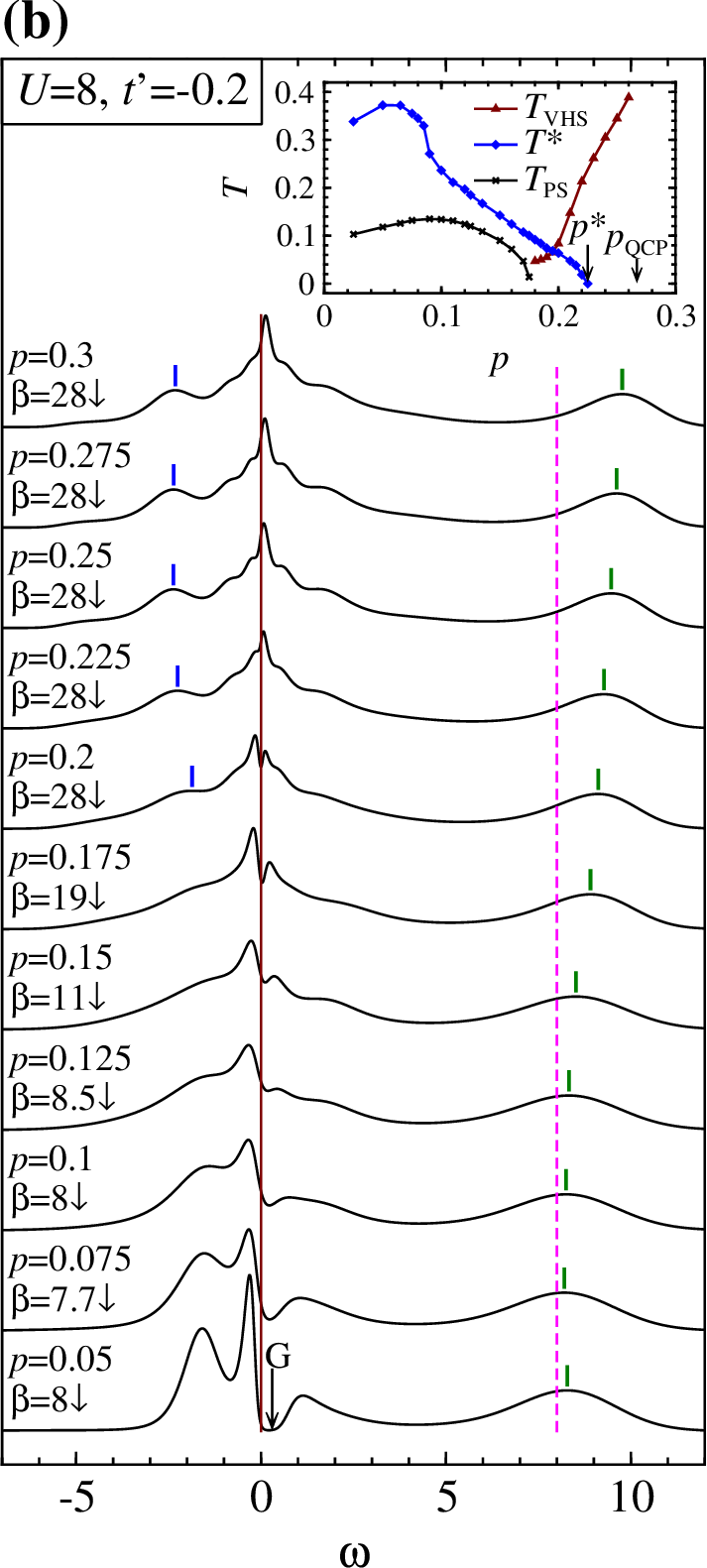}
\includegraphics[width=5.5cm]{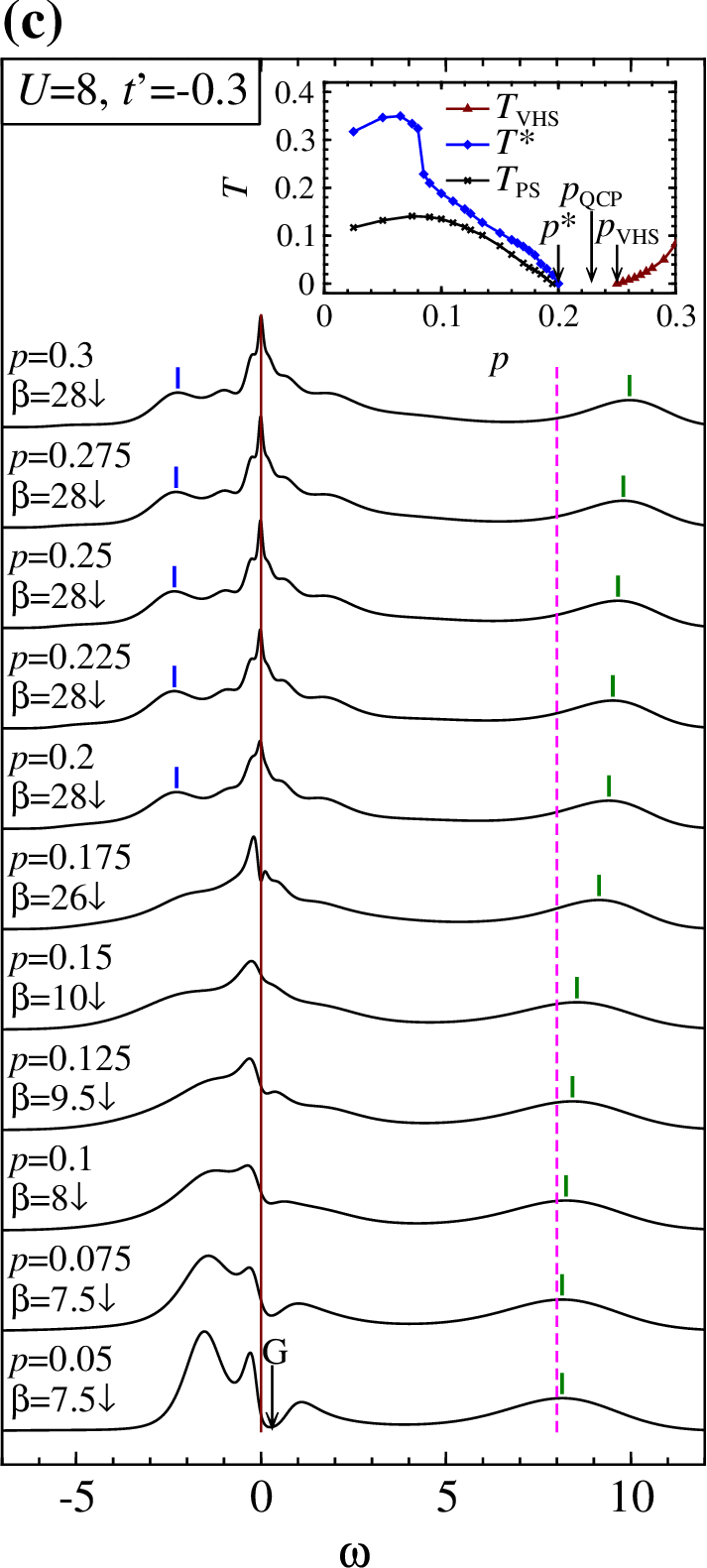}
\caption{\label{DOSp}$p$ dependence of DOS for $U=8$ with $t'=0$ (a), $t'=-0.2$ (b) and $t'=-0.3$ (c). The red vertical line indicates the Fermi level ($\omega=0$) and the magenta dotted line is placed at $\omega=8$. The maximum peak location of the upper Hubbard band in each DOS is shown by the green vertical line. The position of the shoulder structure appeared in the lower Hubbard band with $p\ge 0.2$ is indicated by the blue vertical line. The corresponding $p$-$T$ phase diagram is depicted in the inset in each panel (for details see the figure caption of Fig.~\ref{pTphase}).}
\end{figure*}
To discuss $p$ dependence of DOS defined as
\begin{align}
  \rho(\omega)\equiv -\frac{1}{\pi N}\sum_{\bm{k},\sigma}{\rm Im}[G_{\bm{k},\sigma}(\omega)],\label{DOSdef}
\end{align}
those for $U=8$ with different $t'$ are shown in Fig.~\ref{DOSp}. As seen in the figure, each of DOS consists of two major structures, the so-called lower Hubbard band located around $\omega =0$ and the upper Hubbard band placed around $\omega =8$ and their peak to peak distance is about $U$. The peak top of the upper Hubbard band are indicated by the green vertical lines. As was discussed in the CDMFT study \cite{BKyung2006}, at half filling an insulating gap, the so-called Mott gap, opens between these two structures. Upon hole doping, the Fermi level is immediately shifted to the lower Hubbard band top and the Mott gap is filled with thin ingap states except for a small gap in the vicinity of the lower Hubbard band top (the gap labeled G in each DOS of $p=0.05$). The gap is rapidly smeared and reduced to a pseudogap with further doping. 

The $p$ dependence of the shape of the lower Hubbard band is also strongly affected by the value of $t'$ and resultant change in the energy of the VHS point at the X point $\tilde{\varepsilon}^*_{\bm{X}}$ relative to the Fermi level, which can be seen as the $T_{\rm VHS}$ lines in Fig.~\ref{pTphase}. The main peak, i.e., the maximum intensity peak of the lower Hubbard band, is approximately positioned at $\tilde{\varepsilon}^*_{\bm{X}}$. The main peak arises dominantly from the states nearby the X point in the Brillouin zone, where the VHS exists and thus large DOS is expected. As will be discussed in Sec.~\ref{epsX}, once $\tilde{\varepsilon}^*_{\bm{X}}$ is placed near the Fermi level, the AFM fluctuations are enhanced due to the large DOS and this leads to the pinning of $\tilde{\varepsilon}^*_{\bm{X}}$, i.e., the main peak, near the Fermi level within a certain range of $p$ either by the formation of the pseudogap with $p\lesssim p_{\rm MS}$ or the reduction of $z_{\bm{k}}$ with $p\gtrsim p_{\rm MS}$. Indeed, with $t'=0$ in Fig.~\ref{DOSp}(a), both the main peak pinned at $\omega\approx -0.14$ and the pseudogap are prominent with $p\le 0.14$ compared to those with $t'=-0.2$ and $-0.3$. With further doping the intensity above the Fermi level is increased and developed into the sharp peak around $\omega=0.2$ at $p_{\rm MS}\sim 0.2$ accompanied by a tiny pseudogap at the Fermi level. The main peak continuously moves to the higher energy with increasing $p$ and is reached at $\omega=0.3$ at $p=0.3$. On the other hand, with $t'=-0.2$ in Fig.~\ref{DOSp}(b), the pseudogap is vanished at $p^*=0.225$ and the main peak is pinned at $\omega\approx 0.04$ within $p^*\le p\le p_{\rm QCP}=0.265$ for higher doping. Similarly, with $t'=-0.3$ in Fig.~\ref{DOSp}(c), the pseudogap is vanished at $p^*=0.2$ and the main peak is pinned at $\omega\approx 0$ within $p^*\le p\le p_{\rm VHS}=0.25$ for higher doping.  

In addition to above features, at $p_{\rm MS}$ distinctive change occurs: the main peak of the lower Hubbard band is sharpened and the shoulder structure appeared at $\omega \sim -2$ (indicated by the blue vertical lines in the figure). These change in DOS is connected to the formation of the narrow quasiparticle band in the spectral function discussed in Sec.~\ref{Spectra}.

\begin{figure*}
\includegraphics[width=17cm]{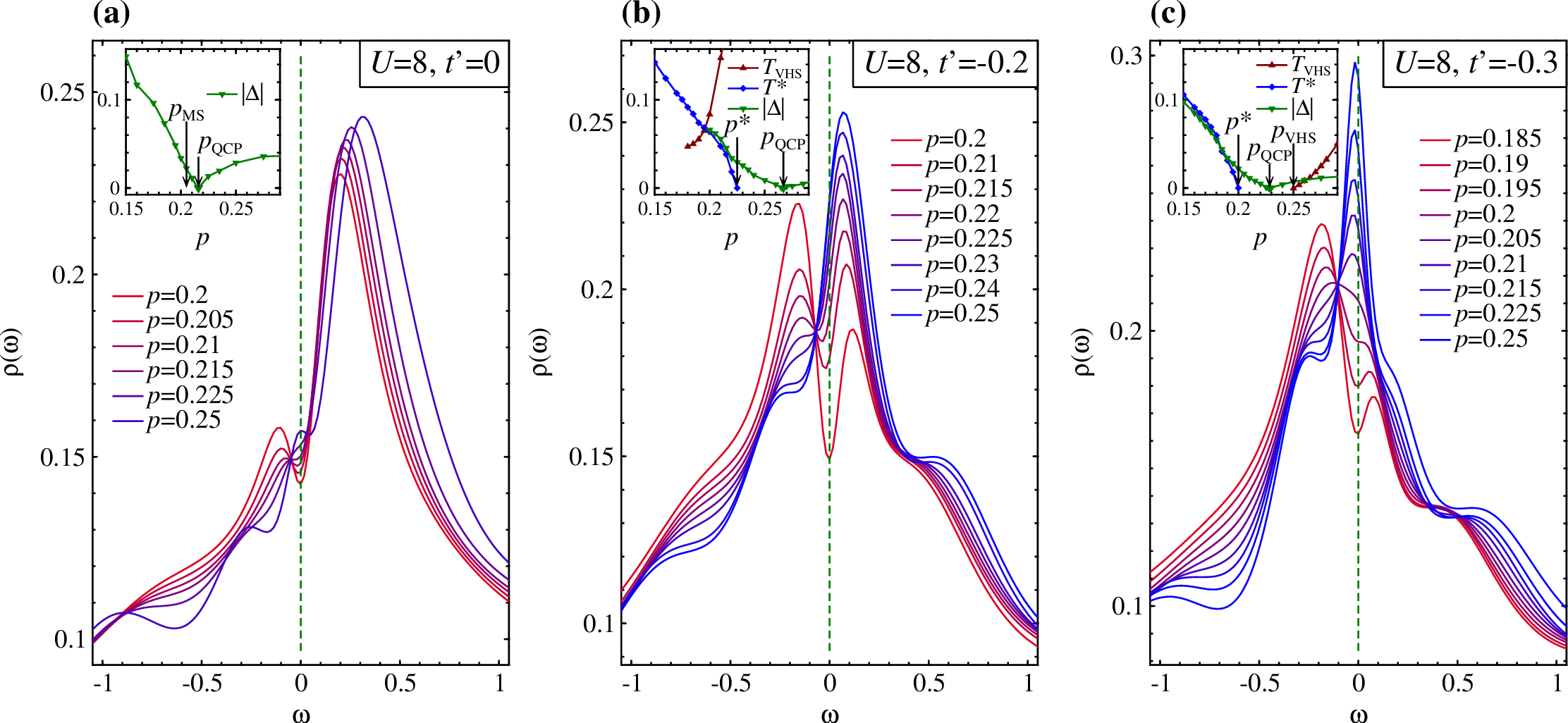}\\
\includegraphics[width=17cm]{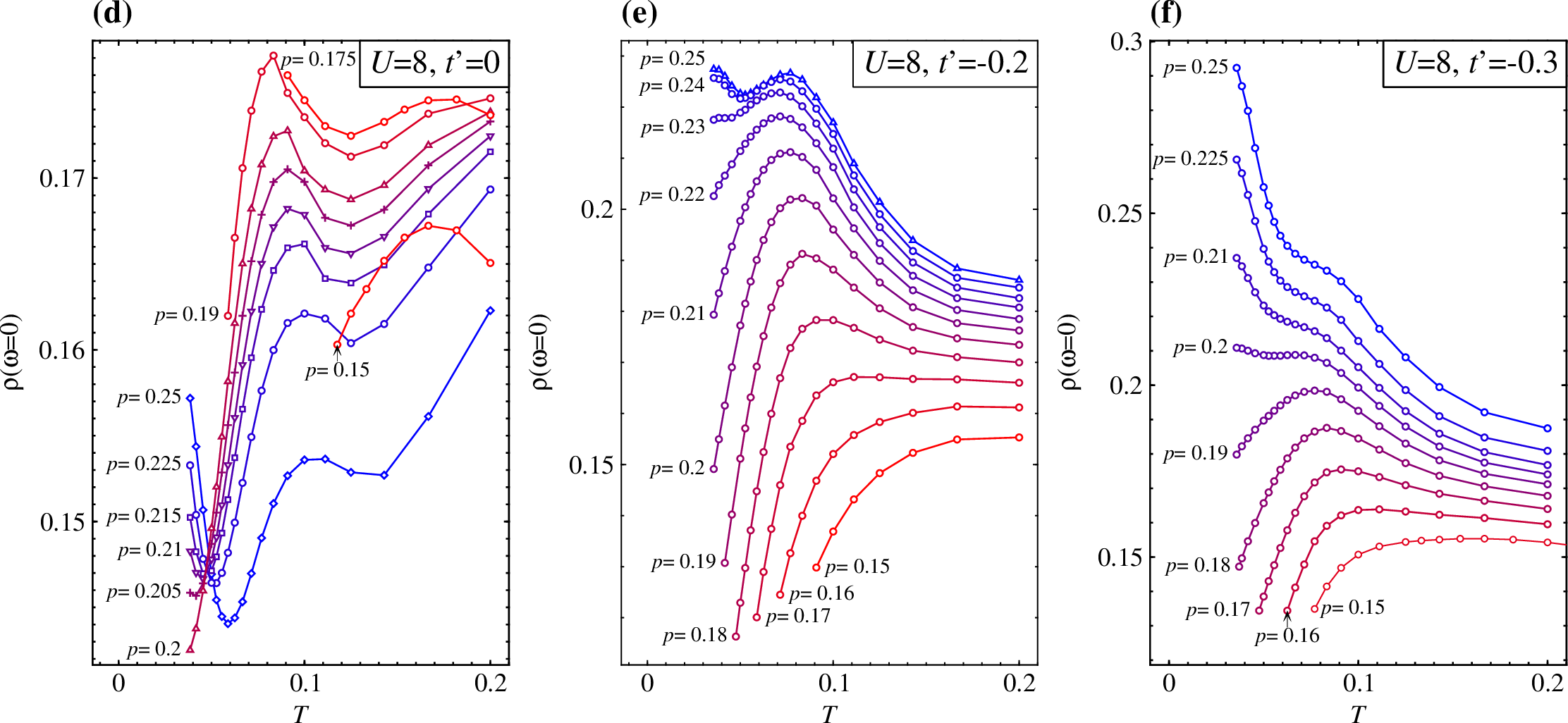}
  \caption{\label{DOST}Upper panels: DOS $\rho(\omega)$ near the Fermi level with fixed $T$ drawn for various values of $p$ for $U=8$ with $t'=0$ at $\beta=26$ (a), $t'=-0.2$ at $\beta=28$ (b) and $t'=-0.3$ at $\beta=28$ (c); in each panel  the green dotted line indicates the Fermi level and the inset shows the corresponding $p$-$T$ phase diagram (for details see the figure caption of Fig.~\ref{pTphase}) together with the spin gap $|\Delta|$ discussed in Sec.~\ref{epsX}. Lower panels: temperature dependence of DOS at the Fermi level $\rho(\omega=0)$ depicted for various values of $p$ for $U=8$ with $t'=0$ (d), $t'=-0.2$ (e) and $t'=-0.3$ (f).}
\end{figure*}
To discuss development of the pseudogap in DOS more in detail, $p$ dependence of DOS near the Fermi level ($-1\le \omega\le 1$)  with fixed $T$ for various values of $p$ are presented in Figs. \ref{DOST}(a)-(c). With $t'=0$, as already mentioned, the main peak is placed at $\omega\sim 0.2$ and shifted toward higher energy upon doping as shown in Fig.~\ref{DOST}(a). A tiny pseudogap can be found in the vicinity of the Fermi level. As will be discussed in Sec.~\ref{Fsurf}, this pseudogap is not caused by the nesting effects of the AFM fluctuations near the X point but those around the $\Gamma$--M symmetry line. In contrast, with $t'=-0.2$, the main structure is located nearby the Fermi level and its prominent modification across $p^*$, which is about five times as large effects as that with $t'=0$, is found in Fig.~\ref{DOST}(b); it has a large pseudogap with the width $\sim 0.1$ at the Fermi level at $p=0.2$ and as the intensity of the peak above the Fermi level is increased upon doping, the pseudogap is shrunken and diminished at $p^*=0.225$, resulting in a single peak at $\omega\sim 0.05$. Similarly, with $t'=-0.3$, the main structure, whose center of the gravity is $\omega\sim -0.1$, has a pseudogap at the Fermi level at $p=0.185$. The pseudogap is reduced and vanished at $p^*=0.2$ upon doping and the main structure is developed into a single peak located slightly below the Fermi level.

The $T$ dependence of DOS at the Fermi level is shown in Fig.~\ref{DOST}(d)-(f). The intensity at the Fermi level is increased as $T$ decreases and then, decreased when the pseudogap is formed. The temperature at which the local maximum of the intensity as a function of $T$ is located can be regarded as a characteristic temperature of the pseudogap, similar to $T^*$, which is defined in terms of the spectral function in Sec. \ref{Overview}. The characteristic temperatures of the pseudogap obtained from the figures are from $T=0.08$ to $0.1$. As seen in the figures, the intensity is increased again below $T\sim 0.05$, if $p$ is larger than a certain value. This occurs between $p=0.2$ and $p=0.205$ for $t'=0$, between $p=0.22$ and $p=0.23$ for $t'=-0.2$ and around $p=0.2$ for $t'=-0.3$. The latter two are, indeed, where the pseudogaps are vanished in Figs.~\ref{DOST}(b) and (c).

These $p$'s at which the pseudogaps in DOSs are vanished coincide with the end points of the pseudogap phase $p^*=0.225$ with $t'=-0.2$ in Fig.~\ref{pTphase}(b) and $p^*=0.2$ with $t'=-0.3$ in Fig.~\ref{pTphase}(c). In the latter, the end point of the PG phase is defined as $p$ where all the pseudogaps of the spectral functions $A_{\bm{k}}(\omega)$ at the Fermi level are vanished. 

\subsection{Doping dependence of spectral functions\label{Spectra}}
In this section doping dependence of spectral function across $p^{*}$ is discussed.
\begin{figure*}
\includegraphics[width=5.5cm]{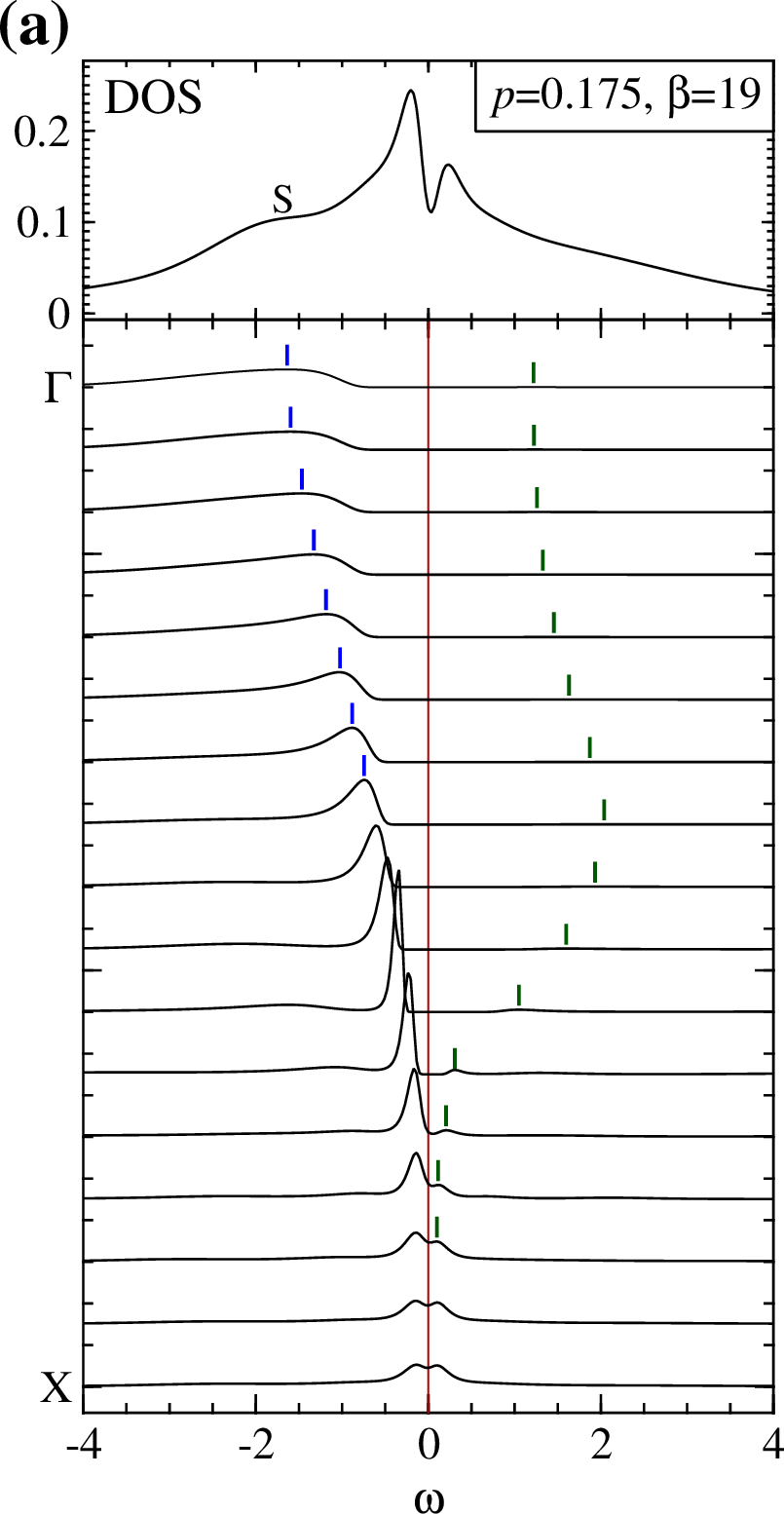}
\includegraphics[width=5.5cm]{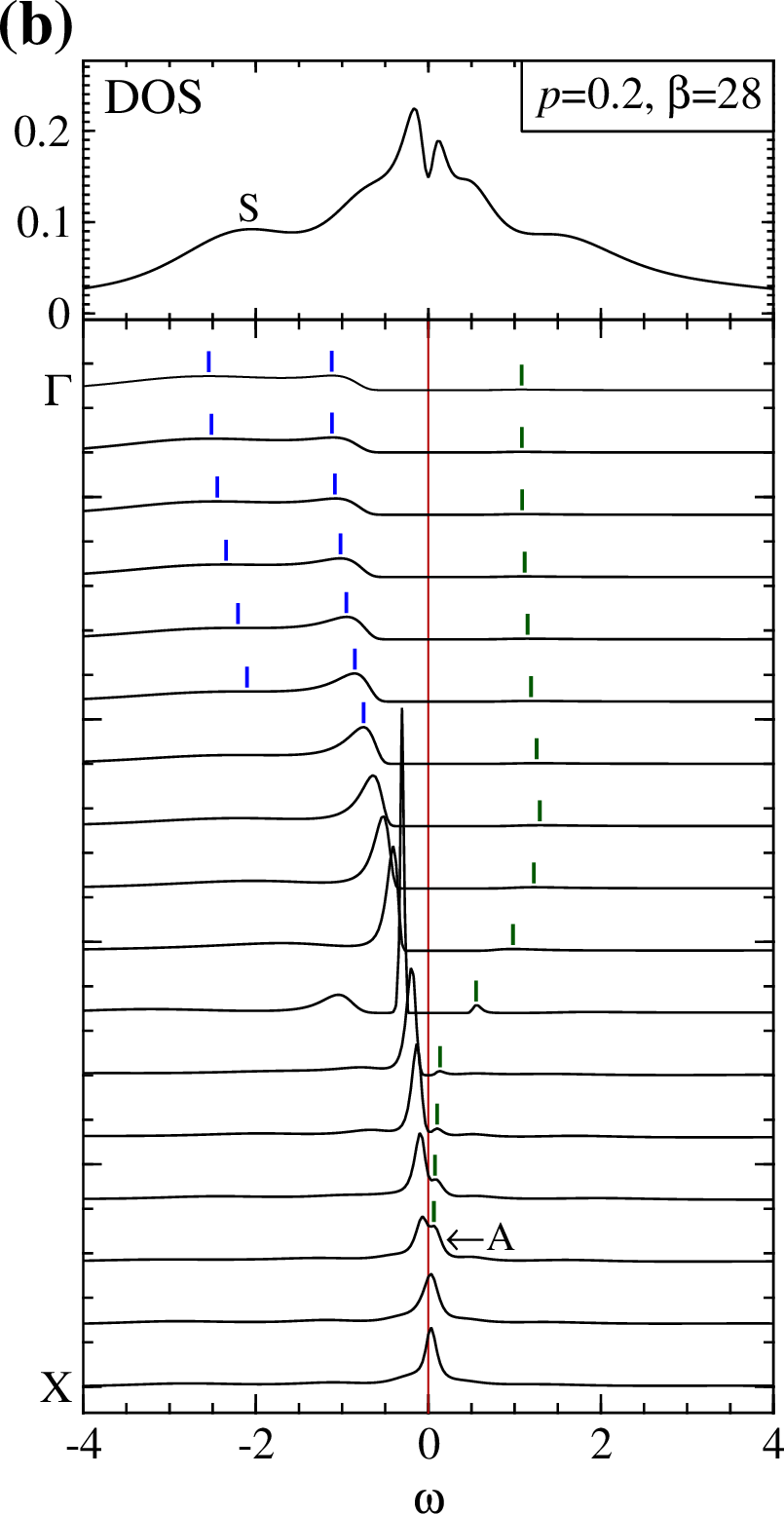}
\includegraphics[width=5.5cm]{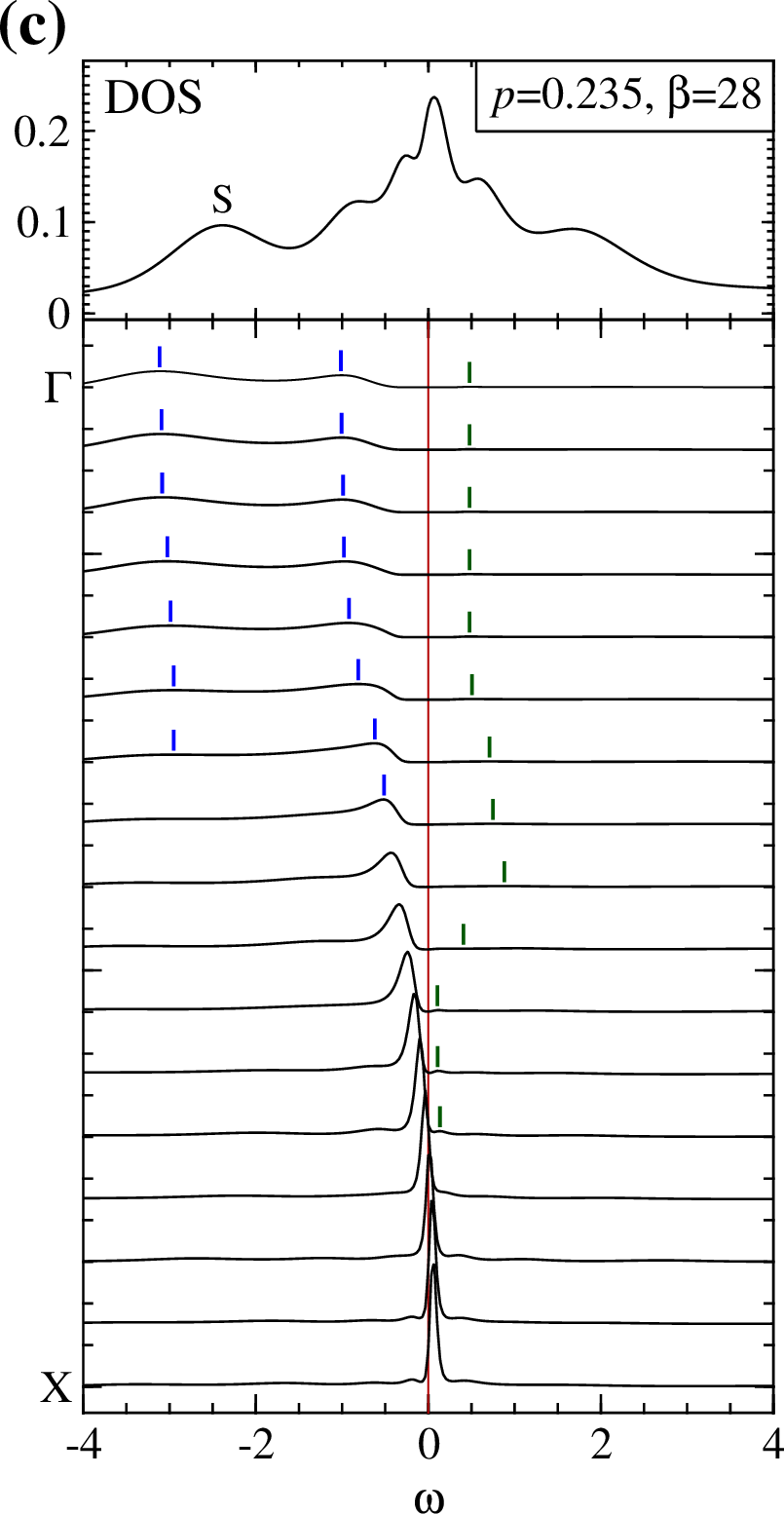}
  \caption{\label{spec_h02}Spectral functions along the $\Gamma$--X symmetry line (the lower panels) and DOSs (the upper panels) for $U=8$ and $t'=-0.2$ with $p=0.175$ at $\beta=19$ (a) and with $p=0.2$ at $\beta=28$ (b) in the PG phase; in (c), those with $p=0.235$ at $\beta=28$ in the SM phase are depicted.}
\end{figure*}
There are distinctive differences in the spectral function particularly nearby the X point between the PG and SM phases. To show this, in Fig.~\ref{spec_h02} the spectral functions along the $\Gamma$--X symmetry line and DOSs for those in the PG phase at $p=0.175$ (a) and $p=0.2$ (b) and in the SM phase at $p=0.235$ (c) are presented.

The most prominent feature of the spectral function in the PG phase is the formation of pseudogap in the vicinity of X point as shown in Fig.~\ref{spec_h02}(a), where double peak structures can be seen at the Fermi level. Because of this peculiar structure, no indication of band dispersion nor the presence of the Fermi surface expected in the vicinity of X point. The formation of the pseudogap in the vicinity of X point is also reflected in the clear depression of the intensity at the Fermi level in DOS presented in the upper panel of Fig.~\ref{spec_h02}(a). As $p$ approaches to $p^*$, the area of the $\bm{k}$-space where pseudogap exists is reduced as can be seen in the lower panel of Fig.~\ref{spec_h02}(b): the double peak structure at the X point is diminished and those nearby the X point along the $\Gamma$--X symmetry line are less clear except for the peak labeled A. In contrast, in the SM phase just above $p^{*}$ in Fig.~\ref{spec_h02}(c) sharp quasiparticle peaks can be found in the vicinity of the X point along the $\Gamma$--X symmetry line indicating restoration of the Fermi surface. The peak at the X point is located at $\omega=0.04$ and the quasiparticle band in the vicinity of the Fermi level is extremely flatten because of the reduction of $z_{\bm{k}}$ due to the strong AFM fluctuations, in particular, around the X point, where $z_{\bm{k}}=0.34$ at $T=1/28$.

In Sec.~\ref{DOS}, it has been found that upon doping the structures of DOS changed across $p_{\rm MS}\sim 0.2$, where the main peak of the lower Hubbard band is sharpened and clear satellite structure appears (see Fig.~\ref{DOSp}). These variation of the structures of DOS can be linked to the change in the structures in the spectral function near the $\Gamma$ point (locations of them are indicated by the blue vertical lines in the lower panels). In Fig.~\ref{spec_h02}(a) at $p=0.175$ the broad structure peaked at $\omega = -1.7$ is found at the $\Gamma$ point and toward the X point along the $\Gamma$--X symmetry line the structure is sharpened and connected to the peaks located near the X point at the Fermi level. In Fig.~\ref{spec_h02}(b) at $p=0.2$, however, this broad structure is separated into two peaks: one is located at $\omega= -1.2$ at the $\Gamma$ point and is connected to the peaks near the X point at the Fermi level along the $\Gamma$--X symmetry line and the other is faint structure located at $\omega= -2.6$ at the $\Gamma$ point and is faded away toward the X point along the $\Gamma$--X symmetry line. Accordingly, the shoulder structure labeled S of DOS with $p=0.175$ in the upper panel of Fig.~\ref{spec_h02}(a) is developed into the satellite structure with $p=0.2$ as in Fig.~\ref{spec_h02}(b). The energy separation of these two structures in the spectral function at the $\Gamma$ point is further increased in the SM phase in Fig.~\ref{spec_h02}(c). The peak located at $\omega =-1.0$ at the $\Gamma$ point now belongs to the narrow quasiparticle band.

In addition to these structures, faint bands are found above the Fermi level in the lower panels in Fig.~\ref{spec_h02} (locations of them are indicated by the green vertical lines). These faint bands are considered to be the so-called shadow bands, which originate from the short-range AFM order and appear as if there were the Brillouin zone folding caused by the long-range AFM order. Indeed, these faint bands are connected to the upper peaks of the double peaks near the X point, e.g., peak A in Fig.~\ref{spec_h02}(b), and the pseudogap structure can be regarded as the results of hybridization of the bands due to the Brillouin zone folding, although this zone folding should be instantaneous. Hence, it is conceivable that electrons near the X point are not mobile without well-defined quasiparticle peaks. Instead, they can tunnel among the fragmented Fermi surfaces, i.e., the Fermi arcs, in the PG phase as shown in Figs.~\ref{spec_h02}(a) and (b).

Since the short-range AFM order is incommensurate in the range of $p$ here we discuss, e.g., $\bm{Q}=(\pi,\,\pm\delta)$, the intensity of these shadow bands is weak and would not be clearly seen in experiments. However, shadow bands have been observed with commensurate $\bm{Q}=(\pi,\,\pi)$ Brillouin zone folding at low doping ($x\le 1/8$) in the ARPES experiments on La$_{2-x}$Sr$_{x}$CuO$_4$ \cite{ERazzoli2010}. This is consistent with the present study, since the AFM fluctuations are commensurate $\bm{Q}=(\pi,\,\pi)$ in low doping (see the $T_{\rm C-IC}$ line in Fig.~\ref{pTphase}(b)).

\begin{figure*}
\includegraphics[width=5.5cm]{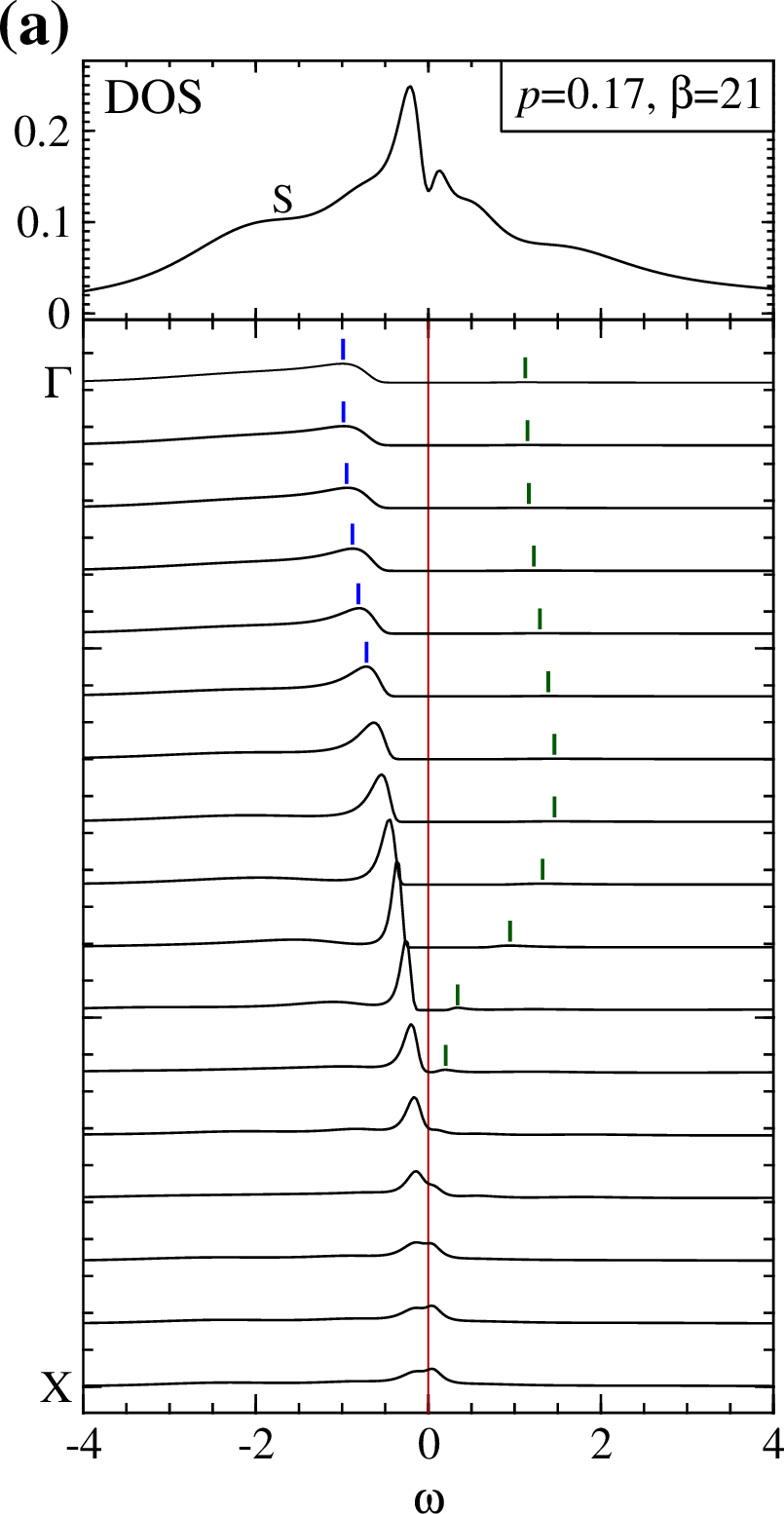}
\includegraphics[width=5.5cm]{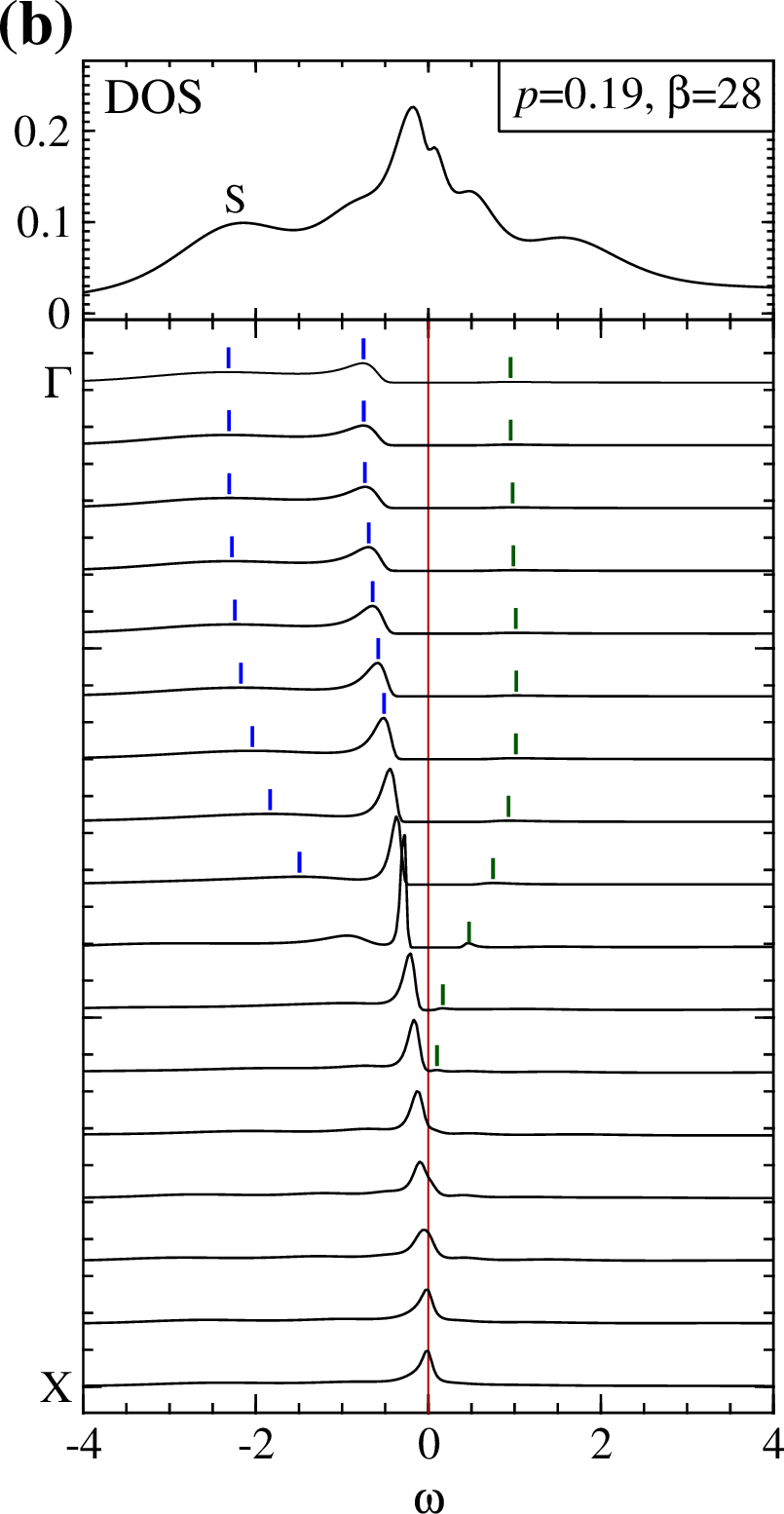}
\includegraphics[width=5.5cm]{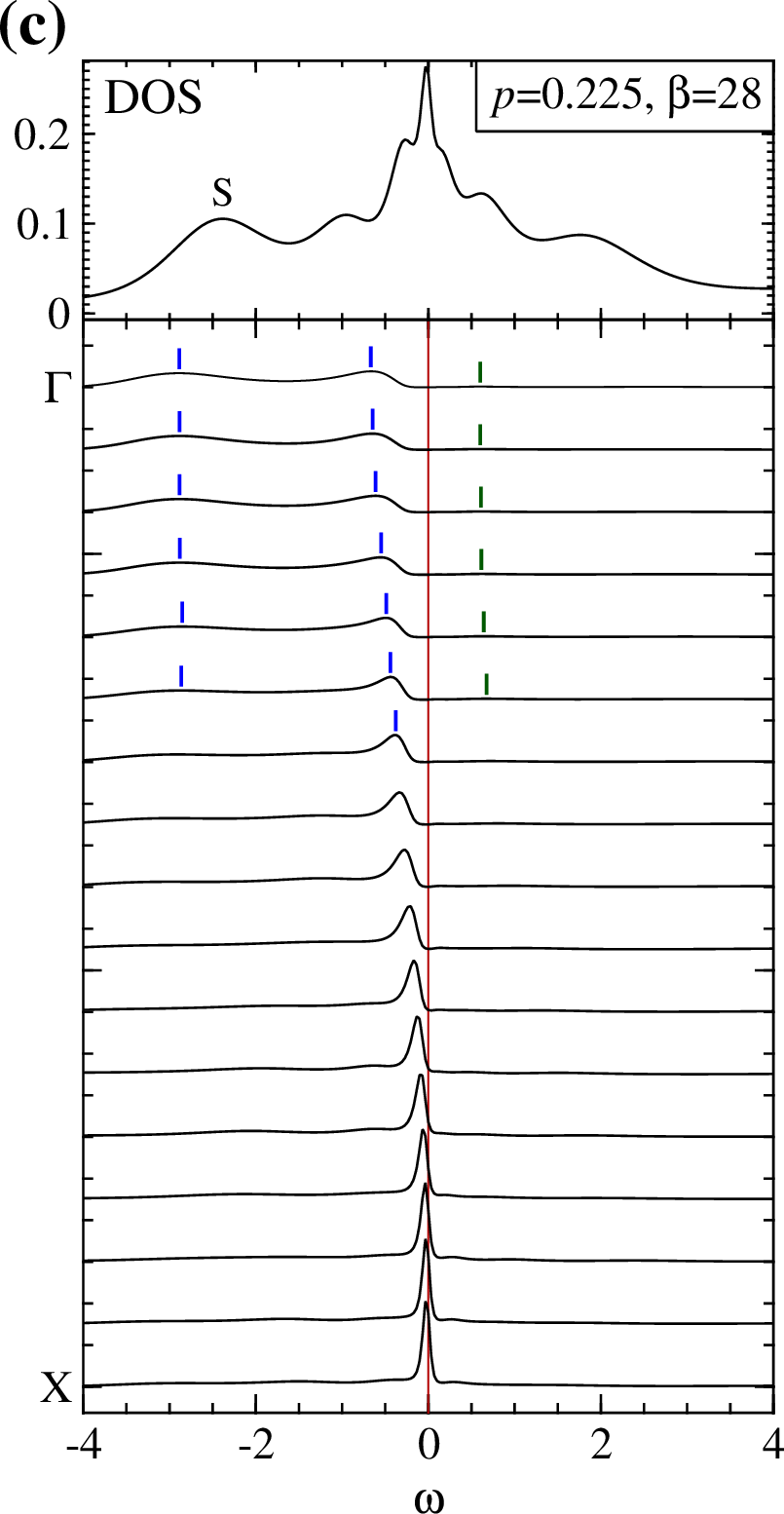}
  \caption{\label{spec_h03}Spectral functions along the $\Gamma$--X symmetry lines (the lower panels) and DOSs (the upper panels) for $U=8$ and $t'=-0.3$ with $p=0.17$ at $\beta=21$ (a) and with $p=0.19$ at $\beta=28$ (b) in the PG phase; in (c), those with $p=0.225$ at $\beta=28$ (c) in the SM phase is shown.}
\end{figure*}
Figure~\ref{spec_h03} shows the spectral functions along the $\Gamma$--X symmetry line and DOSs for $U=8$ and $t'=-0.3$ in the PG phase at $p=0.17$ (a) and $p=0.19$ (b) and in the SM phase at $p=0.225$ (c). The $p$ dependence of the spectral functions are similar to those for $t'=-0.2$. However, since $p_{\rm VHS}=0.25$ is located far above $p_{\rm MS}\sim 0.2$, the formation of the pseudogap is less prominent compared to those for $t'=-0.2$. The clear pseudogap structures can be found near the X point in the vicinity of the Fermi level with $p=0.17$ located deep inside of the PG phase in Fig.~\ref{spec_h03}(a). These pseudogap structures nearby the X point are smeared and less clear with $p=0.19$ in the PG phase near $p^*=0.2$ in Fig.~\ref{spec_h03}(b). All the pseudogap structures near the X point are reduced to the sharp single peaks with $p=0.225$ in the SM phase in Fig.~\ref{spec_h03}(c), resulting in the restoration of the whole Fermi surface. The quasiparticle band near the Fermi level is flatten because of the reduction of $z_{\bm{k}}$ caused by the strong AFM fluctuations. In particular, the single peaks near the X point located slightly below the Fermi level within their life-time broadening width form extremely flat band in low temperatures, where $z_{\bm{k}}=0.33$ at $T=1/28$ at the X point. As will be discussed in the next section, this flatting of the band near the X point is considered to be the realization of the Fermi condensation, which was also found in the LDFA study with the Hubbard model on the triangular lattice \cite{DYuin2014}.

\subsection{Doping and temperature dependence of the VHS point and its relation to spin fluctuation\label{epsX}}
\begin{figure*}
\includegraphics[width=17cm]{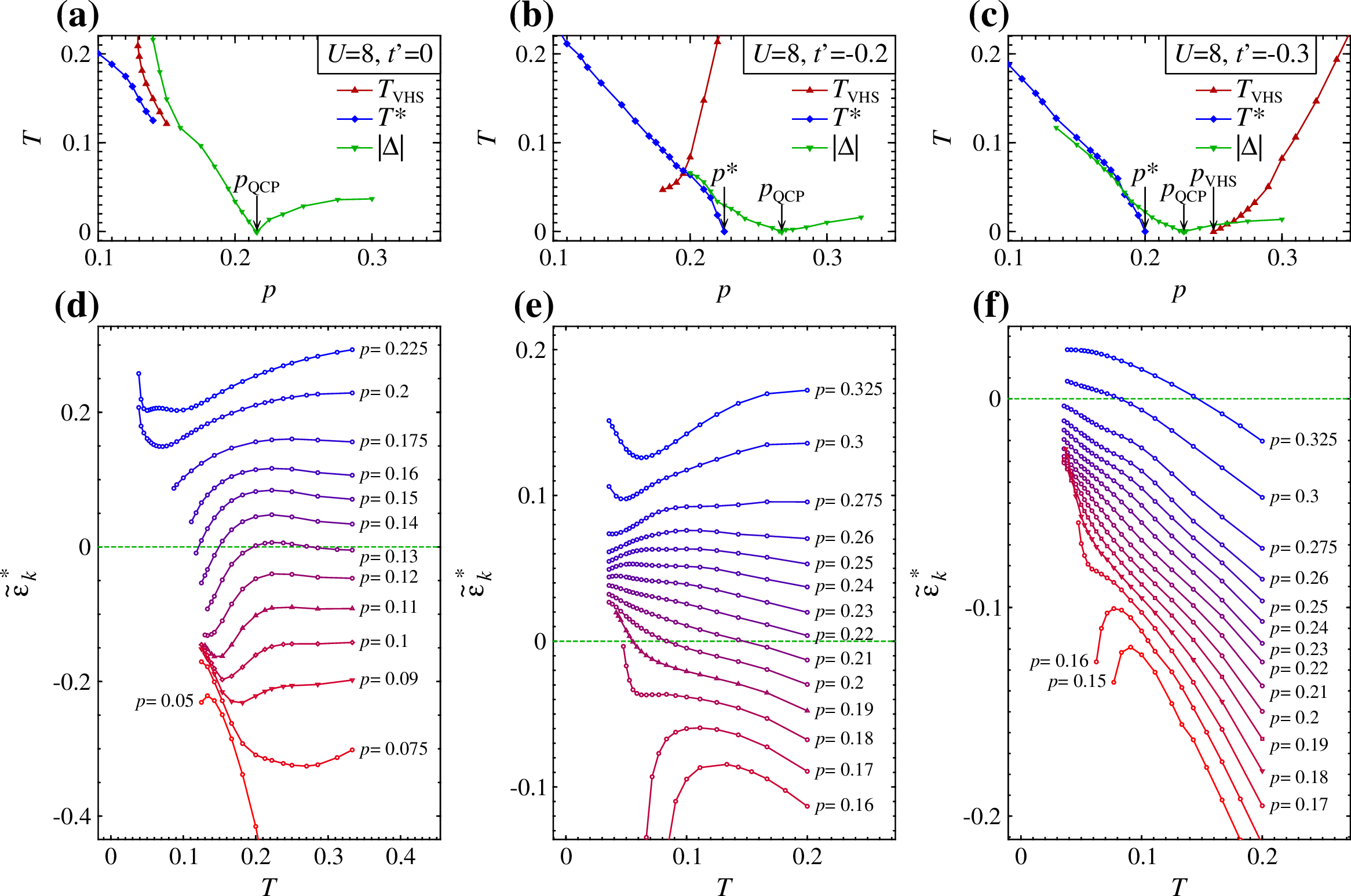}
\caption{\label{eps0T1}Upper panels: The temperature of pseudogap formation $T^*$ (the blue lines), the VHS point at the Fermi level $T_{\rm VHS}$ (the red lines) and the spin gap $|\Delta|$ (the green lines) as a function of $p$ for $U=8$ with $t'=0$ (a), $t'=-0.2$ (b) and $t'=-0.3$ (c). Lower panels: Temperature dependence of the VHS point at the X point $\bm{k}=(\pi,0)$ $\tilde{\varepsilon}^*_{\bm{X}}$ depicted with various values of $p$ for $U=8$ with $t'=0$ (d), $t'=-0.2$ (e) and $t'=-0.3$ (f); in each panel, the green dotted line indicates the Fermi level.}
\end{figure*}
Because of the presence of VHS, the effective one-body states nearby the X point in the Brillouin zone forms intense DOS nearby the VHS point as we have already seen in Sec.~\ref{DOS}. When the VHS point is in the vicinity of the Fermi level, this enhances scattering of electrons caused by the AFM fluctuations. In particular, if the Fermi surface satisfies nesting condition for the wave vector $\bm{Q}$ of the dominant AFM fluctuations and with $p\lesssim p_{\rm MS}$, the pseudogap is formed as the AFM correlation length grows in low temperatures ($T\lesssim t^2/U$), which is reminiscent of the gap formation by the Brillouin zone folding due to the long-range AFM ordering. On the other hand, with $p\gtrsim p_{\rm MS}$ and also $p\lesssim p_{\rm MS}$ at high temperatures ($T\gtrsim T^*$), the pseudogap formation does not occur and, instead, the effective mass is increased, i.e., the quasiparticle residue $z_{\bm{k}}$ is reduced, because of the large scattering of electrons due to the AFM fluctuations. The reduction of $z_{\bm{k}}$ causes further enhancement of the AFM fluctuations due to the flattening of the band and this leads to the pinning of $\tilde{\varepsilon}^*_{\bm{X}}$ near the Fermi level.

To examine the relationship among the pseudogap formation, the VHS point and the AFM fluctuation more closely, in Figs.~\ref{eps0T1}(a)-(c), the $T^*$ and $T_{\rm VHS}$ lines on the $p$-$T$ phase diagram are depicted along with the spin gap $|\Delta|$ as a function of $p$ near $p_{\rm QCP}$, which is obtained by the least squares fit assuming the static spin susceptibility as
\begin{align}\label{chi_static}
  \chi^{\rm sp}(\bm{q}=\bm{Q},\omega=0)\approx\frac{C}{T}\exp(\Delta/T),
\end{align}
where $\bm{Q}$ is the wave vector $\bm{q}$ at which $\chi^{\rm sp}(\bm{q},\omega=0)$ takes the maximum value. As seen in Fig.~\ref{eps0T1}(a), for $U=8$ and $t'=0$, $p_{\rm QCP}=0.215$ is located near $p_{\rm MS}\approx 0.2$. The fan-shaped area of the quantum critical regime above $|\Delta|$ is consistent with what is expected in the ordinary Hertz--Millis--Moriya theory. In this case, the VHS point is located far above the Fermi level ($\tilde{\varepsilon}^*_{\bm{X}}>0.1$) around $p\sim p_{\rm MS}$ as shown in Fig.~\ref{eps0T1}(d). On the other hand, for $t'=-0.2$ and $t'=-0.3$ in Figs.~\ref{eps0T1}(e) and (f), the VHS point lies near the Fermi level around $p\sim p_{\rm MS}$. The positions of $p_{\rm QCP}$ shift toward the higher hole concentrations and the quantum critical regime extends to both lower temperatures and wider ranges of $p$ as seen in Figs.~\ref{eps0T1}(b) and (c) compared to Fig.~\ref{eps0T1}(a). It is also found in Figs.~\ref{eps0T1}(b) and (c) that the upper boundary of the PG phase $T^*$ approximately coincides with the $|\Delta|$ line with $p<p^*$. This indicates that the PG phase is magnetically in the renormalized classical regime and $T^*$ corresponds to the crossover temperature from the quantum critical to renormalized classical regimes within $p<p^*$. A detailed discussion of the pseudogap formation of individual spectral functions near the X point in the vicinity of $p^*$ is presented in Appendix~\ref{pGap}.

For $t'=0$, enhancement of the scattering due to the presence of VHS is particularly significant with $0<p<0.14$ and the pseudogap phase appears below $T^*$. However, the $T^*$ line is ended at $p=0.14$ just above $T_{\rm PS}$ in the $p$--$T$ phase diagram in Fig.~\ref{pTphase}(a). As already mentioned in Sec.~\ref{DOS}, the sharp peak at $\tilde{\varepsilon}^*_{\bm{X}}\sim -0.14$ and deep pseudogap appear $p< 0.14$ in Fig.~\ref{DOSp}(a) and accordingly, as shown in Fig.~\ref{eps0T1}(d), $\tilde{\varepsilon}^*_{\bm{X}}$ approaches $\tilde{\varepsilon}^*_{\bm{X}}=-0.14$ at low temperatures within $0.09 < p < 0.19$ showing strong influence of VHS. Further doping ($p>0.19$), the influence of VHS is reduced and the $\tilde{\varepsilon}^*_{\bm{X}}$ curve is changed from downward to upward at the low temperatures between $p=0.175$ and $p=0.2$ as seen in the figure. As will be discussed in Sec.~\ref{Fsurf}, this is because whereas the nesting condition of the AFM fluctuations is satisfied on the whole Fermi surface with $p<0.19$, the nesting vector $\bm{Q}$ is changed with $p>0.19$ so to be more favorable toward the nesting limited to the straight portions of the Fermi surface near the $\Gamma$--M symmetry line as the AFM fluctuations around the X point reduced with increasing $p$.

For $t'=-0.2$, $\tilde{\varepsilon}^*_{\bm{X}}$ is located nearby the Fermi level around $p_{\rm MS}\sim 0.2$ at low temperatures (see Fig.~\ref{eps0T1}(e)). As has been discussed in Sec.~\ref{DOS}, this strongly enhances the AFM fluctuations around the X point and stabilize the PG phase compared to that with $t'=0$ and $t'=-0.3$: $T^*$ increases and the end point of the PG phase is located at $p^*=0.225$. Because of the formation of the pseudogap in the spectral function at the X point with $p\le 0.195$ in Fig.~\ref{spec_h02}(a), the sharp upward ($p>0.175$) or downward ($p<0.175$) shift of $\tilde{\varepsilon}^*_{\bm{X}}$ in low temperatures can be seen in Fig.~\ref{eps0T1}(e), depending on whether $\tilde{\varepsilon}^*_{\bm{X}}$ placed at the higher peak or the lower peak of the double peak structure of the spectral function, respectively. On the other hand, in the SM phase ($p>p^*=0.225$), because of strong enhancement of the AFM fluctuations, $z_{\bm{k}}$ is reduced and $\tilde{\varepsilon}^*_{\bm{X}}$ is shifted downward.
Together with the upward shift in the PG phase, $\tilde{\varepsilon}^*_{\bm{X}}$ is converged to $\tilde{\varepsilon}^*_{\bm{X}}=0.04$ as $T$ decreases within the range $0.195\le p\le 0.265$. As will be discussed in Sec.~\ref{Fsurf}, the Fermi surface in the SM phase is nearly unchanged within the range $0.225\le p\le p_{\rm QCP}=0.265$ and the nesting condition of the AFM fluctuations is fulfilled almost everywhere on Fermi surface. The upward shift of $\tilde{\varepsilon}^*_{\bm{X}}$ occurs at the low temperatures with $p>p_{\rm QCP}$ is due to the variation of the nesting vector $\bm{Q}$ more favorable toward the straight portions of the Fermi surface around the $\Gamma$--M symmetry line as the AFM fluctuations decrease.

For $t'=-0.3$, $\tilde{\varepsilon}^*_{\bm{X}}$ is located at the Fermi level at $p_{\rm VHS}=0.25$ at $T=0$ (see Fig.~\ref{eps0T1}(f)), where the Fermi-surface topology is changed from hole-like to electron-like with increasing $p$, i.e., the Lifshitz transition. Similar to that of $t'=-0.2$, $\tilde{\varepsilon}^*_{\bm{X}}$ is converged to the Fermi level with $T\to 0$ within the range $0.175<p<0.25$ as can be seen in Fig.~\ref{eps0T1}(f). In this case, the flatting of the band around the X point is occurred at the Fermi level in the SM phase ($p > p^*=0.2$) with $T\to 0$, i.e., the Fermi condensation, as we have already seen in Fig.~\ref{spec_h03}(c). This is also evident from the steep increase of $\rho(\omega =0)$ with $p > p^*=0.2$ as the main peak approaches to the Fermi level with decreasing $T$ indicated in Fig.~\ref{DOST}(f). Similar to that of $t'=-0.2$, the Fermi surface in the SM phase is nearly unchanged within the range $0.2<p<0.25$ and the nesting condition of the AFM fluctuations is fulfilled almost everywhere on Fermi surface.

\subsection{Doping dependence of the Fermi surface and AFM nesting condition\label{Fsurf}}
\begin{figure*}
\includegraphics[width=5.5cm]{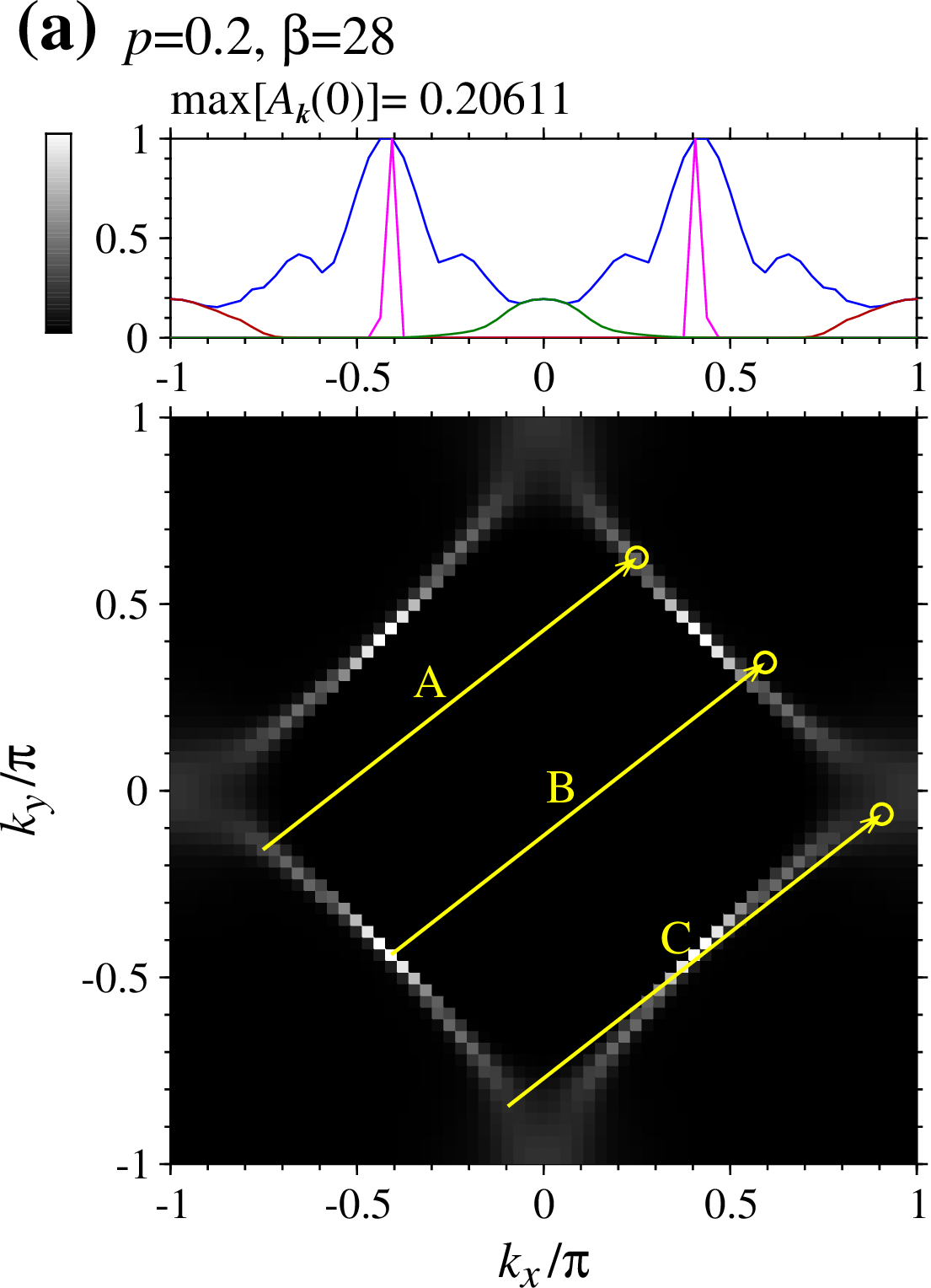}~~
\includegraphics[width=5.5cm]{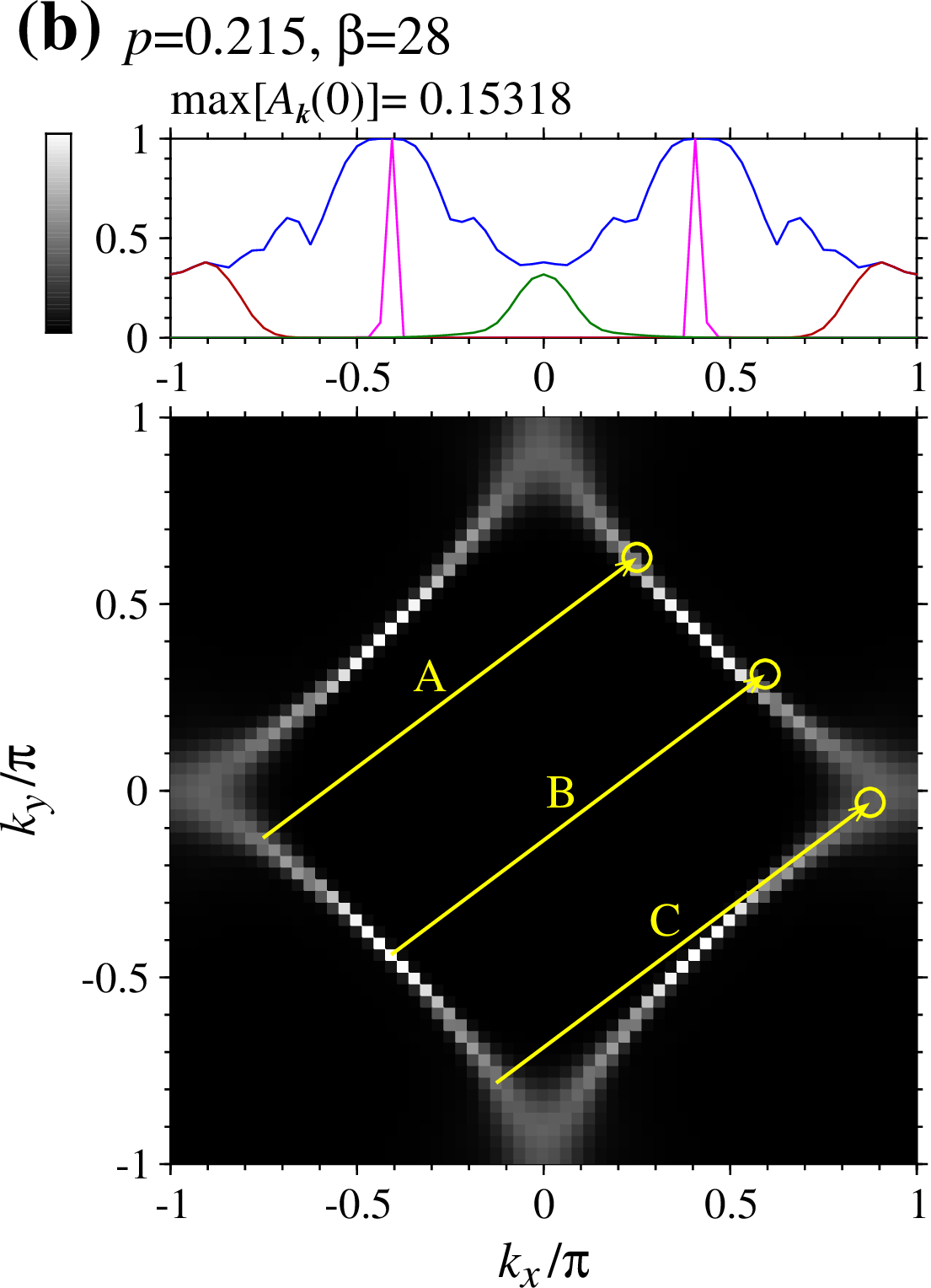}~~
\includegraphics[width=5.5cm]{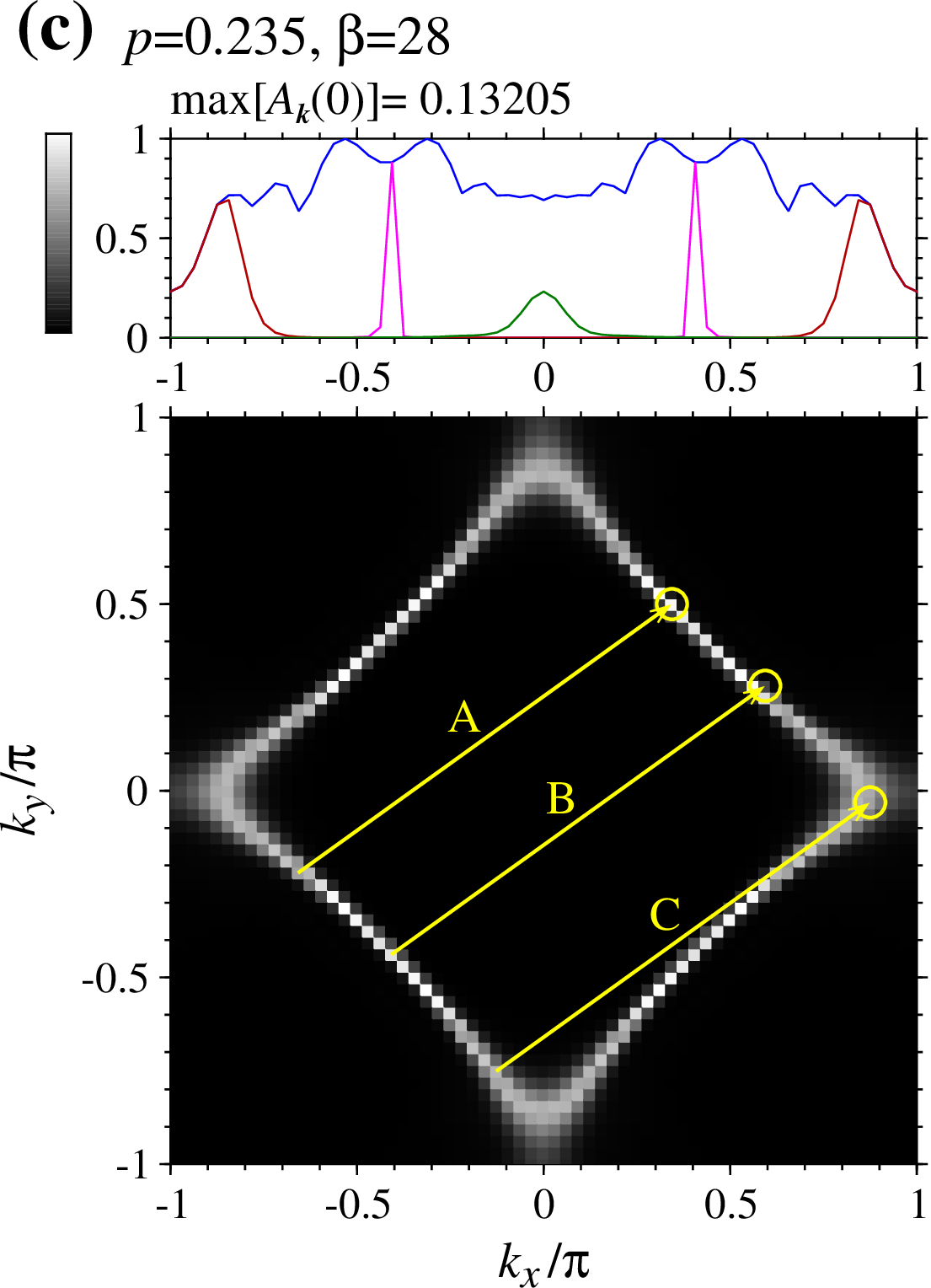}\\
\includegraphics[width=5.5cm]{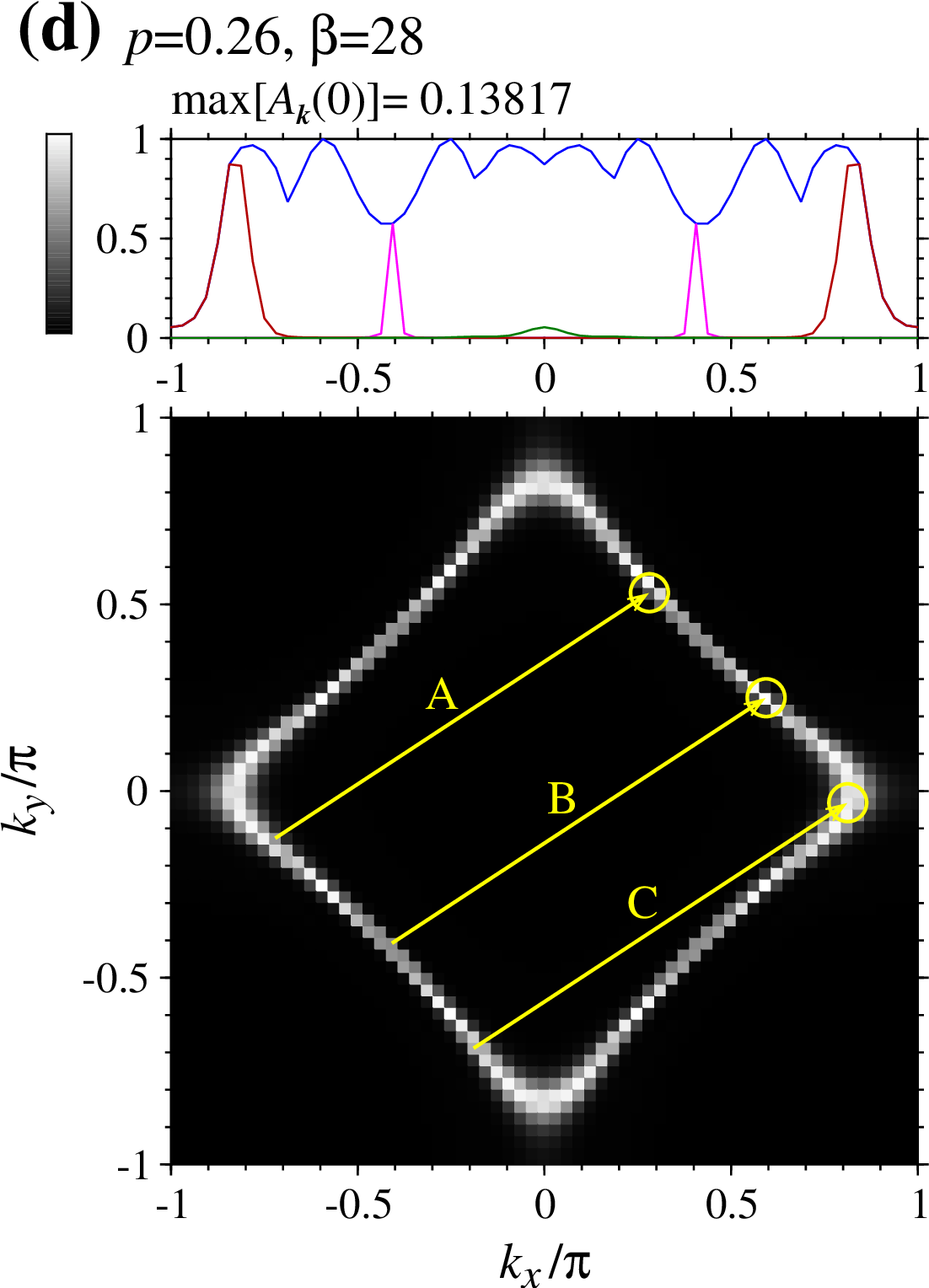}~~
\includegraphics[width=5.5cm]{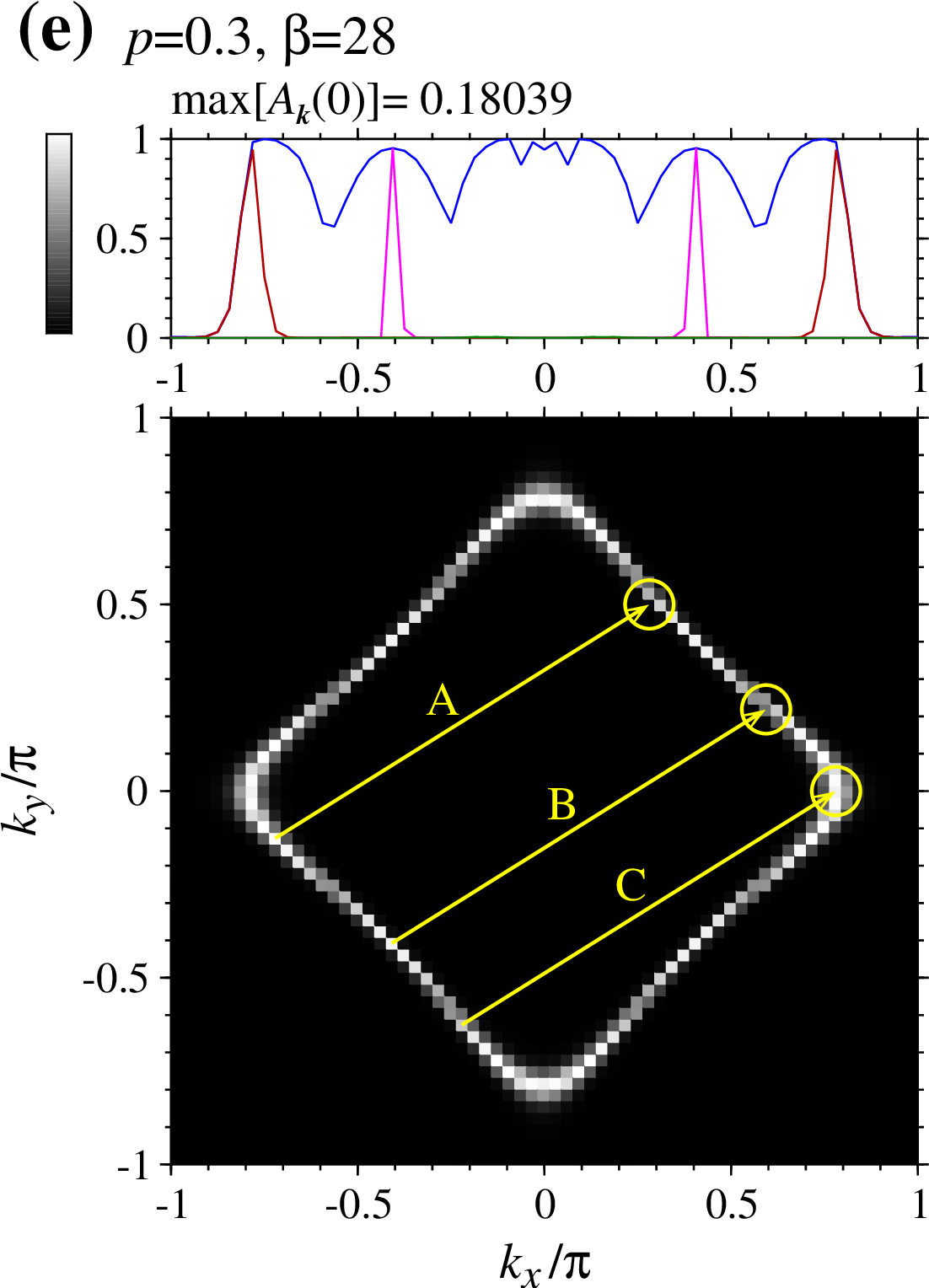}~~
\includegraphics[width=5.5cm]{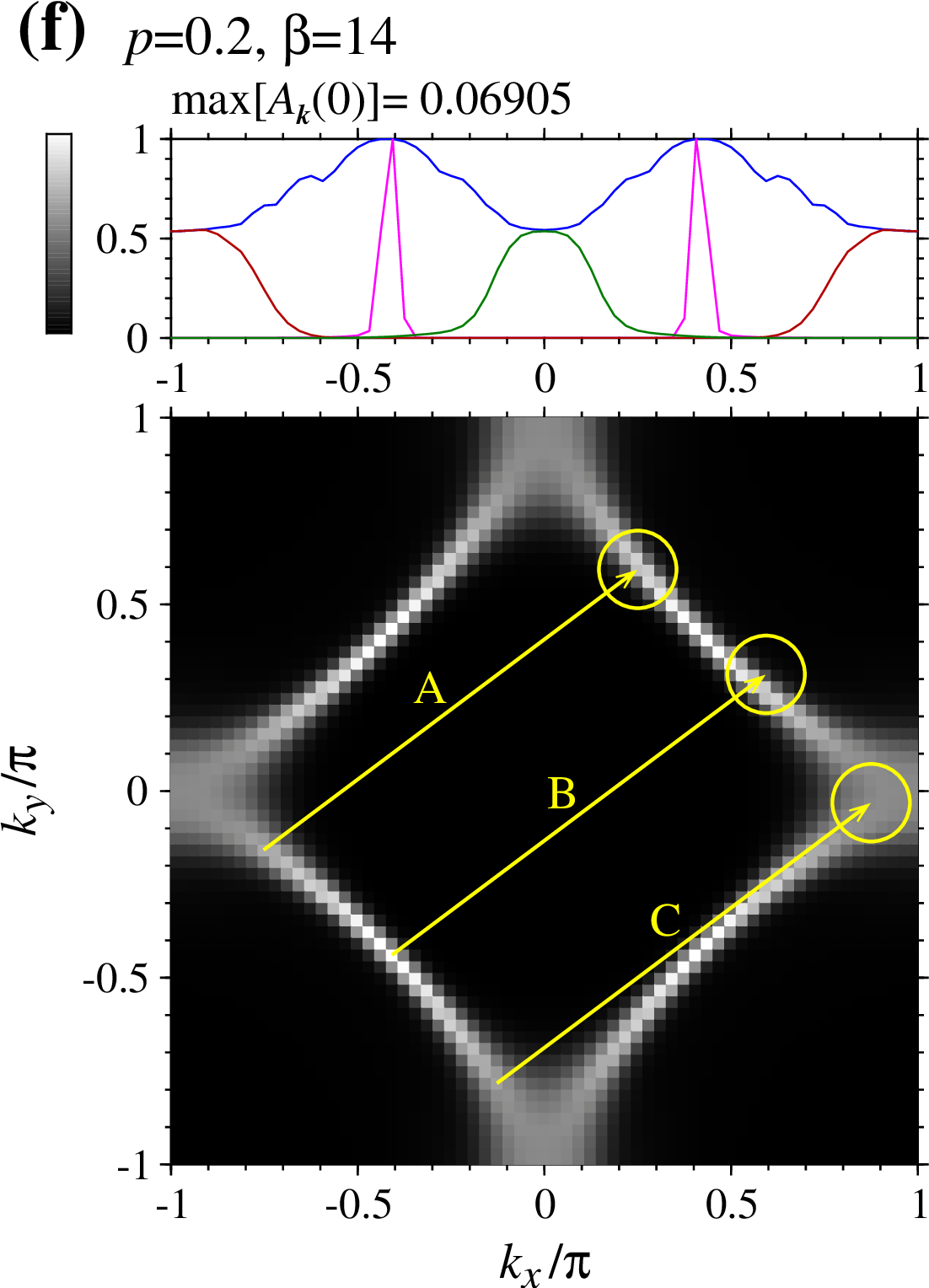}
  \caption{\label{FSh02}Intensities of the spectral functions at the Fermi level $A_{\bm{k}}(\omega=0)$ for $U=8$ and $t'=-0.2$ at $\beta=28$ with $p=0.2$ (a) and $p=0.215$ (b) in the PG phase and with $p=0.235$ (c), $p=0.26$ (d) and $p=0.3$ (e) in the SM phase; in (f), those with $p=0.2$ at $\beta=14$ above $T^*$ in the SM phase are presented. In the lower panel in each of figures (a)--(f), the intensities in the first Brillouin zone are represented by gray scale; to gauge the nesting condition, the wave vector $\bm{Q}=(\pi,\,\delta)$ placed at three different locations of the Fermi surface are represented by the arrows labeled A, B and C; in the upper panel the intensities along the X--$\Gamma$--X,  M--Y--M and M--$\Gamma$--M symmetry lines, i.e., $A_{(k_x,0)}(0)$, $A_{(k_x,\pi)}(0)$ and $A_{(k_x,k_x)}(0)$, are depicted as functions of $k_x$ by the red, green and magenta lines, respectively and the maximum value of $A_{\bm{k}}(0)$ in each fixed value of $k_x$, i.e., $A_{\rm max}(k_x)\equiv\displaystyle\max_{k_y}\{A_{(k_x,k_y)}(0)\}$ is denoted by the blue line.}
\end{figure*}
To discuss the variation of the Fermi surface between the PG and SM phases in relation to the nesting condition of the AFM fluctuations, the intensities at the Fermi level of the spectral functions for $U=8$ and $t'=-0.2$ with various values of $p$ and $T$ are shown in Fig.~\ref{FSh02}. In the PG phase the pseudogaps are formed in the spectral functions around the X and Y points as we have found in Sec.~\ref{Spectra} (see Figs.~\ref{spec_h02}(a) and (b)), where, contrary to the FL theory, the electron scattering rate $\gamma_{\bm{k}}\equiv -{\rm Im}\,\Sigma_{\bm{k}}(\omega)$ increases as temperature decreases and no well-defined quasiparticles exist as will be discussed in Sec.~\ref{SelfE}. As a result, the intensities around the X and Y points are strongly reduced and no clear Fermi surface found (see Figs.~\ref{FSh02}(a) and (b)). This is particularly evident in the upper panel of Fig.~\ref{FSh02}(a), where the intensities at the X and Y points is about 20\% of the maxima located around $k_x/\pi=\pm 0.42$ (the blue line). This bell-shaped profile of the intensity, whose maximum intensity located on the $\Gamma$-M symmetry line, and the absence of a distinctive Fermi surface around the X and Y points are the hallmarks of the PG phase. This bell-shaped profile of the intensity in the PG phase has been observed in the ARPES experiments of lightly doped La$_{2-x}$Sr$_x$CuO$_4$ \cite{TYoshida2003}.

To understand the relation between the partial distraction of the Fermi surface in the PG phase and the AFM fluctuation, the wave vector $\bm{Q}=(\pi,\,\delta)$ where the static spin susceptivity $\chi^{\rm sp}(\bm{q},\omega =0)$ takes its maximum value is indicated as the yellow arrows labeled A, B and C in the lower panel. In addition, the width of the magnetic spot at $\bm{Q}$ $\Delta Q$ in the half width at half maximum (HWHM), which will be discussed in detail in Sec.~\ref{chi}, is represented by the circle with radius $\Delta Q/\pi=0.027$ at each of the arrow tip (see the lower panel of Fig.~\ref{FSh02}(a)).
When the arrow end is placed at $\bm{k}$ and a part of the Fermi surface is inside the circle centered at $\bm{k}+\bm{Q}$, the nesting condition is satisfied and the formation of the pseudogap is expected at $\bm{k}$. Indeed, each of arrows A and C connects wave vectors $\bm{k}$ nearby the X point to $\bm{k}+\bm{Q}$ on the Fermi surface, resulting in the formation of the pseudogap and the reduction of the intensity. On the other hand, any of the wave vectors $\bm{k}$ on the four bright segments of the Fermi surface, the so-called Fermi arcs, cannot be connected to those with $\bm{k}+\bm{Q}$ on the Fermi surface by the arrows such as B and thus the spectral function is single peaked at the Fermi level without AFM nesting effects.

In contrast, in the SM phase at the vicinity of $p^{*}$ shown in Fig.~\ref{FSh02}(c), although the Fermi surface around the X and Y points is still blurred as seen in the intensity mapping in the lower panel, the broad edge of the Fermi surface around $k_x/\pi =\pm 0.85$ (the red line) and the rather flatten intensity with broad peaks around $k_x/\pi=\pm 0.45$ of the Fermi surface (the blue line) in the upper panel clearly indicates that the formation of the electron-like Fermi surface occurs at $p=0.235$. In addition, the blurred corners of the square-shaped Fermi surface are connected to the nearby X or Y point by the ridge structures with weak intensity, which can be seen as the tails of the Fermi surface edge with finite intensity at $k_x/\pi=\pm 1$ (the red line) and the finite intensity around the Y point (the green line) in the upper panel in Fig.~\ref{FSh02}(c). These ridge structures originate from the tails of the quasiparticle peaks on the extremely flatten band just above the Fermi level $\omega\sim 0.04$ in Fig.~\ref{spec_h02}(c). 
Much clear electron-like Fermi surface can be seen deep inside the SM phase at $p=0.26$ in the lower panel of Fig.~\ref{FSh02}(d) as well as the sharp Fermi surface edge at $k_x/\pi =\pm 0.85$ and the flatten intensity of the Fermi surface can be seen in the top panel. The intensity of the ridge structure is also reduced as the AFM fluctuations nearby the X and Y points is decreased with increasing $p$ and vanished at $p_{\rm QCP}=0.265$, where the nesting vector $\bm{Q}$ is changed to be more favorable toward the straight portions of the Fermi surface around the $\Gamma$--M symmetry line and as a result, the flat band nearby the X point departs from the Fermi level with $p>p_{\rm QCP}$ (see Fig.~\ref{eps0T1}(e)). The same distinctive changes in the intensity at the Fermi level of the spectral function between the PG and SM phases are also found for $U=6$ and $t'=-0.2$ at $p^{*}=0.215$ (not indicated).  

Contrary to the PG phase, in the SM phase the correlation length of the AFM fluctuations is short because of its extended quantum critical behavior as will be discussed in Sec.~\ref{chi} and thus $\Delta Q$ of the magnetic spot is larger: $\Delta Q/\pi=0.041$ and $0.051$ with $p=0.235$ and $p=0.26$, respectively. Hence, the nesting condition is not strict as compared to the PG phase and fulfilled everywhere on the Fermi surface in the SM phase as in Figs.~\ref{FSh02}(c) and (d). Because of this feature, the electron scattering rate $\gamma_{\bm{k}}\equiv -{\rm Im}\,\Sigma_{\bm{k}}(\omega=0)$ is nearly uniform and $T$-linear on the whole Fermi surface in the SM phase, as will be discussed in Sec.~\ref{SelfE}. This almost perfect nesting is caused by the presence of the VHS point $\tilde{\varepsilon}_{\bm{X}}^*$ in the vicinity of the Fermi level: the nearly square Fermi surface, where flat band near the X and Y points enhances the AFM fluctuations. Since the VHS point stays at $\tilde{\varepsilon}_{\bm{X}}^*\sim 0.04$ at low temperatures within $0.225\le p\le 0.265$ (see Fig.~\ref{eps0T1}(e)), the shape of the Fermi surface is almost unchanged with the extended quantum critical AFM fluctuations within this range of $p$, which is the origin of the marginal Fermi liquid property of the SM phase as will be detailed in Sec.~\ref{Discussion}. The quantum critical behavior of the Fermi surface can be found also with $p=0.2$ at $T=1/14$ just above $T^*$ as in Fig.~\ref{FSh02}(f). Because of large $\Delta Q/\pi=0.104$, the nesting condition is fulfilled everywhere on the Fermi surface. Since the temperature $T=1/14$ is near $T_{\rm VHS}$ (see Fig.~\ref{pTphase}(b)), the shape of the Fermi surface is not distinctively electron-like or hole-like.

The above-mentioned blurred corners of the square-shaped Fermi surface connected to the nearby X or Y point by the ridge structures in the SM phase as in Figs.~\ref{FSh02}(c) and (d) has been observed in the ARPES experiments of overdoped La$_{2-x}$Sr$_x$CuO$_4$ with $x=0.22$ \cite{TYoshida2001,ERazzoli2010}. The partial reduction of the intensity of the Fermi surface caused by the formation of the pseudogap nearby the X and Y points in the PG phase and the appearance of the electron-like Fermi surface in the SM phase are consistent with the ARPES study on the doping dependence of the Fermi surface of La$_{2-x}$Sr$_x$CuO$_4$ \cite{TYoshida2006,ERazzoli2010}.

\begin{figure*}
\includegraphics[width=5.5cm]{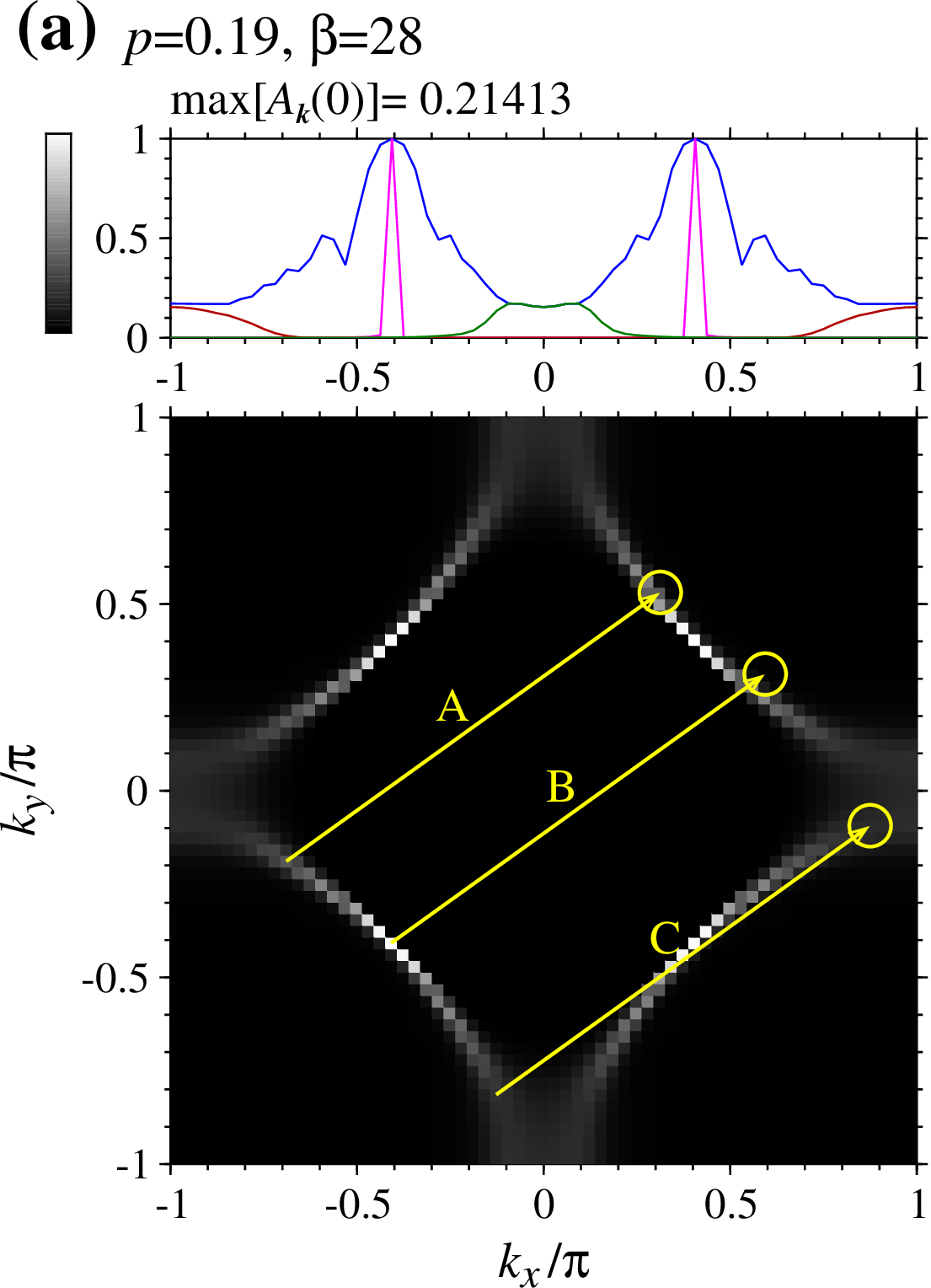}~~
\includegraphics[width=5.5cm]{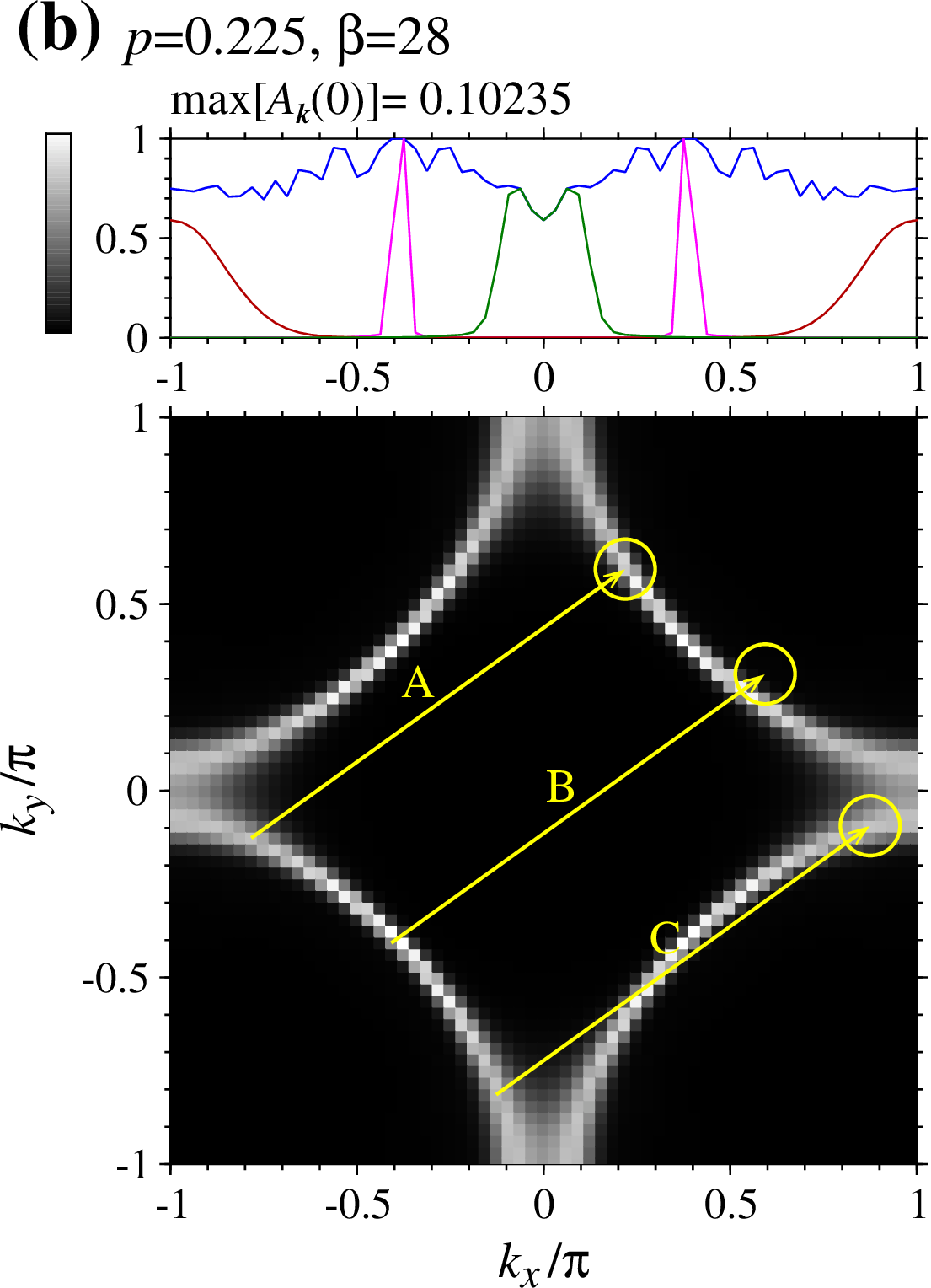}~~
\includegraphics[width=5.5cm]{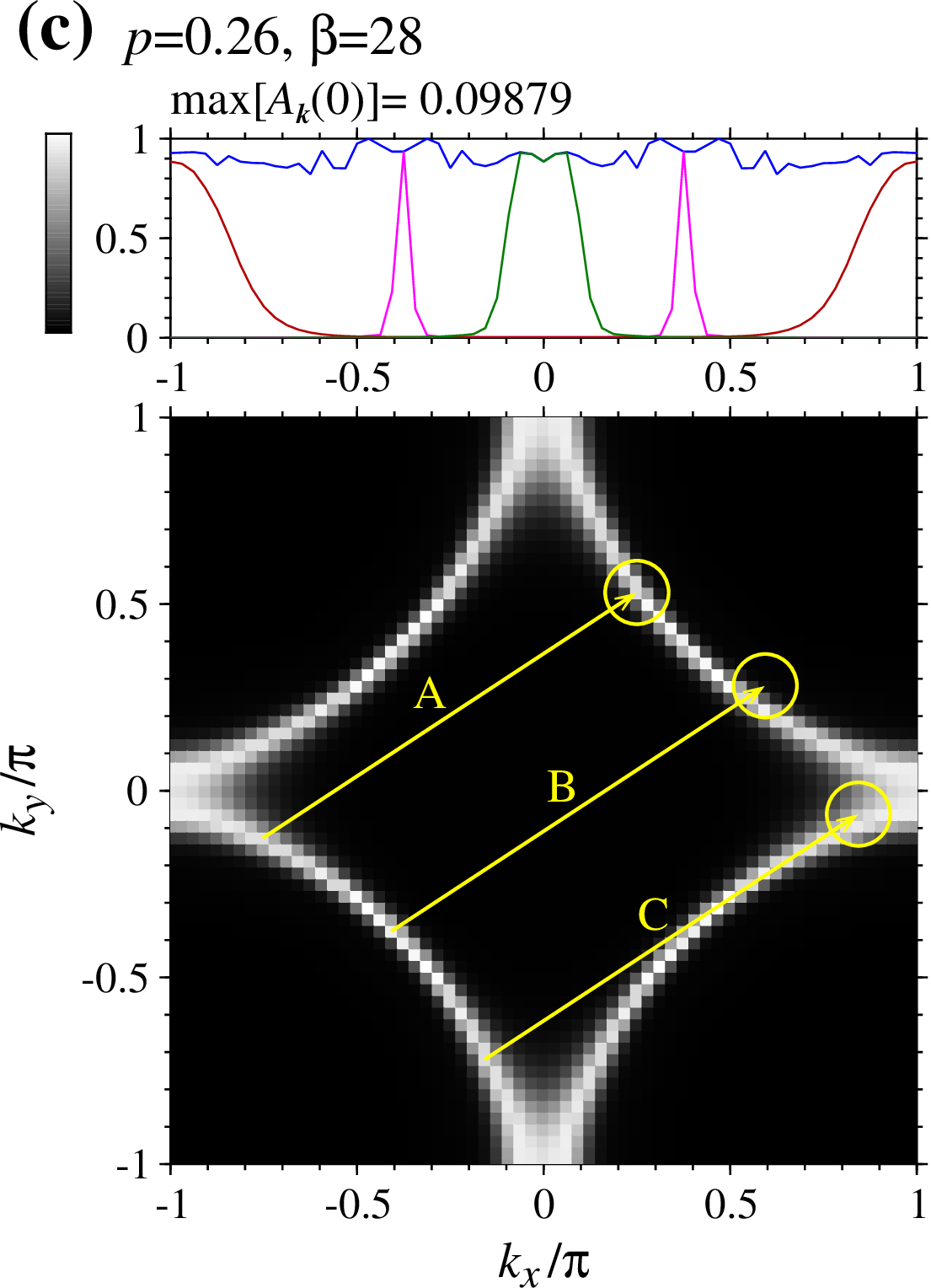}
  \caption{\label{FSh03}The same as Fig.~\ref{FSh02} but for $U=8$ and $t'=-0.3$ at $\beta=28$ with $p=0.19$ (a) in the PG phase and with $p=0.225$ (b) and $p=0.26$ (c) in the SM phase.}
\end{figure*}
Similar change in $A_{\bm{k}}(\omega=0)$ also occurs at $p^{*}=0.2$ for $U=8$ and $t'=-0.3$. Again, the bell shaped profile of the intensity of the spectral function at the Fermi level with the peak maximum on the $\Gamma$-M symmetry line and the strong suppression of the intensity around the X and Y points due to the formation of the pseudogap can be seen in the PG phase at $p=0.19$ in Fig.~\ref{FSh03}(a). On the other hand, a hole-like Fermi surface, whose structure near the X and Y points are blurred due to the presence of the extremely flatten band near the X and Y points (see Fig.~\ref{spec_h03}(c)), can be found in the SM phase at $p=0.225$ in Fig.~\ref{FSh03}(b). This is also evidenced by the flatten peak top and the two broad ridge structures around $k_x/\pi =\pm 0.08$ in the upper panel, which vertically cut through the Brillouin zone boundary near the Y point as shown in the lower panel. Although the Lifshitz transition is expected to occur around $p_{\rm VHS}=0.25$ at $T=0$, this only happens at low temperatures (see Fig.~\ref{pTphase}(c)) and the Fermi surface with $p=0.26$ at $T=1/28$ in Fig.~\ref{FSh03}(c) is still hole-like, where the Fermi surface and the nesting condition are nearly unchanged except for slight increase of the intensity near the X point.

The same trend to that with $t'=-0.2$ is found for the spectral functions at the Fermi surface with $t'=-0.3$ in the PG phase in Fig.~\ref{FSh03}(a): while the nesting condition is fulfilled for those near the X and Y points (arrows A and C) and the pseudogaps are formed, the nesting condition is not met for those on the Fermi arcs (arrow B) and the spectral functions remain single peaked. In contrast, in the SM phase the width $\Delta Q/\pi=0.08$ of the magnetic spot at $\bm{Q}$ of $\chi^{\rm sp}(\bm{Q},\omega =0)$ in Fig.~\ref{FSh03}(b) is broader than that in the PG phase $\Delta Q/\pi=0.056$ in Fig.~\ref{FSh03}(a) and the nesting condition is fulfilled everywhere at the Fermi level. Since the position of the VHS point is converged to $\tilde{\varepsilon}_{\bm{X}}^*=0$ as $T\to 0$ within $0.175\le p\le 0.25$ (see Fig.~\ref{eps0T1}(f)), the shape of the Fermi surface is almost unchanged with nearly quantum critical AFM fluctuations, resulting the MFL property within this range of $p$, similar to the case with $t'=-0.2$.

\begin{figure*}
\includegraphics[width=5.5cm]{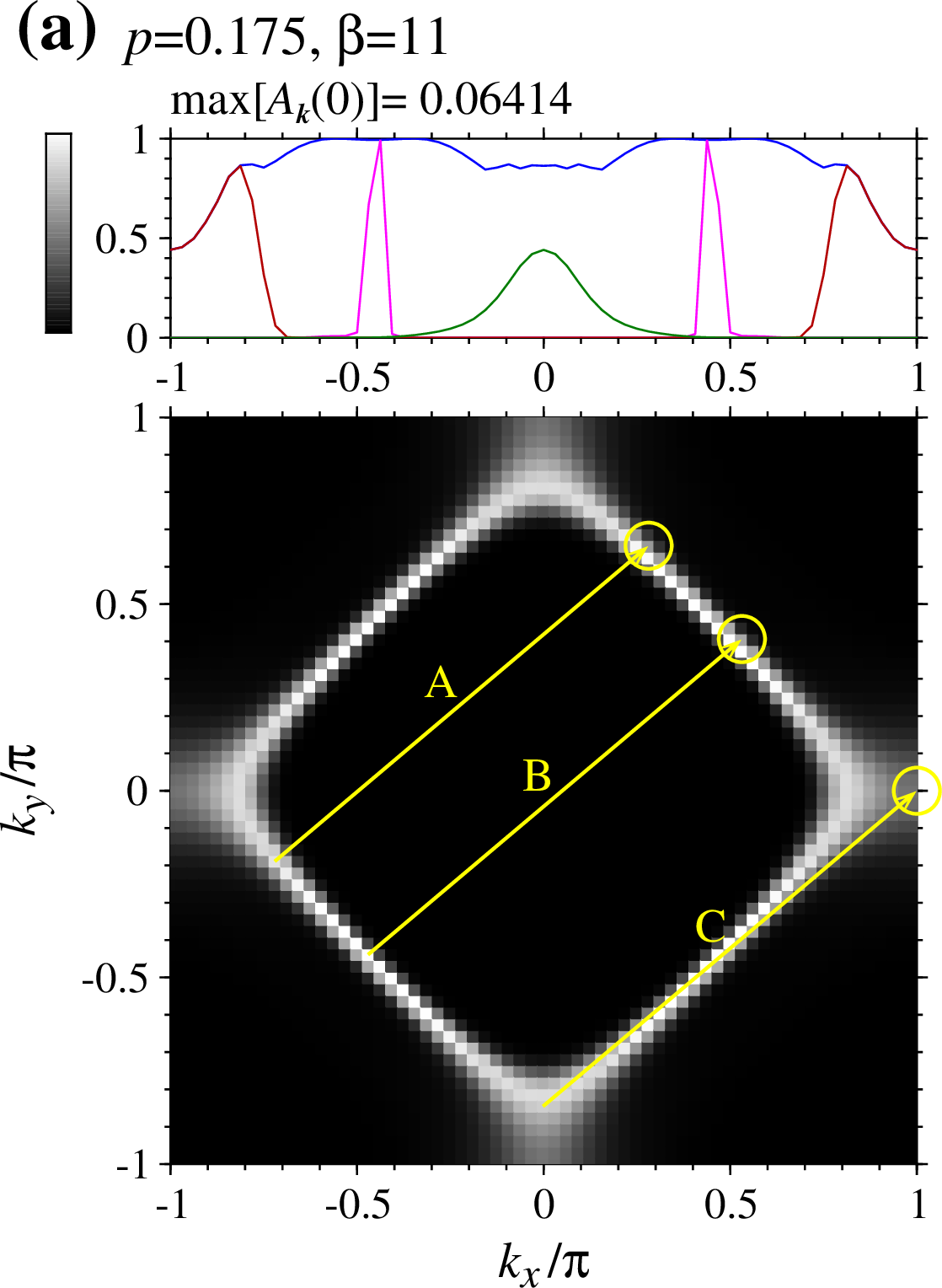}~~
\includegraphics[width=5.5cm]{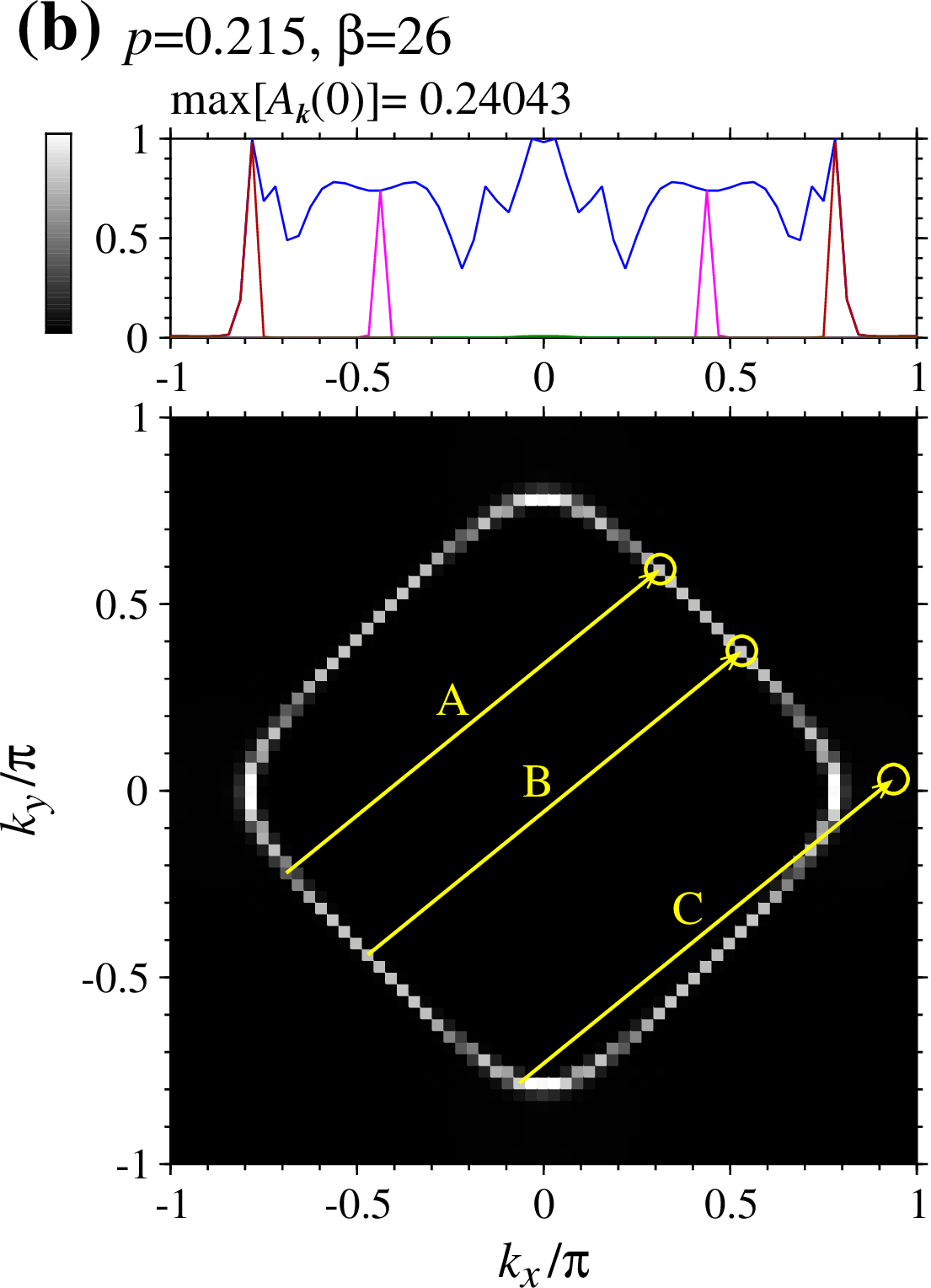}~~
\includegraphics[width=5.5cm]{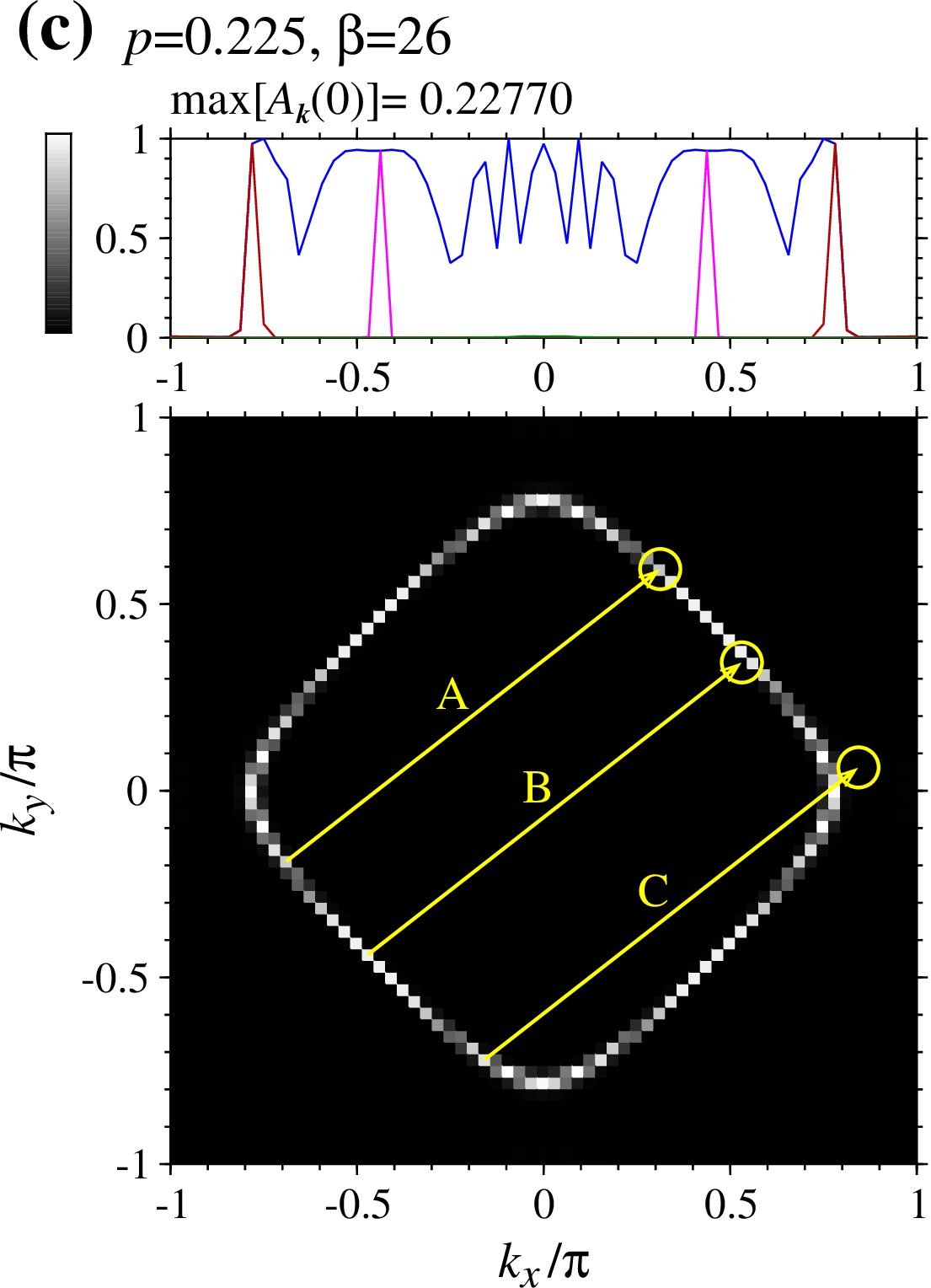}
  \caption{\label{FSh0}The same as Fig.~\ref{FSh02} but for $U=8$ and $t'=0$  with $p=0.175$ at $\beta=11$ (a), $p=0.215$ at $\beta=26$ (b) and $p=0.225$ at $\beta=26$ (c).}
\end{figure*}
The Fermi surface for $U=8$ and $t'=0$ with $p=0.175$ presented in Fig.~\ref{FSh0}(a) bears close resemblance to that for $U=8$ and $t'=-0.2$ in the SM phase in Fig.~\ref{FSh02}(c). Because of the ridge structures, which connect the corners of the square-shaped Fermi surface to the nearby X or Y points, the AFM nesting condition is satisfied on the whole Fermi surface similar to the extended quantum critical regime in the SM phase. 
However, as $\tilde{\varepsilon}_{\bm{X}}^*$ departs from the Fermi level with increasing $p$ as in Fig.~\ref{eps0T1}(d), this property of the AFM nesting condition is changed around $p=0.19$ and as seen in Fig.~\ref{FSh0}(b), the Fermi surface is sharp, clear and electron-like at $p=0.215$. Because the Fermi surface is square-shaped with rounded corners, the AFM nesting occurs only on its straight portions around $\Gamma$-M symmetry line (arrows A and B) and does not happen on its corners (arrow C) as can be seen in the lower panel. For this reason, the reduction of intensity occurs other than near the X and Y points and clear Fermi surface edges can be seen at its corners $k_x/\pi =\pm 0.78$ in the upper panel (the red line). The effects of the nesting is reduced with increasing $p$ from $p=0.215$ and the Fermi surface intensities at the corners and those on the $\Gamma$-M symmetry line are almost even at $p=0.225$ in Fig.~\ref{FSh0}(c).

Although both the shapes of the Fermi surface of $t'=0$ with $p=0.215$ in Fig.~\ref{FSh0}(b) and $t'=-0.2$ with $p=0.235$ in the SM phase in Fig.~\ref{FSh02}(c) are electron-like, in the former the VHS point is far from the Fermi level (see Fig.~\ref{eps0T1}(d)) and there is no flat band near the X or Y point in the vicinity of the Fermi level as in the latter (see Fig.~\ref{spec_h02}(c)). Hence, no enhancement of scattering near the X or Y point for the former. This is why the scattering due to the AFM fluctuations is weak with $p>p_{\rm MS}\sim 0.2$ with $t'=0$ and the $T$-linear resistivity is limited in the vicinity of $p_{\rm QCP}=0.215$ at low temperatures with $t'=0$ as will be discussed in Sec.~\ref{Resistivity}.

\subsection{MFL scaling relation of self-energy and $T$-linearity of scattering rate\label{SelfE}}
To understand differences in the electronic state between the PG and SM phases more in detail, in this section, we discuss the imaginary part of the electron self-energy ${\rm Im}\,\Sigma_{\bm{k}}(\omega)$.
\begin{figure*}
\includegraphics[width=8cm]{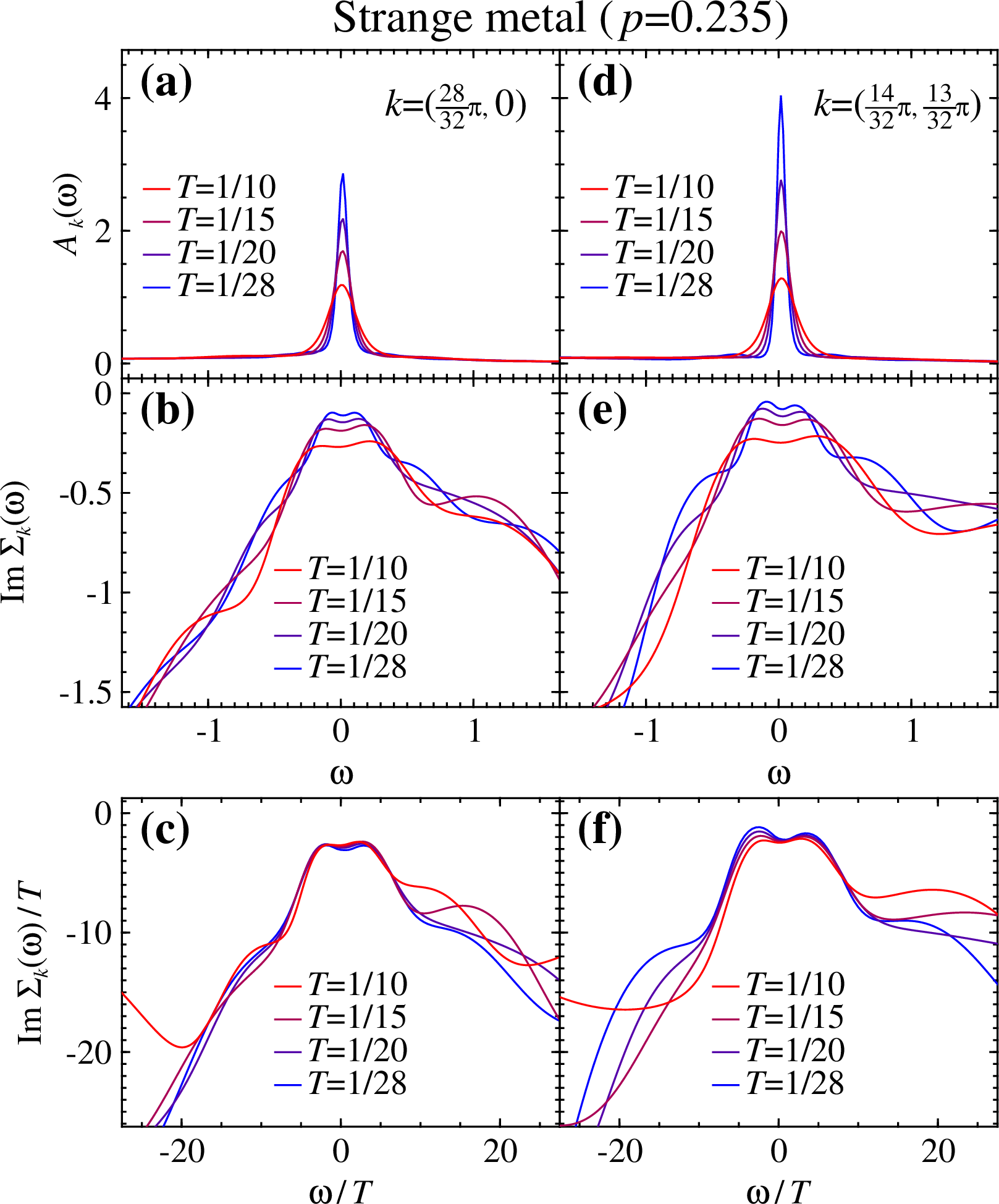}~~
\includegraphics[width=8cm]{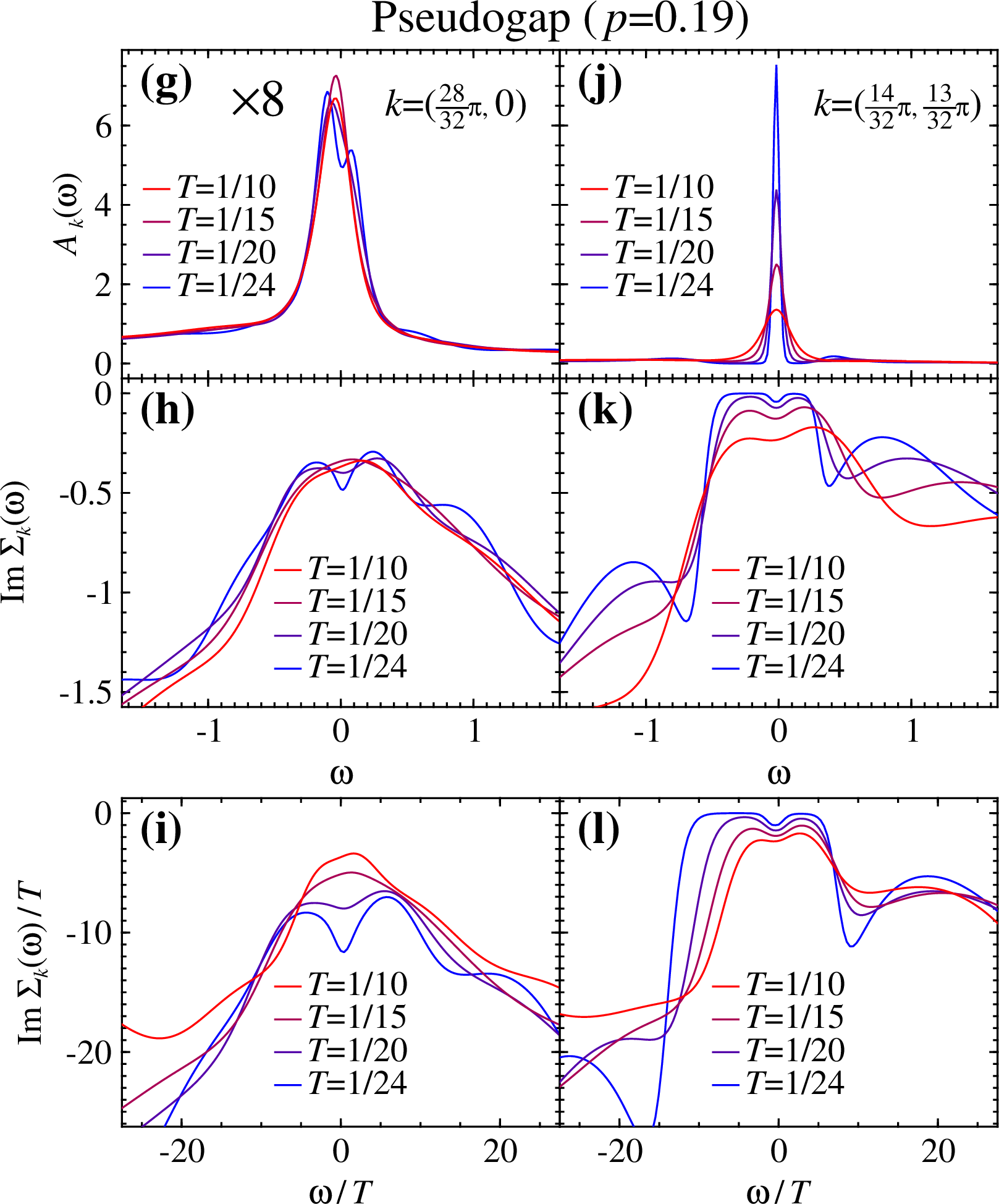}
  \caption{\label{ImSGM}Spectral functions $A_{\bm{k}}(\omega)$ (upper panels), the imaginary part of self-energies ${\rm Im}\,\Sigma_{\bm{k}}(\omega)$ at the Fermi surface as functions of $\omega$ (middle panels) and ${\rm Im}\,\Sigma_{\bm{k}}(\omega)/T$ versus $\omega/T$ plots (lower panels) are shown at various temperatures for $U=8$ and $t'=-0.2$: the six panels on the left (a)--(f) are for the SM phase with $p=0.235$ and those on the right (g)-(l) for the PG phase with $p=0.19$. Those at $\bm{k}=(\tfrac{28}{32}\pi,\,0)$ on the $\Gamma$--X symmetry line are shown in (a)--(c) and (g)--(i) and those at $\bm{k}=(\tfrac{14}{32}\pi,\,\tfrac{13}{32}\pi)$ near the $\Gamma$--M symmetry line are depicted in (d)--(f) and (j)--(l).}
\end{figure*}
Figure \ref{ImSGM} compares $T$ dependence of the spectral functions $A_{\bm{k}}(\omega)$ (upper panels) and ${\rm Im}\,\Sigma_{\bm{k}}(\omega)$ (middle panels) at the Fermi surface with $U=8$ and $t'=-0.2$ for the SM state at $p=0.235$ (panels (a)--(e)) and the PG state at $p=0.19$ (panels (g)--(k)). Furthermore, to discuss ${\bm{k}}$ dependency, those at $\bm{k}=(\tfrac{28}{32}\pi,\,0)$ on the $\Gamma$--X symmetry line and  $\bm{k}=(\tfrac{14}{32}\pi,\,\tfrac{13}{32}\pi)$ near the $\Gamma$--M symmetry line are presented side by side.

It has been claimed that the SM state of cuprates behaves as MFL \cite{CMVarma1989}, where the resistivity is not proportional to $T^2$ as in the Fermi liquid but proportional to $T$ at low temperatures. In MFL, the imaginary part of the electron self-energy has scaling relation as
\begin{align}\label{MFLScaling}
{\rm Im}\,\Sigma_{\bm{k}}(\omega)=Tf(\omega/T).
\end{align}
To assess the veracity of the scaling relation, ${\rm Im}\,\Sigma_{\bm{k}}(\omega)/T$ versus $\omega/T$ plots are also shown in the lower panels in Fig.~\ref{ImSGM}.

In the SM phase, the spectral functions are single peaked as shown in Figs.~\ref{ImSGM}(a) and (d). Their intensities at the peak top are inverse proportional to $T$ and widths are proportional to $T$. These properties of the spectral functions in the SM phase are originated from the peculiar behavior of the imaginary part of the self-energy in Figs.~\ref{ImSGM}(b) and (e), where all the ${\rm Im}\,\Sigma_{\bm{k}}(\omega)$ curves are nearly similar and their sizes are proportional to $T$. This is what expected in MFL and indeed these curves collapse onto universal curves in Figs.~\ref{ImSGM}(c) and (f), indicating the scaling relation of Eq.~(\ref{MFLScaling}) is well satisfied. The same MFL scaling relation in ${\rm Im}\,\Sigma_{\bm{k}}(\omega)$ is also found for $U=8$ with $t'=-0.3$ in the SM phase. As will be discussed in Sec.~\ref{chi}, this MFL scaling relation stems from the extended quantum critical property of the AFM fluctuations in the SM phase, where ${\rm Im}\,\chi^{\rm sp}(\bm{q},\omega)$ obeys $\omega/T$ scaling relation.
The distinctive upside-down pail-like structure of ${\rm Im}\,\Sigma_{\bm{k}}(\omega)$ and its MFL scaling is consistent with the imaginary part of the self-energy inferred from the ARPES experiments \cite{TVella}.

In the PG phase, on the other hand, the properties of the spectral functions are clearly different from those in the SM phase. As already discussed in Sec.~\ref{Spectra}, the spectral functions near the X point develop into those with the double peak structures as $T$ degreases below $T^*\sim 0.074$ as shown in Fig.~\ref{ImSGM}(g) and other than this feature, the hight and width of the peak are nearly unchanged. Accordingly, a dip appears at $\omega\sim 0$ at low temperatures in ${\rm Im}\,\Sigma_{\bm{k}}(\omega)$ in Fig.~\ref{ImSGM}(h), indicating enhancement of the AFM fluctuations within $|\omega|\lesssim 0.1$. In contrast, the spectral function near the $\Gamma$--M line remains single peaked even at low temperatures as in Fig.~\ref{ImSGM}(j). The increase of the peak-top intensity is much rapid with decreasing $T$ as in the SM phase. This is because ${\rm Im}\,\Sigma_{\bm{k}}(\omega=0)$ in Fig.~\ref{ImSGM}(k) approaches to zero more rapidly below $T^*$ as expected in FL with $\propto T^2$ than that in the SM phase with the MFL behavior $\propto T$, indicating the presence of well-defined quasiparticle on the Fermi arc. Moreover, ${\rm Im}\,\Sigma_{\bm{k}}(\omega=0)$ within the range $-0.5<\omega<0.2$ is practically zero at $T=1/24$ except for the dip corresponding to the peak in Fig.~\ref{ImSGM}(j), indicating clear suppression of the electron scattering caused by the AFM fluctuations.
As discussed in Sec.~\ref{epsX}, the PG phase is considered to be magnetically in the renormalized classical regime, where the AFM correlation length increases exponentially with decreasing $T$. This makes magnetic spots at $\bm{Q}$ in $\chi^{\rm sp}(\bm{q},\omega)$ more sharp and thus the nesting condition is more strict with decreasing $T$ as discussed in Sec.~\ref{Fsurf}. Because of this, electron scattering due to the AFM fluctuations increases near the X point, where the nesting condition is satisfied, and decreases near the $\Gamma$--M line, where the condition is not fulfilled (see Fig.~\ref{FSh02}(a)).
These results are consistent with the strong nodal–antinodal dichotomy observed in the ARPES experiments in the PG phase \cite{MHashimoto2014}. 

\begin{figure*}
\includegraphics[width=17cm]{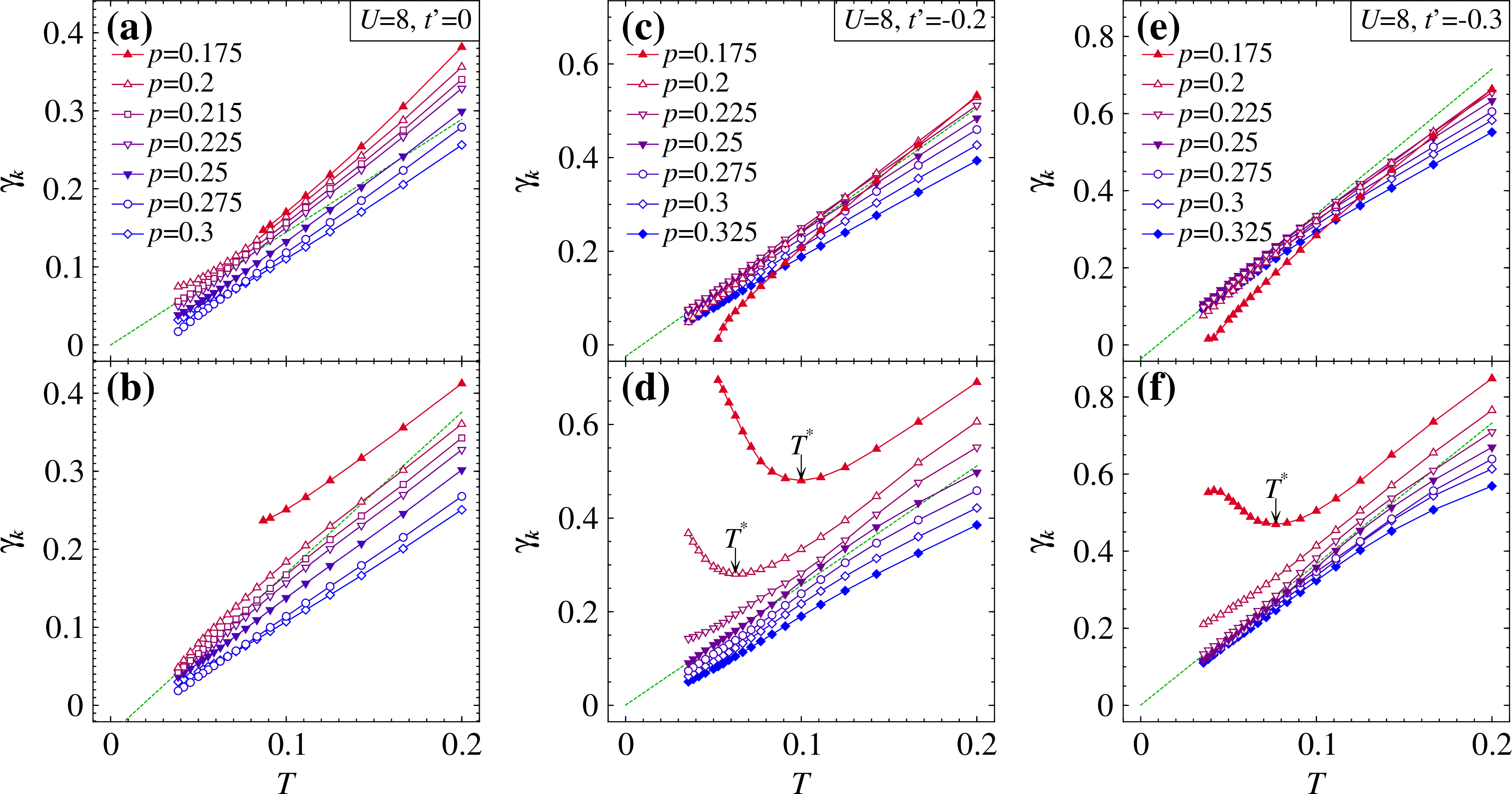}
  \caption{\label{sgm0k}Temperature dependence of the electron scattering rate $\gamma_{\bm{k}}\equiv -{\rm Im}\,\Sigma_{\bm{k}}(\omega =0)$ for $U=8$ with $t'=0$ (a)-(b), $t'=-0.2$ (c)-(d) and $t'=-0.3$ (e)-(f) for various values of $p$. Those in the vicinity of the Fermi level along the $\Gamma$--M symmetry line (the upper panels) and the $\Gamma$--X symmetry line (the lower panels) are shown.}
\end{figure*}
To examine $T$ dependence of the electron scattering rate $\gamma_{\bm{k}}\equiv -{\rm Im}\,\Sigma_{\bm{k}}(\omega =0)$, which is closely related to the resistivity discussed in the next section, those for $U=8$ with $t'=0$, $t'=-0.2$ and $t'=-0.3$ for various values of $p$ are presented in Fig.~\ref{sgm0k}. In upper panels, those with $\bm{k}$ in the vicinity of the Fermi level near the $\Gamma$--M symmetry line and in the lower panels those with $\bm{k}$ along the $\Gamma$--X symmetry line are depicted. The green dashed lines are the linear least squares fit made for $\gamma_{\bm{k}}=a_0+a_1T$ near the AFM QCP point $p_{\rm QCP}$.
As seen in Fig.~\ref{sgm0k}(a), for $t'=0$, the linear approximation is reasonable only in low temperatures at $p\sim p_{\rm QCP}=0.215$ with $\bm{k}$ near the $\Gamma$--M symmetry line, where the nesting condition is satisfied. The fitting is satisfactory only for $p$ near the $p_{\rm QCP}$, where $\gamma_{\bm{k}}$ is proportional to $T$, and not for $\bm{k}$ along the $\Gamma$--X symmetry line (see Fig.~\ref{sgm0k}(b)).
This property of $\gamma_{\bm{k}}$ has a similarity to the Hertz--Millis--Moriya theory, where $T$-proportionality of $\gamma_{\bm{k}}$ is expected in the vicinity of $p_{\rm QCP}$ of 2D antiferromagnets at low temperatures with limited nesting $\bm{k}$ points (the hot spots).

On the other hand, unlike those with $t'=0$, the linear relation is fold for wide range of $p$ in the SM phase with $t'=-0.2$ and $-0.3$.
The green dashed lines are those of the linear least squares fit at $p=0.25$ for $t'=-0.2$ in Figs.~\ref{sgm0k}(c) and (d) and $p=0.225$ for $t'=-0.3$ in Figs.~\ref{sgm0k}(e) and (f). The linear relation is well satisfied within $T<0.1$ with $\bm{k}$ not only near the $\Gamma$--X line but also the $\Gamma$--M symmetry line. Furthermore, in the SM phase, the $\bm{k}$ and $p$ dependence of $\gamma_{\bm{k}}$ is remarkably small. The coefficient $a_1$ of the linear fitting is $a_1=2.65$ and $a_1=2.50$ with $\bm{k}$ along the $\Gamma$--M and $\Gamma$--X symmetry lines, respectively and the variation of these values are only 3\% within $0.235 \le p \le 0.275$ for $t'=-0.2$. Similarly, $a_1=3.69$  and $a_1=3.76$ with $\bm{k}$ along the $\Gamma$--M and $\Gamma$--X  symmetry lines, respectively and the variation of these values are only 5\% within $0.215 \le p \le 0.25$ for $t'=-0.3$. 
These results are in accordance with the fact that the AFM nesting condition is fulfilled everywhere on the Fermi surface in the SM phase due to the VHS point in the vicinity of the Fermi level (see Figs.~\ref{eps0T1}(e) and (f)) as discussed in Sec.~\ref{Fsurf}. These results are also consistent with isotropic and $T$-linear inelastic scattering rate found in the angle dependent magnetoresistance experiments on La$_{1.6-x}$Nd$_{0.4}$Sr$_x$CuO$_4$ \cite{GGrissonnanche2021}. The $T$-linear $\gamma_{\bm{k}}$, which is isotropic on the Fermi surface, in the overdoped regime in the 2D Hubbard model also has been found in a previous study with DCA \cite{WWu2022}. 

Note that $\gamma_{\bm{k}}$ along the $\Gamma$--X symmetry line in the vicinity of $p^*$ in the range $0.225 \le p \le 0.235$ with $t'=-0.2$ is substantially deviated from the linear relation and has a finite value even at $T=0$. The fact, however, does not affect the $T$-proportionality of resistivity as will be discuss in Sec.~\ref{Resistivity}. The similar deviation from the linear relation is also found in $\gamma_{\bm{k}}$ along the $\Gamma$--X symmetry in the range $0.2 \le p \le 0.215$ with $t'=-0.3$.

In contrast, in the PG phase, there are clear difference between $\gamma_{\bm{k}}$ with $\bm{k}$ near the $\Gamma$--M symmetry line, where the AFM nesting condition is not satisfied, and along the $\Gamma$--X symmetry line, where the AFM nesting condition is fulfilled. The $\gamma_{\bm{k}}$ curves of $t'=-0.2$ for $p=0.175$ and $p=0.2$ with $\bm{k}$ along the $\Gamma$--X symmetry line in Fig.~\ref{sgm0k}(d) are deviated upward below $T^*$ from the $T$--linear line as they enter the PG phase and are deviated downward below $T^*$ with $\bm{k}$ near the $\Gamma$--M symmetry line in Fig.~\ref{sgm0k}(c). The same trend can be seen for the $\gamma_{\bm{k}}$ curve of $t'=-0.3$ for $p=0.175$ in Figs.~\ref{sgm0k}(e) and (f).
 
It has been argued that the transport scattering rate $1/\tau_{\rm tr}$ of electrons in the SM phase is near the Planckian time scale $\tau_{\rm P}=\hbar/k_{\rm B}T$, which is the shortest time scale of inelastic relaxation \cite{JZaanen2004}.
If the upside-down pail-like structure of ${\rm Im}\,\Sigma_{\bm{k}}(\omega)$ in Figs.~\ref{ImSGM}(c) and (f) is approximated by  
\begin{align}
 {\rm Im}\,\Sigma_{\bm{k}}(\omega)\approx -{\rm max}(a_1T,|\omega|),
\end{align}
the real part of the self-energy with $|\omega| \ll T$ can be express with the similar derivation in Ref.~\onlinecite{CMVarma1989} as
\begin{align}
  {\rm Re}\,\Sigma_{\bm{k}}(\omega)-{\rm Re}\,\Sigma_{\bm{k}}(0)=-\frac{2}{\pi}\omega\ln\left(\frac{\Lambda}{a_1T}\right),
\end{align}
where $\Lambda\sim 1$ is a cutoff energy. Using Eq.~(\ref{zk}),
\begin{align}
  \frac{1}{\tau_{\rm tr}}\approx \gamma_{\bm{k}}^*\equiv z_{\bm{k}}\gamma_{\bm{k}}\approx \frac{a_1}{1+\frac{2}{\pi}\ln\left(\frac{\Lambda}{a_1T}\right)}T
\end{align}
is obtained. As mentioned above, $a_1$ is order of unity and thus $1/\tau_{\rm tr}$ is at the Planckian limit. However, when $1/\tau_{\rm tr}$ is written as $1/\tau_{\rm tr}=\alpha k_{\rm B}T/\hbar$, the coefficient $\alpha$ is $T$ dependent and reduced as $T$ decreases, because of the presence of the logarithmic term \cite{CMVarma2020}. 

\subsection{$T$-proportionality of resistivity\label{Resistivity}}
In this section, to understand the linear-in-$T$ resistivity in the SM phase found in the experiments, the temperature dependence of the resistivity is discussed in relation to the electron scattering ratio $\gamma_{\bm{k}}$ presented in the previous section. 
\begin{figure*}
\includegraphics[width=17cm]{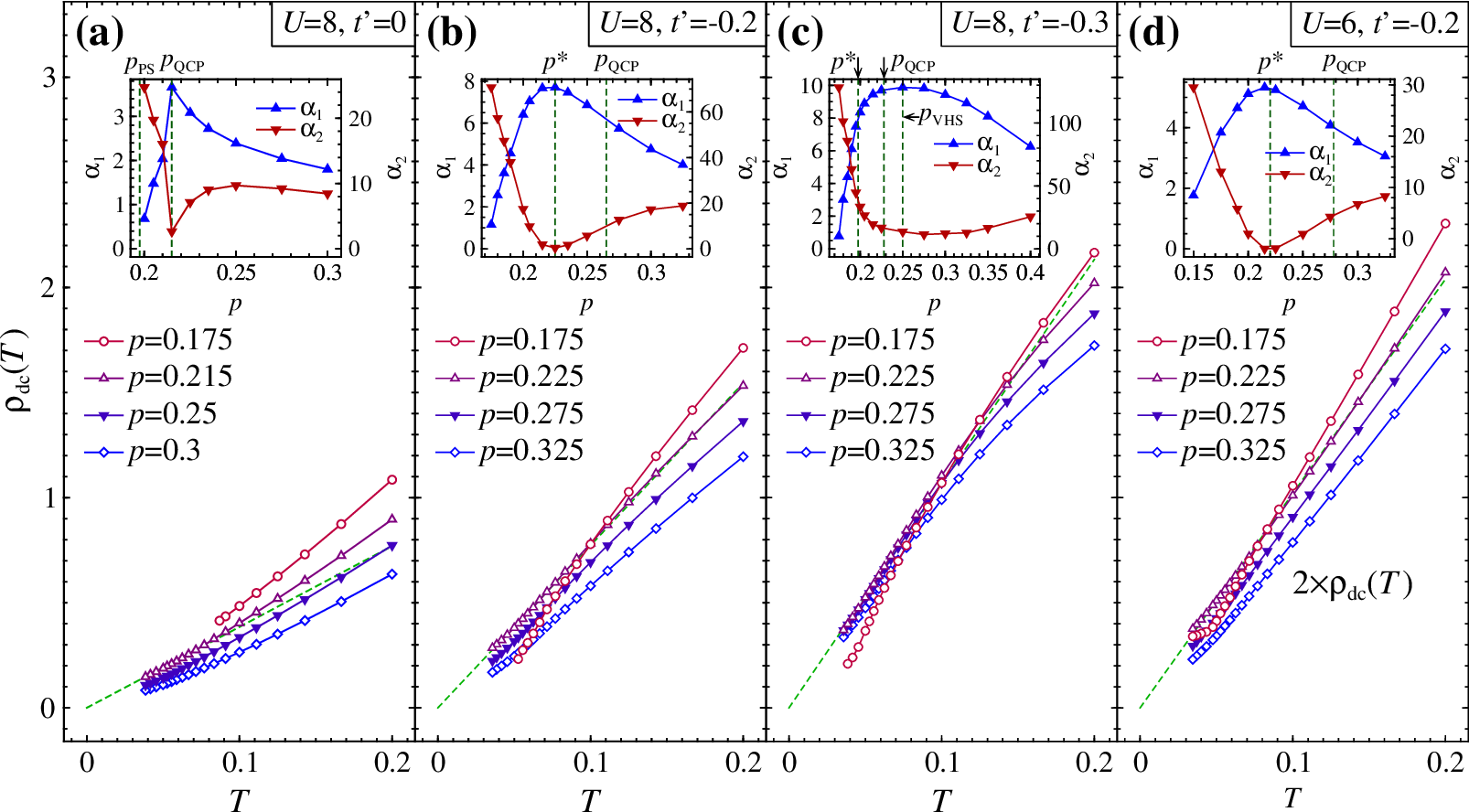}
  \caption{\label{rho_dc}Temperature dependence of the resistivity $\rho_{\rm dc}(T)$ for various hole concentration $p$ for $U=8$ with $t'=0$ (a), $t'=-0.2$ (b) and $t'=-0.3$ (c); in panel (d) two times magnified values of $\rho_{\rm dc}(T)$ for $U=6$ with $t'=-0.2$ are shown. Inset: coefficients $\alpha_1$ and $\alpha_2$ obtained from the least squares fit assuming $\rho_{\rm dc}(T)=\alpha_1T+\alpha_2T^2$ are indicated.}
\end{figure*}
The resistivity $\rho_{\rm dc}=1/\sigma_{\rm dc}$ is calculated using the Kubo formula for the conductivity $\sigma_{\rm dc}$ from the spectral function $A_{\bm{k}}(\omega)$ as \cite{GDMahan1990}
\begin{align}
  \sigma_{\rm dc}=\frac{2\pi}{N}\sum_{\bm{k}}\frac{|\nabla\varepsilon_{\bm{k}}|^2}{2}\int\frac{\beta}{4\cosh^2\big(\frac{\beta\omega}{2}\big)}\big(A_{\bm{k}}(\omega)\big)^2\,d\omega.\label{sigma_dc}
\end{align}
Here, it is assumed that the major contribution to $\rho_{\rm dc}(T)$ comes from the umklapp scattering process due to the AFM fluctuations, which has large momentum transfer. The effects of the back-flow, i.e., the vertex corrections as a requirement for momentum conservation \cite{KYamada2004}, are not consider in the present study, since as detailed in Appendix~\ref{vertex}, the vertex corrections are absent within LDFA, similar to DMFT itself \cite{AKhurana1990,AGeorges1996}. Note, however, that a substantial contribution of the vertex corrections to the resistivity at high temperatures $T\gtrsim t$ has been found in a recent study of the square-lattice Hubbard model \cite{JVucicevic2019}.

The $T$ dependence of the resistivity $\rho_{\rm dc}(T)$ for various hole concentration $p$ are shown in Fig.~\ref{rho_dc} for four different parameter sets: $U=8$ with $t'=0$ (a), $t'=-0.2$ (b) and $t'=-0.3$ (c) and $U=6$ with $t'=-0.2$ (d). The coefficients $\alpha_1$ and $\alpha_2$ obtained from the least squares fit assuming $\rho_{\rm dc}(T)=\alpha_1T+\alpha_2T^2$ within $T<0.1$ are also presented in the inset in each panel (for $p<p_{\rm QCP}$ in (a), $\rho_{\rm dc}(T)=\alpha_0+\alpha_1T+\alpha_2T^2$ is assumed instead). Clearly, there are distinctive differences in their $p$ and $T$ dependence among those with $t'=0$, $-0.2$ and $-0.3$.
For $t'=0$ in Fig.~\ref{rho_dc}(a), $\rho_{\rm dc}(T)$ is not linear function in $T$ at hight temperatures. However, $\rho_{\rm dc}(T)$ is proportional to $T$ only around $p_{\rm QCP}$ within low temperatures $T<0.1$ similar to the Hertz--Millis--Moriya theory.

On the other hand, for $t'=-0.2$ in Figs.~\ref{rho_dc}(b) and (d), $\rho_{\rm dc}(T)$ is not proportional to $T$ at $p_{\rm QCP}$ but at $p^*$ with wider range of $T$ at least up to $T=0.2$. The range in which $\rho_{\rm dc}(T)$ is proportional to $T$ is more limited within low temperatures as $p$ departs from $p^*$. This is also in accordance with the fact that the ratio $\alpha_1/\alpha_2$ decreases as the distance between $p$ and $p_{\rm QCP}$ increases as shown in the insets of Figs.~\ref{rho_dc}(b) and (d). This peculiar feature of the resistivity is consistent with the in-plane resistivity $\rho_{ab}(T)$ of La$_{2-x}$Sr$_x$CuO$_4$ experiments in high magnetic fields, where the $T$-linear coefficient $\alpha_1$ is, indeed, peaked at $p^*$ \cite{RACooper2009}. 

For $t'=-0.3$ in Fig.~\ref{rho_dc}(c), $\rho_{\rm dc}(T)$ is nearly proportional to $T$ around $0.225\le p\le 0.275$ within $T<0.1$, where $\rho_{\rm dc}(T)$ curves for $p=0.225$ and $p=0.275$ overlap. This feature is also evident in $p$ dependence of $\alpha_1$ and $\alpha_2$ in the inset: from $p=1.75$ to $p=0.2$ $\alpha_1$ sharply increases followed by the plateau centered around $p\sim p_{\rm VHS}=0.25$, where $\alpha_2$ takes minimum value accordingly, and then $\alpha_1$ gradually decreases. Again, the maximum linearity of $\rho_{\rm dc}(T)$ is found not at $p_{\rm QCP}$ but $p\sim 0.25$, where the VHS point is placed at the Fermi level.

\begin{figure}
\includegraphics[width=3.5cm]{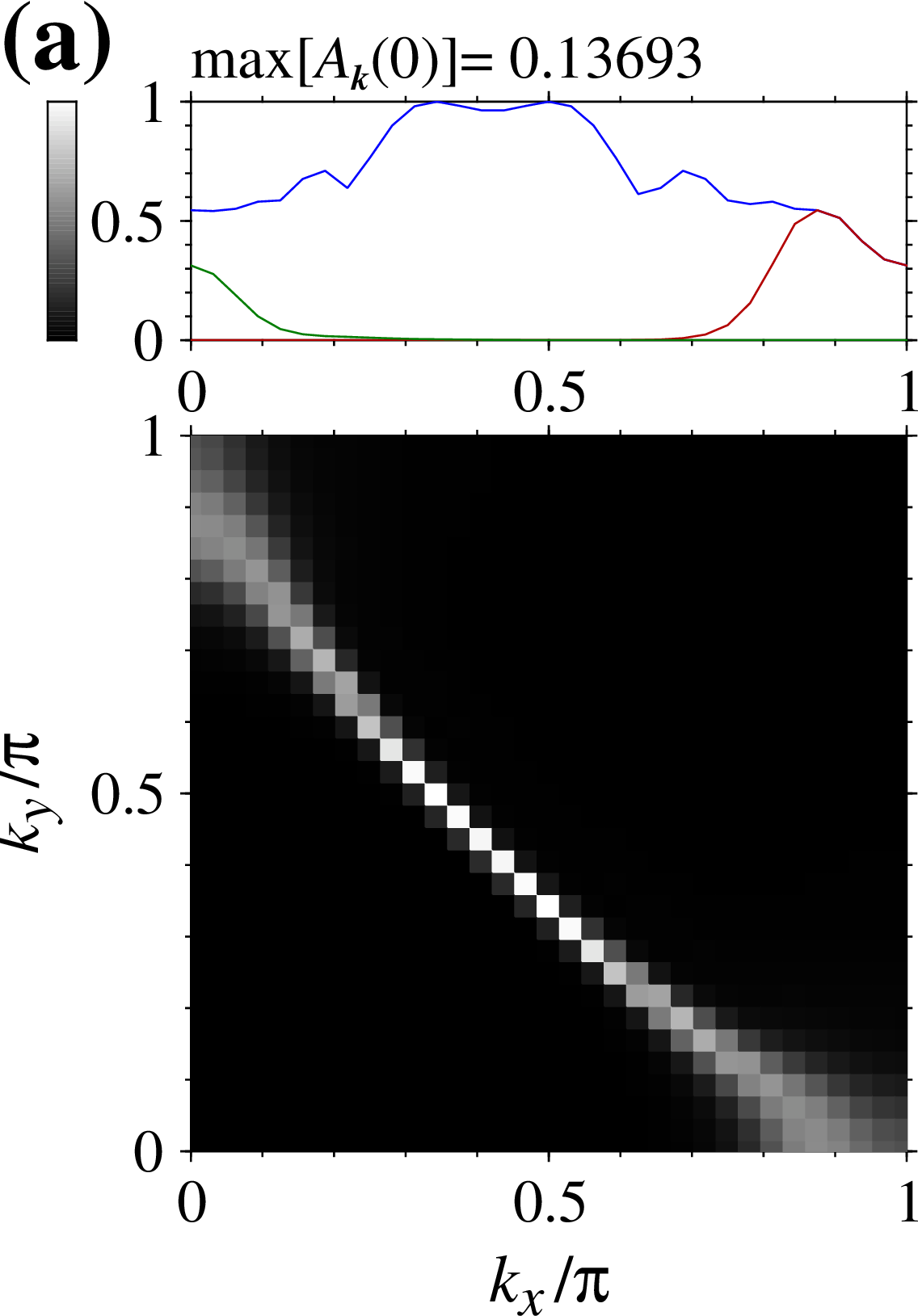}~~
\includegraphics[width=3.5cm]{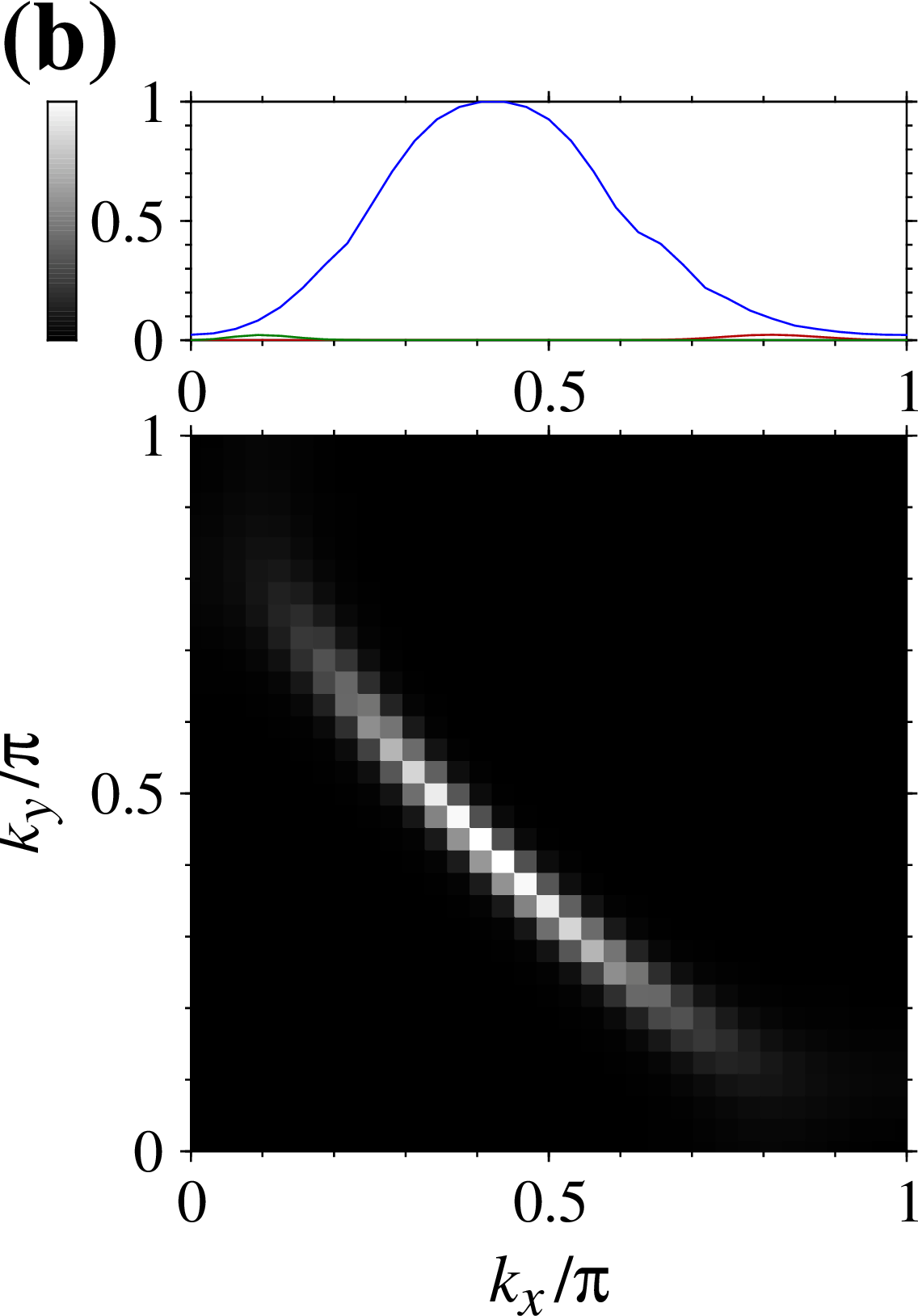}
  \caption{\label{dc_map}Intensities of the spectral functions at the Fermi level $A_{\bm{k}}(\omega=0)$ (a) are compared with the $\bm{k}$ components on the right hand side of Eq.~(\ref{sigma_dc}) (b) for $U=8$ and $t'=-0.2$ at $\beta=28$ with $p=0.225$. In the lower panels the quantities are mapped onto the top right quadrant of the first Brillouin zone with gray scale. In the upper panels the quantities along the $\Gamma$--X and Y--M symmetry lines are depicted by the red and green lines, respectively and the maximum values in each of fixed value of $k_x$ are denoted by the blue lines.}
\end{figure}
These clear differences of $\rho_{\rm dc}(T)$ between that of $t'=0$ and the others can be explained as follows. $\sigma_{\rm dc}$ in Eq.~(\ref{rho_dc}) can be approximated by \cite{GDMahan1990}
\begin{align}
  \sigma_{\rm dc}\approx \frac{1}{N}\sum_{\bm{k}}\frac{\beta}{4\cosh^2\big(\beta\tilde{\varepsilon}_{\bm{k}}/2\big)}\frac{|\nabla\varepsilon_{\bm{k}}|^2}{2\gamma_{\bm k}}\label{sigma_dc2}.
\end{align}
As discussed in Sec.~\ref{Fsurf}, for $t'=-0.2$ and $-0.3$, the nesting condition is satisfied everywhere on the Fermi surface in the SM phase, because of the presence of the VHS point near the Fermi level. In contrast, due to the round corner square shape of the Fermi surface for $t'=0$, the nesting condition is only satisfied on the straight portions of the Fermi surface around the $\Gamma$--M symmetry line with $p>p_{\rm MS}\sim 0.2$. Since the factor $|\nabla\varepsilon_{\bm{k}}|^2$ in Eq.~(\ref{sigma_dc2}) is small in the vicinity of the X point, $1/\gamma_{\bm{k}}$ only around the $\Gamma$--M symmetry line mainly contribute $\sigma_{\rm dc}$. To demonstrate this, in Fig.~\ref{dc_map}(b) on the lower panel, instead of making summation over $\bm{k}$, each of the $\bm{k}$ components on the right hand side of Eq.~(\ref{sigma_dc}) is mapped onto the quadrant of the first Brillouin zone for $U=8$ with $t'=-0.2$ at $p^*=0.225$. In contrast to the Fermi surface in Fig.~\ref{dc_map}(a), the intensity is concentrated around the $\Gamma$--M symmetry line and is lacking near the X point in Fig.~\ref{dc_map}(b). Indeed, clear resemblance can be found between $\rho_{\rm dc}(T)$ of $U=8$ with $t'=0$, $-0.2$ and $-0.3$ presented in Figs.~\ref{rho_dc}(a)-(c) and corresponding $\gamma_{\bm{k}}(T)$ along the $\Gamma$--M symmetry line shown in Figs.~\ref{sgm0k}(a), (c) and (e). Because of this, the deviation from the $T$-linear dependence of $\gamma_{\bm{k}}$ in the SM phase near $p^*$ on the $\Gamma$--X symmetry line as mentioned in Sec.~\ref{SelfE} (see also Figs.~\ref{sgm0k}(d) and (f)) does not directly affect the resistivity and this makes the resistivity of $U=8$ with $t'=-0.2$ and $-0.3$ in the SM phase extremely linear in $T$ in the wide range of $p$. As discussed in Sec.~\ref{SelfE}, for $\gamma_{\bm{k}}$ of $U=8$ with $t'=0$, the linear approximation is reasonable only in low temperatures at $p\sim p_{\rm QCP}=0.215$ with $\bm{k}$ near the $\Gamma$--M symmetry line and this feature is exactly what found in Fig.~\ref{rho_dc}(a).

\subsection{Dynamical spin susceptibility\label{chi}}
\begin{figure*}
\includegraphics[width=17cm]{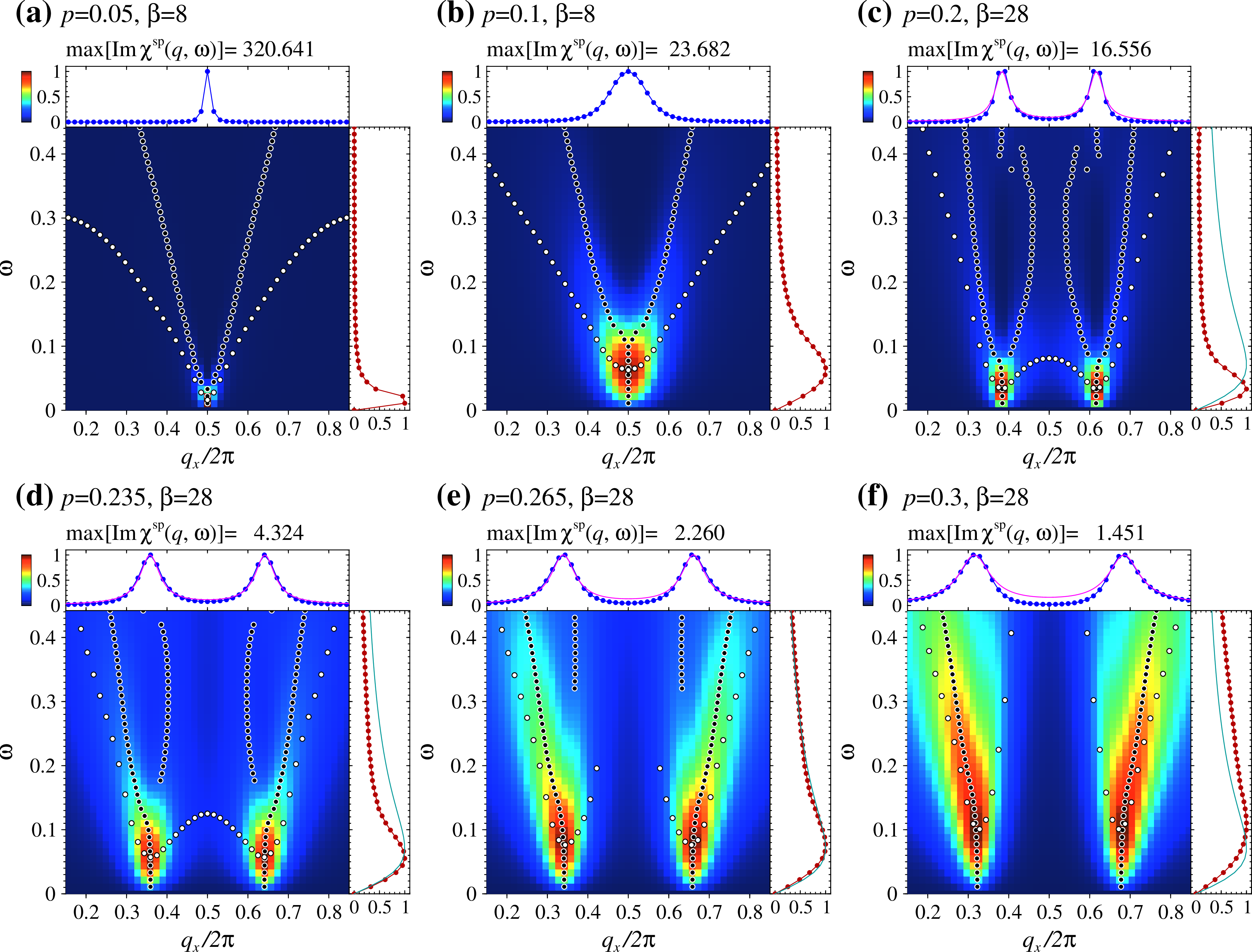}~
  \caption{\label{chiqome1}${\rm Im}\,\chi^{\rm sp}(\bm{q},\omega)$ as a function of wave vector $\bm{q}=(q_x,\pi)$ and frequency $\omega$ for $U=8$ and $t'=-0.2$; in (a) and (b) those at $\beta =8$ with $p=0.05$ (a) and $p=0.1$ (b) are shown; in (c)--(f), those at $\beta =28$ with $p=0.2$ (c), $p=0.235$ (d), $p=0.265$ (e) and $p=0.3$ (f) are indicated. In each of the figures (a)-(f), in addition to the color plot of the intensity of ${\rm Im}\,\chi^{\rm sp}(\bm{q},\omega)$ in the main panel, in the panel on the right the maximum intensity of fixed $\omega$, i.e., along horizontal line in the main panel are shown by the red dots and the cyan line in (c)--(f) depicts the model function in Eq.~(\ref{chi_max}) and in the upper panel, the intensity as a function of $q_x$ along the $\omega=\omega_{\rm max}$ line which cuts through the maximum of ${\rm Im}\,\chi^{\rm sp}(\bm{q},\omega)$ is shown by the blue dots and the magenta line in (c)--(f) depicts the model function in Eq.~(\ref{chi_Q}). In the main panel, the white dots denote the maximum points along fixed $q_x$ lines and the black dots are the local maximum points along fixed $\omega$ lines. }
\end{figure*}
To discuss the property of the dynamical spin susceptibility in the SM phase and its relation to the linear-in-$T$ resistivity, the imaginary part of the dynamical spin susceptibility ${\rm Im}\,\chi^{\rm sp}(\bm{q},\omega)$ as a function of wave vector $\bm{q}=(q_x,\pi)$ and frequency $\omega$ for $U=8$ and $t'=-0.2$ for various values of $p$ are shown in Fig.~\ref{chiqome1}. In the main panel the intensities of ${\rm Im}\,\chi^{\rm sp}(\bm{q},\omega)$ are shown as the color plots on $q_x$-$\omega$ plane for those in the PG phase in Figs.~\ref{chiqome1}(a)-(c) and for those in the SM phase in Figs.~\ref{chiqome1}(d)-(f). 

In each panel on the right the maximum intensity of fixed $\omega$ is indicated by the red dots. The cyan line represents corresponding maximum intensity assuming the overdamped spin wave at the Planckian dissipation limit  as
\begin{align}\label{chi_max}
  {\rm Im}\,\chi^{\rm sp}_{\rm max}(\omega)&\equiv\max_{\bm{q}}\, {\rm Im}\,\chi^{\rm sp}(\bm{q},\omega)\nonumber\\
  &\approx\frac{2}{\pi}\frac{\Gamma\omega}{\omega^2+\Gamma^2}\chi^{\rm sp}(\bm{Q},\omega=0)\nonumber\\
  &=\frac{2}{\pi}\frac{2C\omega}{\omega^2+(2T)^2},
\end{align}
where $\Gamma$ denotes the relaxation rate with $\Gamma=2T$ and Eq.~(\ref{chi_static}) at $p_{\rm QCP}$, i.e., $\Delta=0$, is assumed, which is a good approximation at $p_{\rm QCP}$ for all the four representative parameter sets as will be shown below. ${\rm Im}\,\chi^{\rm sp}_{\rm max}(\omega)$ reaches its maximum at $\omega =2T$.

In the upper panel, the intensity of ${\rm Im}\,\chi^{\rm sp}(\bm{q},\omega_{\rm max})$ as a function of $q_x$, where $\omega=\omega_{\rm max}$ line which cuts through the maximum of ${\rm Im}\,\chi^{\rm sp}(\bm{q},\omega)$ is shown by the blue dots. The magenta line depicts ${\rm Im}\,\chi^{\rm sp}(\bm{q},\omega_{\rm max})$ of a model dynamical spin susceptibility with four maximum $\bm{q}$ points corresponding to the incommensurate AFM fluctuations as
\begin{align}\label{chi_Q}
  {\rm Im}\,\chi^{\rm sp}(\bm{q},\omega)\approx \frac{2}{\pi}\sum_{i=1}^4\frac{2C\omega}{\omega^2+(2T)^2+[(\bm{q}-\bm{Q}_i)/W]^2},
\end{align}
where $\bm{Q}_1=(\delta,\,\pi)$, $\bm{Q}_2=(-\delta,\,\pi)$, $\bm{Q}_3=(\pi,\,\delta)$ and $\bm{Q}_4=(\pi,\,-\delta)$. The width of the scattering peak centered at $\bm{Q}_i$ is given by $\Delta Q=2WT$ in HWHM. In each of Figs.~\ref{chiqome1}(c)--(f), the parameters $W$ and $\delta$ are adjusted so to fit the LDFA results.

Figure \ref{chiqome1}(a) shows ${\rm Im}\,\chi^{\rm sp}(\bm{q},\omega)$ with $p=0.05$, in which clear spin-wave dispersion originating from the commensurate $\bm{Q}=(\pi,\pi)$ AFM fluctuations can be seen (white dots). Similar to the results of the inelastic neutron scattering study \cite{RColdea2001} of La$_2$CuO$_4$, the dispersion relation of the spin wave is linear near the M point $\bm{Q}=(\pi,\pi)$ and the intensity is concentrated in the vicinity of the M point indicating the presence of long-range AFM correlation.
However, with $p=0.1$ in Fig.~\ref{chiqome1}(b), low energy structure of the spin-wave dispersion below $\omega \sim 0.06$ is smeared showing the strong damping of the spin wave. The broadening of the peak in the top panel arises from the reduction of the AFM spin correlation length. The black dots, which are the local maximum points along fixed $\omega$ lines, in the main panel shape like the letter Y. The Y-shaped spectrum of ${\rm Im}\,\chi^{\rm sp}(\bm{q},\omega)$ arising from the commensurate AFM fluctuations has been observed in underdoped ($p\approx 0.095$) HgBa$_2$CuO$_{4+\delta}$ in the inelastic neutron-scattering experiments \cite{MKChan} and is also consistent with the previous LDFA study \cite{JPFLeBlanc2019}.

On the other hand, with $p=0.2$, the AFM fluctuations are incommensurate $\bm{Q}=(\pm\delta,\pi)$, $(\pi,\pm\delta)$. Accordingly, two peaks appear at $q_x/(2\pi)=0.38$ and $0.62$ corresponding to $\bm{Q}=(\pm\delta,\,\pi)$ in the top panel in Fig.~\ref{chiqome1}(c). In the SM phase, the $\omega$ dependence of the maximum intensities shown in the right panels in Figs.~\ref{chiqome1}(d)-(f) are well approximated by that of the extremely overdamped spin-wave model at the Planckian dissipation limit $\Gamma =2T$ in Eq.~(\ref{chi_max}). In particular, in Fig.~\ref{chiqome1}(e) the excellent agreement can be seen at the AFM QCP $p=0.265$. Indeed, ${\rm Im}\,\chi^{\rm sp}_{\rm max}(\omega)$ with $\Gamma=2T$ obeys $\omega/T$ scaling expected in the quantum critical regime. 

These results are in agreement with the inelastic neutron scattering experiments of LSCO \cite{GAeppli1998}; the findings in the experiments that the peaks caused by the magnetic fluctuations in the normal state in the low frequency limit ${\rm Im}\,\chi^{\rm sp}_{\rm max}(\omega)/\omega$ have amplitude that decrease as $1/T^2$ and widths that increase in proportion with $T$ is exactly what expected in Eq.~(\ref{chi_max}). The extremely overdamped spin wave with relaxation rate at the Planckian limit $\hbar\Gamma\sim k_{\rm B}T$, which is consistent with the present results, has been also reported in the LSCO experiments \cite{MZhu2023}.

The scaling relation of $\chi^{\rm sp}(\bm{q},\omega)$ in Eq.~(\ref{chi_Q}) at QCP can be expressed with $\bm{q}$ located near one of $\bm{Q}_i$'s as \cite{SSachdev2011}
\begin{align}\label{chi_scale}
  \chi^{\rm sp}(\bm{q},\omega)=\frac{1}{T^{(2-\eta)/z}}f\left(\frac{|\bm{q}-\bm{Q}_i|}{T^{1/z}},\frac{\omega}{T}\right),
\end{align}
where the dynamical exponent $z=1$ and the anomalous dimension $\eta =1$ or, equivalently, the critical exponents $\gamma$ and $\nu$ are $\gamma=\nu=1$. This critical behavior of $\chi^{\rm sp}(\bm{q},\omega)$ is not compatible with the Hertz--Millis--Moriya theory, where $z=2$ for 2D antiferromagnets \cite{TMoriya1985,PColeman2001}. More discussions on the magnetic correlation length and the critical exponents $\nu$ and $\gamma$ can be found in Appendix~\ref{corr0}.

\begin{figure*}
\includegraphics[width=17cm]{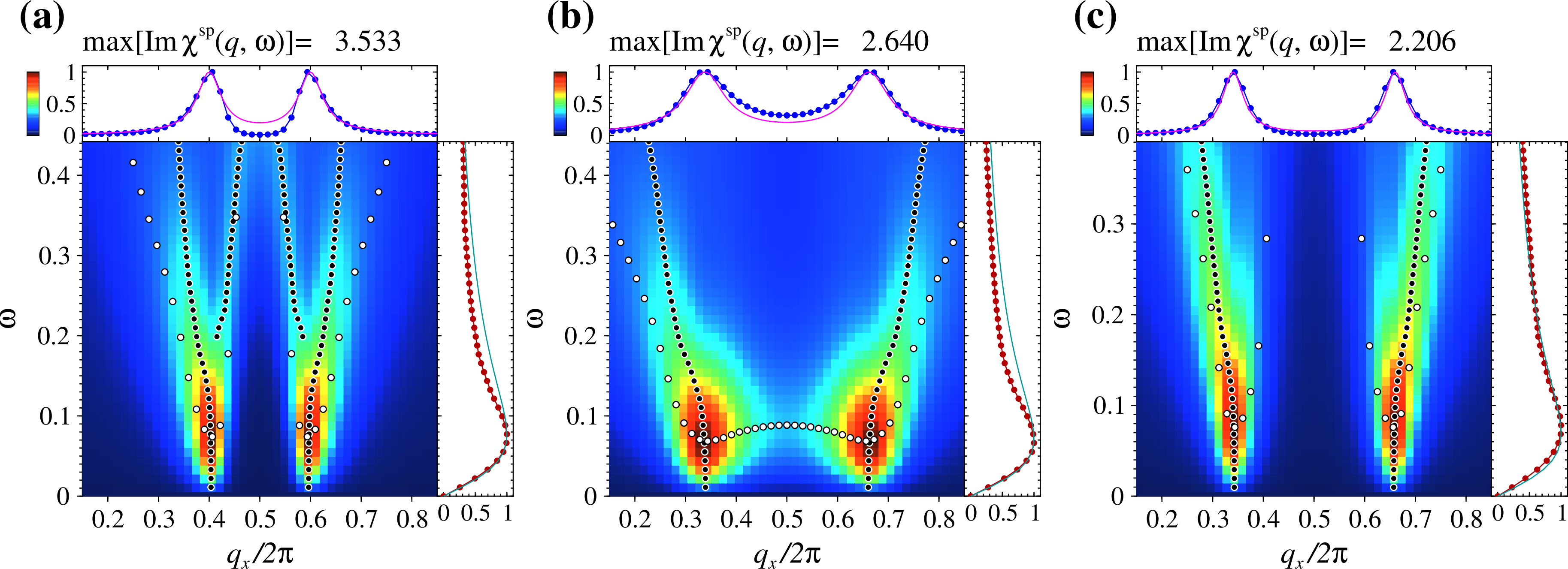}~
\caption{\label{chiqome2}The same as Fig.~\ref{chiqome1} but for $U=8$ and $t'=0$ with $p=0.215$ at $\beta=26$ (a), $U=8$ and $t'=-0.3$ with $p=0.225$ at $\beta=28$ (b) and $U=6$ and $t'=-0.2$ with $p=0.275$ at $\beta=26$ (c)}
\end{figure*}
To see whether the behavior of $\chi^{\rm sp}(\bm{q},\omega)$ as the overdamped spin wave at the Planckian dissipation limit in the vicinity of $p_{\rm QCP}$ found for $U=8$ and $t'=-0.2$ is also valid approximation for those of the other parameter sets, in Fig.~\ref{chiqome2}, ${\rm Im}\,\chi^{\rm sp}(\bm{q},\omega)$'s near $p_{\rm QCP}$ for $U=8$ and $t'=0$ with $p=0.215$ (a), $U=8$ and $t'=-0.3$ with $p=0.225$ (b) and  $U=6$ and $t'=-0.2$ with $p=0.275$ (c) are shown. Again, the reasonably good agreements between the LDFA results and the spin-wave model in Eq.~(\ref{chi_max}) with $\Gamma =2T$ can be seen. The maximum peaks of these magnetic fluctuations with regard to $\omega$ of Eq.~(\ref{chi_max}) is, therefore, placed at $\omega_{\rm max}=2T$ at $p_{\rm QCP}$.

\begin{figure*}
\includegraphics[width=17cm]{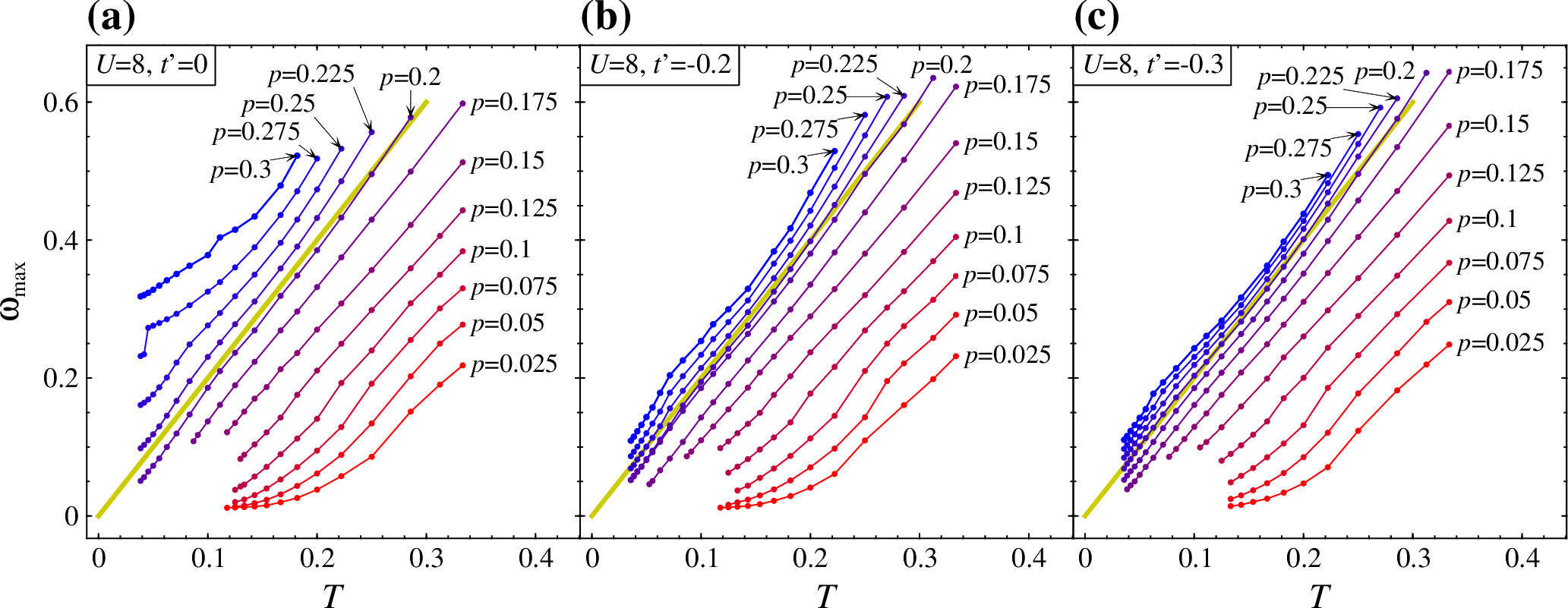}\\
\includegraphics[width=17cm]{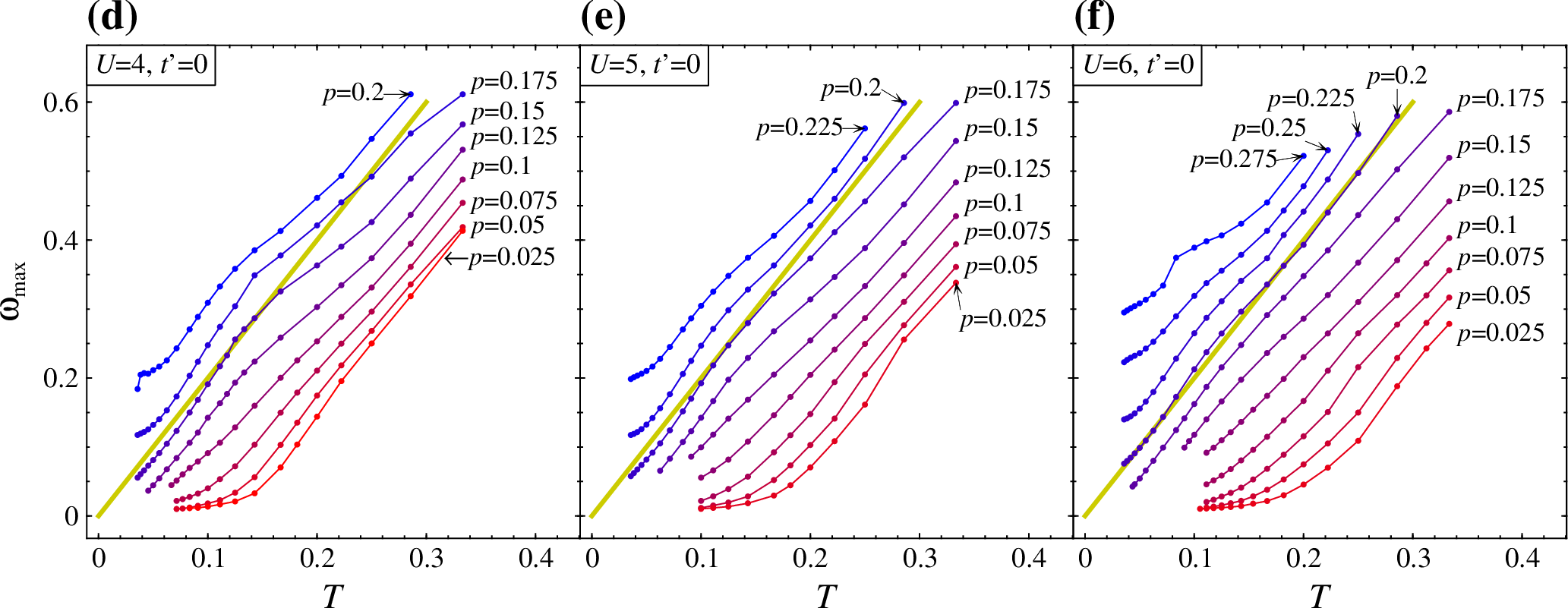}
  \caption{\label{dchi}Temperature dependence of the frequency where ${\rm Im}\,\chi^{\rm sp}(\bm{q},\omega)$ at the magnetic spot $\bm{Q}$ takes maximum value $\omega_{\rm max}$ are plotted for various values of $p$. In upper panels those for $U=8$ with $t'=0$ (a), $t'=-0.2$ (b) and $t'=-0.3$ (c) are presented. In lower panels those for $U=4$ (d), $U=5$ (e) and $U=6$ (f) with $t'=0$ are shown. In each panel $\omega=2T$ line is drawn with dark yellow.}
\end{figure*}
To explain why quantum critical region is extended wide range of $p$ in the SM state, in Fig.~\ref{dchi}, the temperature dependence of $\omega$ where ${\rm Im}\,\chi^{\rm sp}(\bm{q},\omega)$ at the magnetic spot $\bm{Q}$ takes the maximum value $\omega_{\rm max}$ are presented.
The yellow line represents $\omega_{\rm max}=2T$, the relation we found well satisfied at $p_{\rm QCP}$. The linearity still holds with $p$ near to $p_{\rm QCP}$ but shifted by a constant $T_0$ as $\omega_{\rm max}=2(T+T_0)$: $T_0>0$ for $p>p_{\rm QCP}$ and $T_0<0$ for $p<p_{\rm QCP}$. The values of $p_{\rm QCP}$ with $t'=0$ and $U=4$, 5, 6 and 8 are $p_{\rm QCP}=0.154$, $0.180$, $0.195$ and $0.215$, respectively. While for $U=4$ and $U=5$ the linearity holds only within $T <0.15$, for $U=6$ and $U=8$ the relation is satisfied at least up to $T=0.3$. 

On the other hand, the values of $p_{\rm QCP}$ for $U=8$ with $t'=-0.2$ and $t'=-0.3$ are $p_{\rm QCP}=0.265$ and $p_{\rm QCP}=0.228$, respectively and the interval of the lines nearby $p_{\rm QCP}$ are more tighten, i.e., $|T_0|$ are small as compared to those of $t'=0$. As discussed in Sec.~\ref{epsX}, this is because $\tilde{\varepsilon}^*_{\bm{X}}$ converges to a particular energy as $T\to 0$ in the wide range of $p$ including $p_{\rm QCP}$ where the AFM nesting occurs almost the whole Fermi surface in the SM phase and thus its quantum critical behavior is kept nearly unchanged: $\tilde{\varepsilon}^*_{\bm{X}}=0.04$ for $t'=-0.2$ within $0.225\le p\le 0.265$ and  $\tilde{\varepsilon}^*_{\bm{X}}=0$ for $t'=-0.3$ within $0.2\le p\le 0.25$ (see Figs.~\ref{eps0T1}(e) and (f)). This is also consistent with the nearly quantum critical behavior of the spin fluctuations observed in the neutron scattering experiments \cite{GAeppli1998}.
This extended quantum critical behavior in the SM phase is contrasted with the $U=8$ with $t'=0$ results, where the AFM nesting of the Fermi surface at $p_{\rm QCP}=0.215$ is restricted on the straight portions around the $\Gamma$--M symmetry line as in Fig.~\ref{FSh0}(b) and $\tilde{\varepsilon}^*_{\bm{X}}$ is located far from the Fermi level and not fixed at particular energy as in Fig.~\ref{eps0T1}(d). Hence, the quantum critical behavior is limited in the vicinity of $p_{\rm QCP}$ at low temperatures.

\subsection{Pairing susceptibility\label{pair}}
\begin{figure*}
\includegraphics[width=17cm]{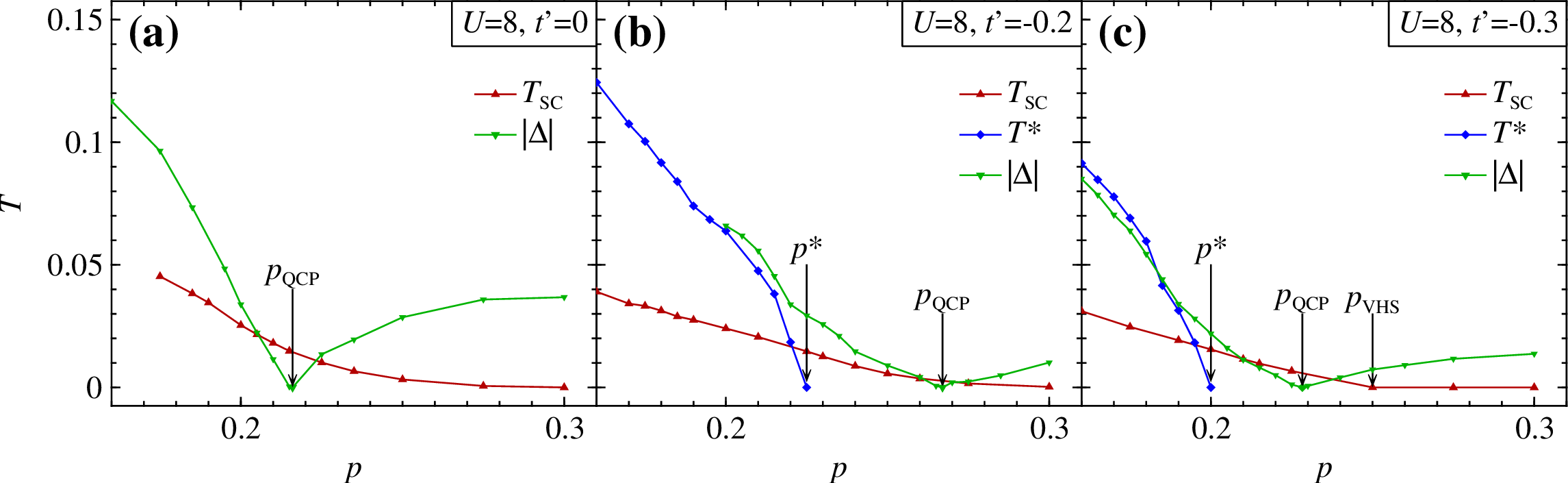}
  \caption{\label{TSC}Superconducting transition temperatures $T_{\rm SC}$ (the red lines) as a function of $p$ compared with the spin gap $|\Delta|$ (the green lines) for $U=8$ with $t'=0$ (a), $t'=-0.2$ (b) and $t'=-0.3$ (c); The pseudogap temperatures $T^*$ (the blue lines) are also shown in (b) and (c).}
\end{figure*}

To estimate the superconducting transition temperature $T_{\rm SC}$, the same method described in Refs.~\onlinecite{HHafermann2009b} and \onlinecite{JOtsuki2014} is employed. Instead of calculating directly the pairing susceptibility of electrons including the ladder approximation with the irreducible vertex of the particle-particle channel, corresponding pairing susceptibility of dual fermions is considered. The maximum eigenvalue $\lambda^{\rm SC}_{\gamma}$ and its eigenstate $\varphi_{\gamma,k}$ which belongs to the irreducible representation $\gamma$ of the ${\rm D_{4h}}$ point group were obtained with the eigenvalue equation 
\begin{align}
  \frac{T}{2N}\sum_{k'}\Gamma^{\rm pp,s}_{k,k';q=0}G^d_{-k'}G^d_{k'}\varphi_{\gamma,k'}=\lambda^{\rm SC}_{\gamma}\varphi_{\gamma,k},
\end{align}
where $\Gamma^{\rm pp,s}_{k,k';q}$ denotes the irreducible vertex of the particle-particle channel for the spin-singlet pair of dual fermions, $G^d_k$ represents the dual-fermion Green's function within LDFA, $N$ indicates the number of discretized $\bm{k}$ points in the first Brillouin zone, and their subscripts $k$, $k'$ and $q$ are combined indices for the wave vector and the Matsubara frequency, e.g., $k=(\bm{k}, i\omega_n)$. The temperature dependence of $\lambda_{\gamma}^{\rm SC}$ with $\gamma={\rm B_1}$, i.e., the $d$ wave with $k_x^2-k_y^2$ symmetry, is well approximated by the function
\begin{align}
  \lambda^{\rm SC}_{\rm B_1}=a\ln^2 \left(\frac{\Lambda}{T}\right)+b,\label{lambda_SC}
\end{align}
where $a$, $b$ and $\Lambda$ are parameters to be determined by the least squares fit and $T_{\rm SC}$ is assumed to be the temperature at which $\lambda^{\rm SC}_{\rm B_1}=1$ is satisfied with the low temperature extrapolation using Eq.~(\ref{lambda_SC}).

Figure \ref{TSC} shows $T_{\rm SC}$ (the red lines) together with the spin gap $|\Delta|$ (the green lines) as a function of $p$ for $U=8$ with $t'=0$ (a), $t'=-0.2$ (b) and $t'=-0.3$ (c). The pseudogap temperatures $T^*$ (the blue lines) are also indicated in (b) and (c). 
$T_{\rm SC}$ decreases monotonically as $p$ increases and is nearly zero at $p_{\rm QCP}$ for $U=8$ with $t'=-0.2$ and at $p_{\rm VHS}$ for $U=8$ with $t'=-0.3$. The same feature to $U=8$ with $t'=-0.2$ is also found for $U=6$ with $t'=-0.2$ presented in Fig.~\ref{pTphase}(d). These extended ranges of finite $T_{\rm SC}$ beyond $p^*$ is exactly where the pinning of the VHS point occurs in the SM phase and this is what expected in the Van Hove scenario \cite{VYIrkhin2001}. However, here, the pinning is caused by not only the presence of VHS point near the Fermi level but also the nesting effects of short-range AFM order. In contrast, there is no distinctive change in $T_{\rm SC}$ at $p_{\rm QCP}$ for $U=8$ with $t'=0$.

\section{Discussion\label{Discussion}}
In Sec.~\ref{chi}, it is found that the dynamical spin susceptibility in the SM phase is well approximated by a overdamped spin wave with the relaxation rate at the Planckian limit in Eq.~(\ref{chi_Q}). In particular, the dynamical critical exponent $z=1$ is not consistent with the Hertz--Millis--Moriya theory, which predicts $z=2$ for 2D antiferromagnets \cite{TMoriya1985,PColeman2001}.
In this section, we discuss how the $T$-linear dependence of ${\rm Im}\,\Sigma_{\bm{k}}(\omega=0)$ and the $T$-proportionality of the resistivity in the SM state can be related with the dynamical spin susceptibility in Eq.~(\ref{chi_Q}).

We consider the first order correction to the electron self-energy by the AFM fluctuations to obtain the scattering rate $\gamma_{\bm{k}}$ as
\begin{align}\label{Gammak1}
  \gamma_{\bm{k}}&\equiv -{\rm Im}\,\Sigma_{\bm{k}}(\omega =0)\nonumber\\
  &=g^2\frac{1}{N}\sum_{\bm{q}}\int^{\infty}_{-\infty}\frac{d\,\omega'}{2\pi}\left[\coth\left(\frac{\omega'}{2T}\right)-\tanh\left(\frac{\omega'}{2T}\right)\right]\nonumber\\
  &~~~~~~~~~~~~~~~~\times {\rm Im}\,\chi^{\rm sp}(\bm{q},\omega') {\rm Im}\,G_{\bm{k}+\bm{q}}(\omega').
\end{align}
Here, we naively assume that the effects of the interaction other than the spin fluctuations are already effectively considered in the one-body energy $\varepsilon_{\bm{k}}$ and neglect the self-energy in  $G_{\bm{k}}(\omega)$: ${\rm Im}\,G_{\bm{k}+\bm{q}}(\omega')\approx \delta(\omega'-\varepsilon_{\bm{k}+\bm{q}})$.
Further assuming Eq.~(\ref{chi_Q}), we obtain
\begin{align}\label{Gammak2}
  &\gamma_{\bm{k}}=g^2\frac{1}{N}\sum_{i=1}^4\sum_{\bm{q}}\int^{\infty}_{-\infty}\frac{d\,\omega'}{2\pi}\left[\coth\left(\frac{\omega'}{2T}\right)-\tanh\left(\frac{\omega'}{2T}\right)\right]\nonumber\\
  &~~~~~~~~\times \frac{2C\omega'}{{\omega'}^2+(2T)^2+[(\bm{q}-\bm{Q}_i)/W]^2}\delta(\omega'-\varepsilon_{\bm{k}+\bm{q}}).
\end{align}
Using the Taylor expansion around $\bm{k}+\bm{Q}_i$
\begin{align}
  \varepsilon_{\bm{k}+\bm{q}}\approx \varepsilon_{\bm{k}+\bm{Q}_i}+\bm{v}_{\bm{k}+\bm{Q}_i}\cdot(\bm{q}-\bm{Q}_i), 
\end{align}
where $\bm{v}_{\bm{k}+\bm{Q}_i}=\bm{\nabla}_{\bm{k}}\varepsilon_{\bm{k}+\bm{Q}_i}$, $\gamma_{\bm{k}}$ can be approximated by
\begin{align}\label{Gammak3}
  &\gamma_{\bm{k}}\approx g^2\sum_{i=1}^4\int^{\infty}_{-\infty}\frac{d\,\omega'}{2\pi}\left[\coth\left(\frac{\omega'}{2T}\right)-\tanh\left(\frac{\omega'}{2T}\right)\right]\nonumber\\
  &~~~~~~\times \frac{CW\omega'}{\sqrt{v_{\bm{k}+\bm{Q}_i}^2[{\omega'}^2+(2T)^2]+[(\omega'-\varepsilon_{\bm{k}+\bm{Q}_i})/W]^2}}.
\end{align}
Utilizing the approximation
\begin{align}
  \int^{\infty}_{-\infty}(\coth x-\tanh x)f(x)\,dx\approx \frac{\pi^2}{4}f'(0),
\end{align}
we finally get 
\begin{align}\label{Gammak4}
   \gamma_{\bm{k}}\approx g^2\frac{\pi}{2}\sum_{i=1}^4\frac{CWT^2}{\sqrt{(2v_{\bm{k}+\bm{Q}_i}T)^2+(\varepsilon_{\bm{k}+\bm{Q}_i}/W)^2}}.
\end{align}
As expected, if the nesting condition is strictly fulfilled, $\varepsilon_{\bm{k}+\bm{Q}_i}$ is zero and $\gamma_{\bm{k}}$ is proportional to $T$. When the nesting condition is not satisfied, $\gamma_{\bm{k}}$ is proportional to $T^2$ with $T\ll |\varepsilon_{\bm{k}+\bm{Q}_i}|/(2v_{\bm{k}+\bm{Q}_i}W)$ and this deviation from the proportionality relation would occur at low temperatures. Note that the condition $T < |\varepsilon_{\bm{k}+\bm{Q}_i}|/(2v_{\bm{k}+\bm{Q}_i}W)$ is exactly the condition that $\varepsilon_{\bm{k}+\bm{Q}_i}$ cannot be reached from the Fermi level within the width $\Delta Q=2WT$ of the scattering peak $\bm{Q}_i$: $|\varepsilon_{\bm{k}+\bm{Q}_i}|>v_{\bm{k}+\bm{Q}_i}\Delta Q$ (see Figs.~\ref{FSh02}-\ref{FSh0}).

So far, we have considered only the first order correction to the self-energy by the AFM fluctuations. However, if we assume the imaginary part of the self-energy of electron ${\rm Im}\,\Sigma_{\bm{k}+\bm{q}}(\omega)$ on the right-hand side of Eq.~(\ref{Gammak1}) is finite because of the higher order correction, the Fermi surface is blurred at $\bm{k}+\bm{Q}_i$ and the condition $|\varepsilon_{\bm{k}+\bm{Q}_i}|<v_{\bm{k}+\bm{Q}_i}\Delta Q$ is not strictly required for the nesting. Such a situation can be possible, if $\bm{k}$ is on the straight portions of the Fermi surface and $\bm{k}+\bm{Q}_i$ is placed near the X point where the scattering is strong due to the present of the VHS point. As was discussed in Sec.~\ref{Fsurf}, with $t'=-0.2$ almost perfect nesting occurs because of the presence of the VHS point in the vicinity of the Fermi level in the SM phase. This is partly because the imaginary part of the self-energy of electron remains finite even at $T=0$ near the X point in the vicinity of $p^{*}$, which we have seen in Sec.~\ref{SelfE} in the discussion of $\gamma_{\bm{k}}$. In fact, the Fermi surface is blurred near the X point, in particular near $p^*$, and this blurring is reduced as $p$ departs from $p^*$ in the SM phase (the ridge structures near the X point in Figs.~\ref{FSh02}(c) and (d)). In the PG phase, the scattering near the X point is suppressed due to the formation of the pseudogap as $p$ departs from $p^*$. These are the reasons why the ratio $\alpha_1/\alpha_2$ of the coefficients in the linear fitting of the resistivity is maximum around $p^*$ with $t'=-0.2$ in Figs.~\ref{rho_dc}(b) and (d) in Sec.~\ref{Resistivity}.

As for $t'=-0.3$, the quasiparticle band nearby the X point is flatten due to the Fermi condensation around $p_{\rm VHS}=0.25$ and placed slightly below the Fermi level. Furthermore, $\tilde{\varepsilon}^*_{\bm{k}}$ near the X point is approximately proportional to $T$ with $T<0.1$ as shown in Fig.~\ref{eps0T1}(f). In this situation, if ${\bm{k}+\bm{Q}_i}$ in Eq.~(\ref{Gammak4}) is near the X point, $\tilde{\varepsilon}_{\bm{k}+\bm{Q}_i}\propto T$ and $\gamma_{\bm{k}}$ in Eq.~(\ref{Gammak4}) is also proportional to $T$. This explains the $T$-proportionality of $\gamma_{\bm{k}}$ in the range $0.215\le p\le 0.275$ in Figs.~\ref{sgm0k}(e) and (f).

In the original theory of MFL by Varma \cite{CMVarma1989}, it is required that the charge and spin excitation occur over a wide range of $\bm{q}$. However, in the present theory the range of $\bm{q}$ is limited within the narrow magnetic spot with $\Delta Q=2WT$. The apparent discrepancy can be understood by the fact that in addition to the straight portions of the Fermi surface, the incommensurate nesting vectors $\bm{Q}=(\pi,\pm\delta)$, $(\pm\delta,\pi)$ in this study can connect $\bm{k}$ points on these straight portions of the Fermi surface and those near the X and Y points, where the scattering is strong due to the present of the VHS point in the vicinity of the Fermi level. This efficiently enhances the scattering due to the AFM fluctuations on whole the Fermi surface as discussed above. In contrast, scattering caused by spin or charge fluctuations at QCP without the VHS point near the Fermi level is usually limited within narrow portions of the Fermi surface (the hot spots) as seen in Fig.~\ref{FSh0}(b) for $U=8$ with $t'=0$ and excitation to the remainder part of the Fermi surface requires some additional mechanism \cite{SAHartnoll2022}.

Since the $T$-proportionality of the resistivity at $p^*$ found in the LSCO experiments \cite{RACooper2009} is well explained within the result with $U=8$ and $t'=-0.2$ as shown in Fig.~\ref{rho_dc}(b) in Sec.~\ref{Resistivity}, it is now clear that $p^*$ is not required to be a QCP of some unknown order anymore. Instead, the property of the resistivity at $p^*$ can be explained by the extended quantum critical behavior of the spin fluctuations found in this study. It has been also pointed out that the critical behavior of the electronic specific heat $C_e/T\sim \ln (1/T)$ found at $p^*$ of cuprates \cite{BMichon2019,CGirod2021} can be explain by the MFL property of the SM phase \cite{CMVarma1989} and therefore consistent with the present study.

The $T$-linear $\gamma_{\bm{k}}$ and the $\omega/T$ scaling of ${\rm Im}\,\Sigma_{\bm{k}}(\omega)$ ascribed to AFM fluctuations were found in the square-lattice Hubbard model with DCA in a NFL phase placed between the PG and FL phases \cite{WWu2022}. Although the range of this NFL phase $0.17<p<0.20$ obtained with $U=7$ and $t'=-0.2$ is substantially different from that of the SM phase in the present study, the properties and magnitude of $\gamma_{\bm{k}}$ and the $\omega/T$ scaling of ${\rm Im}\,\Sigma_{\bm{k}}(\omega)$ are consistent with those of the SM phase (Sec.~\ref{SelfE}). Furthermore, the present study demonstrates that both the presence of the antiferromagnetic QCP and the VHS point near the Fermi level around $p_{\rm MS}$ are required for the resistivity to exhibit $T$-proportionality over a wide range of $T$ (Sec.~\ref{Resistivity}). Recently, the temperature dependence of the scattering rate in the triangular-lattice Hubbard model has been studied using DCA \cite{JFournier2024}, where, unlike in the present study, only short-range AFM fluctuations are present due to geometrical frustration. Two $T$-linear scattering-rate regimes were found depending on hole doping. This indicates that only short-range AFM scattering alone is sufficient to induce a linear-in-$T$ scattering rate, although this linearity does not persist as $T\to 0$.

It has been argued that the PG state appears only with a hole-like Fermi surface \cite{WWu2018,MMeixner2024,MKlett2022}. In the present study, however, this is not a necessary condition for the stability of the PG state, as can be seen in Figs.~\ref{pTphase}(b) and (d), where a part of the $T_{\rm VHS}$ line is placed beneath the $T^*$ line. Here, the hole-like (electron-like) Fermi surface below $T^*$ only indicates $\tilde{\varepsilon}^*_{\bm{X}}<0$ ($\tilde{\varepsilon}^*_{\bm{X}}>0$), since in the PG phase quasiparticles are not well defined around the X and Y points, where $\gamma_{\bm{k}}$ increases with $T\to 0$, as discussed in Sec.~\ref{SelfE}. Furthermore, the regions in the first Brillouin zone where the pseudogap exists are reduced into points as $p$ approaches to $p^*$ and these points just below $p^*$ are not the X and Y points as reported in these previous studies but are located at the corners of the electron-like Fermi surface for $t'=-0.2$ as detailed in Appendix~\ref{pGap}. Similarly, for $U=8$ and $t'=-0.3$, the $\bm{k}$ points where the pseudogaps remain just below $p^*$ are not located at the intersections of the hole-like Fermi surface and the first Brillouin zone boundary but are placed slightly inside the first Brillouin zone along the Fermi surface. While $T^*_{\bm{k}}$ for the latter changes continuously, $T^*_{\bm{k}}$ for the former abruptly drops as $p$ approaches to $p^*$, which is consistent with the abrupt change in the intensity found at the intersection of the Fermi surface and the first Brillouin zone boundary in the ARPES experiments \cite{SDChen2019}.

In a previous parquet approximation study within DFA \cite{FKrien2022}, the pseudogap was ascribed to the damping of antinodal fermions on the Fermi surface, while the Fermi arc was attributed to the antidamping of nodal fermions. These effects stem from the large imaginary part of the spin-fermion vertex in Hedin's equations: a fermion at $\bm{k}$ is damped or antidamped depending on whether $\bm{k}+\bm{Q}$ lies inside or outside of the Fermi surface, with $\bm{Q}=(\pi,\,\pi)$. The present theory provides a more intuitive explanation for the formation of the pseudogap and Fermi arc in the PG phase (Secs.~\ref{Fsurf} and \ref{SelfE}). In the SM phase, the magnetic correlation length is short $\xi\propto 1/T$ and the nesting condition is fulfilled everywhere on the Fermi surface due to a large magnetic spot size $\Delta Q=2WT$, leading to MFL behavior. Conversely, in the PG phase, the magnetic correlation length grows exponentially with decreasing temperature. This makes magnetic spot size $\Delta Q$ smaller than in the SM phase and the nesting condition stricter. Hence, the pseudogap is formed near the X point, where the nesting condition of the AFM fluctuations is satisfied. Here, a large scattering rate $\gamma_{\bm{k}}$ is expected due to strong AFM fluctuations along the magnetic Brillouin zone boundary near the X point. On the other hand, the nesting condition is not fulfilled on the Fermi arc and the AFM fluctuations are suppressed. This allows fermions on the Fermi arc to behave as a FL.

\section{Conclusion\label{Conclusion}}
The pseudogap and strange metal states and their mutual relationship in the Hubbard model on the square lattice have been discussed in connection with the AFM fluctuations, the Mott--Heisenberg and Slater mechanisms of electron localization and the VHS point by means of LDFA. For the purpose, the Lanczos exact-diagonalization technique to obtain accurate local four-point vertex function of the effective impurity Anderson model \cite{ATanaka2019} and a new method for the electron self-energy correction were applied.

It is found that these two states appear when the VHS point is located at the vicinity of the Fermi level at $p$ near $p_{\rm MS}$ due to the enhancement of the AFM fluctuations around the X point. The end point of the pseudogap phase $p^*$ is not placed near $p_{\rm QCP}$ but $p_{\rm MS}$ because of the pinning of the VHS point near the Fermi level. The pseudogap state is magnetically in renormalized classical regime and the pseudogap appears on the Fermi surface where the nesting condition of the short-range AFM order satisfied as the correlation length grows at low temperatures $T\lesssim T^*$ with $p<p^*$. In contrast, the strange metal state is in quantum critical regime and the nesting condition of the AFM fluctuations is fulfilled almost everywhere on the Fermi surface due to the flattening of the band near the X point and because of the pinning of the VHS point, its quantum critical property is extended in the wide range of $p$ at the low temperatures. 

The distinctive properties of the dynamical spin susceptivity having $\omega/T$ scaling with overdamped spin wave with the relaxation rate at the Planckian limit in the strange metal state agrees with the inelastic neutron scattering experiments of LSCO \cite{GAeppli1998,MZhu2023}. Notably, the $\omega/T$ scaling with the dynamical exponent $z=1$ in the dynamical spin susceptivity found in this study is not compatible with the conventional Hertz--Millis--Moriya theory \cite{TMoriya1985,PColeman2001}. Intriguingly, the critical exponents $\gamma=\nu=1$ found in this study coincide with those expected in a study of SDW QCP with Kohn anomaly in the 3D Hubbard model \cite{TSchaefer2017} and in a similar discussion on 2D metals \cite{THolder2014}. More investigation is needed in this direction.  

The marginal Fermi liquid behavior and in particular, the $T$-proportionality of the resistivity in cuprates \cite{RACooper2009}, which extends a wide range of $T$, are explained from these features of the strange metal state in the Hubbard model. Recently, the $\omega/T$ scaling in the dynamical staggered-spin susceptivity and the strange metal behavior in the optical conductivity and resistivity similar to the present study is found in the cellular DMFT studies of the Kondo destruction QCP in heavy-fermion strange metal materials \cite{AGleis2024,AGleis2025}. 
How extent these characteristic behaviors of the strange metal state in the Hubbard model are shared by other strange metal materials, especially, those accompanied with the transformation of the Fermi surface near QCP \cite{MTaupin2022,HHu2024,SSachdev2025} would be interesting problem for future research. It would be also interesting to incorporate the particle-particle ladder diagram in the present approach to explain the superconducting and charge density wave states in cuprates and to clarify their relation to the pseudogap and strange metal states as well as the effects of the vertex corrections on the resistivity at low temperatures.
\begin{acknowledgments}
This work was supported by JSPS KAKENHI Grant Numbers 18K03541, 18H03683, 24K03202 and 24K06942.
The author also would like to thank A. Lichtenstein for fruitful discussions.
\end{acknowledgments}
\appendix
\section{Behavior of the spectral function near the end point of pseudogap phase\label{pGap}}
\begin{figure}
\includegraphics[width=8cm]{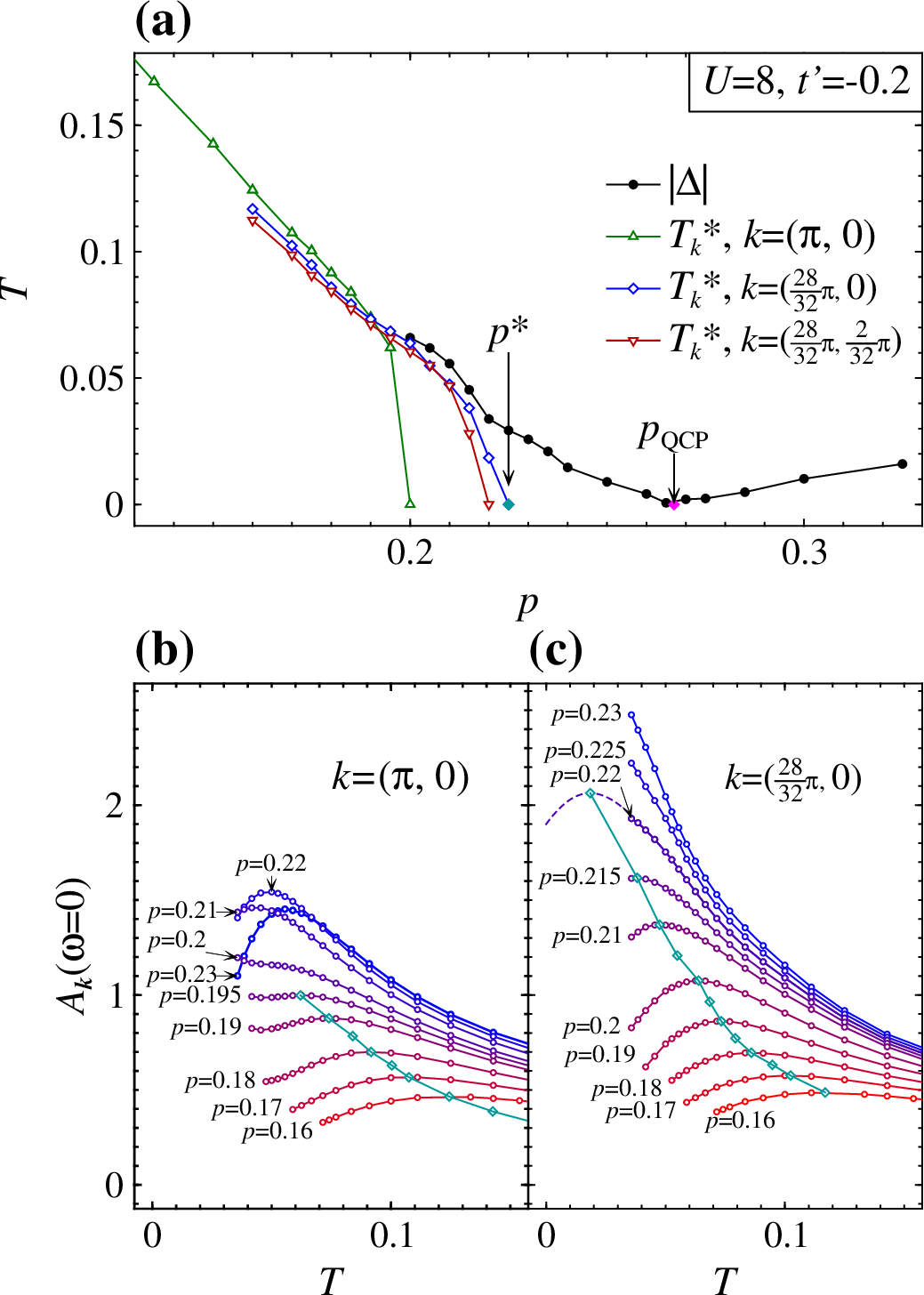}
\caption{\label{pGapChih02}(a) Pseudogap temperature $T_{\bm{k}}^*$ of $A_{\bm{k}}(\omega)$ as a function of $p$ for $U=8$ and $t'=-0.2$ is compared with the spin gap $|\Delta|$; $T_{\bm{k}}^*$'s at $\bm{k}=(\pi,0)$, $(\tfrac{28}{32}\pi,\,0)$ and $(\tfrac{28}{32}\pi,\,\tfrac{2}{32}\pi)$ are represented by the green, blue and red lines, respectively and $|\Delta|$ with black line. In the lower two panels $A_{\bm{k}}(\omega=0)$ as a function of $T$ with various values of $p$ are shown for $\bm{k}=(\pi,0)$ (b) and $\bm{k}=(\tfrac{28}{32}\pi,\,0)$ (c); the maximum points of $A_{\bm{k}}(\omega=0)$'s, whose positions with respect to $T$ define $T_{\bm{k}}^*$'s, are connected with the pale blue lines.}
\end{figure}
To discuss $p$ and $T$ dependence of individual spectral function $A_{\bm{k}}(\omega)$ with $\bm{k}$ in the vicinity of the X point near $p^*$, in Fig.~\ref{pGapChih02}(a) the pseudogap temperatures $T_{\bm{k}}^*$ of the spectral functions $A_{\bm{k}}(\omega)$ and the spin gap $|\Delta|$ are compared for $U=8$ and $t'=-0.2$. $T_{\bm{k}}^*$ is defined as $T$ at which $A_{\bm{k}}(0)$ takes the maximum value. $T_{\bm{k}}^*$'s at three different $\bm{k}$ points are presented: the X point $\bm{k}=(\pi,0)$ and those located on the corner $\bm{k}=(\tfrac{28}{32}\pi,\,0)$ and near the corner $\bm{k}=(\tfrac{28}{32}\pi,\,\tfrac{2}{32}\pi)$ of the rectangular Fermi surface in Fig.~\ref{FSh02}(b). Among them, $\bm{k}=(\tfrac{28}{32}\pi,\,0)$ is the last $\bm{k}$ point, at which the pseudogap structure of spectral function is lost with increasing $p$ and $p^*$ is defined as this vanishing point of the pseudogap. 

There is a clear difference in $p$ dependence of $T_{\bm{k}}^*$'s between that at the X point and the other two $\bm{k}$ points: in the former $T_{\bm{k}}^*$ discontinuously drops to zero between $p=0.195$ and $p=0.2$ with increasing $p$ and in the latter $T_{\bm{k}}^*$'s are continuously but steeply reduced to zero. The reason of the difference can be understand from Fig.~\ref{spec_h02}. The spectral function at the X point is double peaked and centered at $\omega=0$ with $p=0.175$ in Fig.~\ref{spec_h02}(a). As $p$ increases, this peak is shifted toward the higher energy and developed rapidly into the single peak at $\omega=0.04$ with $p=0.235$ in Fig.~\ref{spec_h02}(c). On the other hand, the spectral function at $\bm{k}=(\tfrac{28}{32}\pi,\,0)$, which is labeled A in Fig.~\ref{spec_h02}(b), keeps its center of the gravity at $\omega=0$ from $p=0.175$ to $p=0.235$. In this case, the double peak structure is stable and changed gradually into the single peak with $p=0.235$ in Fig.~\ref{spec_h02}(c). Note that the presence of the maxima in the $A_{\bm{k}}(0)$ curves with $p=0.21$, $0.22$ and $0.23$ in Fig.~\ref{pGapChih02}(b) are not due to the formation of the pseudogap. These are results of the reduction of the width of the single peak with decreasing $T$, whose peak center is away from the Fermi level similar to the peak at the X point in Fig.~\ref{spec_h02}(c).

It is also found in Fig.~\ref{pGapChih02}(a) that, the upper bound of the PG phase $T^*$, i.e., the maximum $T_{\bm{k}}^*$ among $\bm{k}$ where the pseudogap is formed on the Fermi surface, approximately coincides to $|\Delta|$ in the PG phase. As is discussed in Sec.~\ref{epsX}, the PG phase is magnetically in renormalized classical regime and $T^*$ corresponds to the crossover temperature from the quantum critical to renormalized classical regimes with $p<p^*$.

\begin{figure}
\includegraphics[width=8cm]{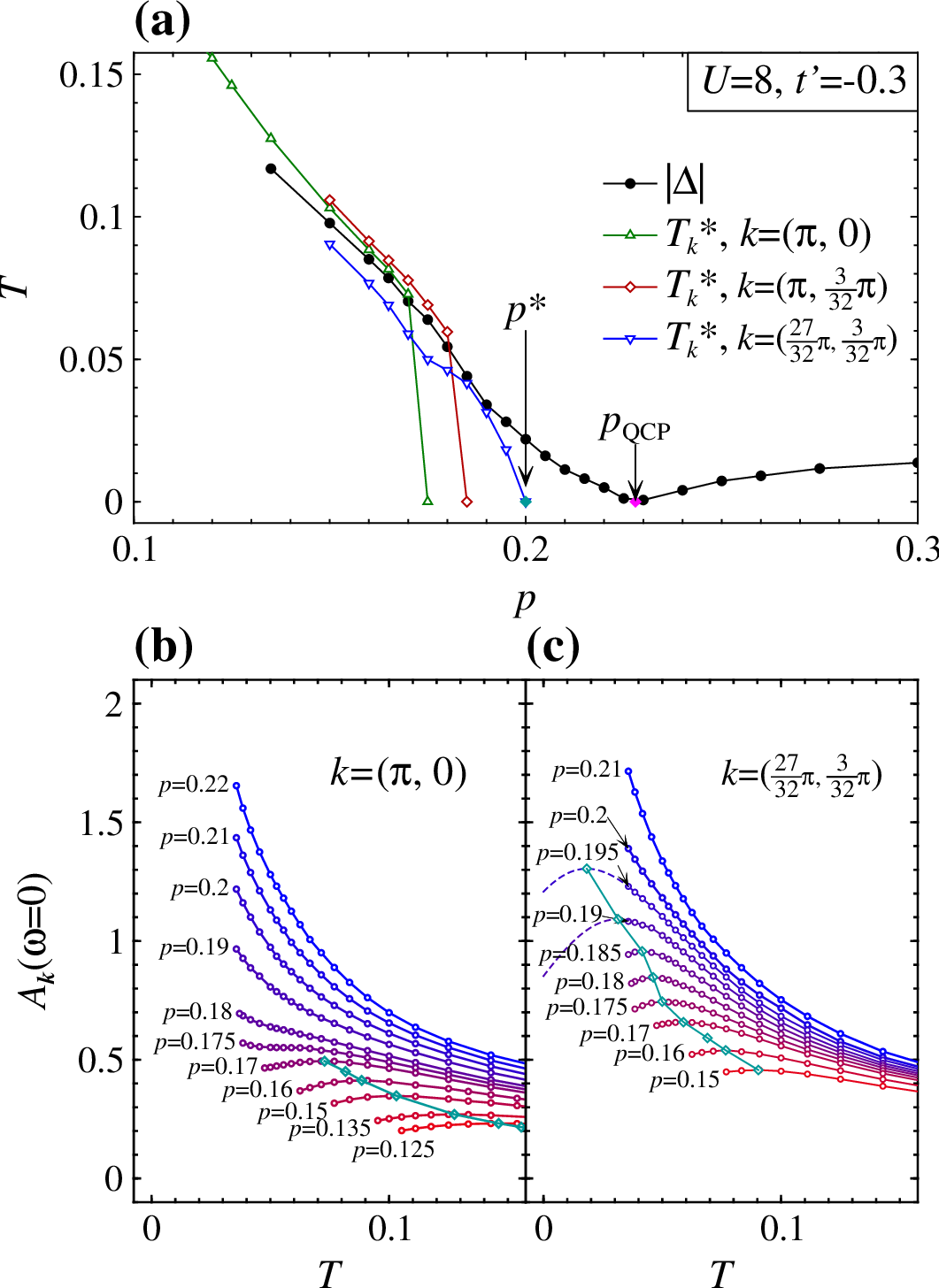}~~
\caption{\label{pGapChih03}The same as Fig.~\ref{pGapChih02} but for $U=8$ and $t'=-0.3$.}
\end{figure}
The similar relation between the pseudogap temperatures $T_{\bm{k}}^*$ and $|\Delta|$ for $U=8$ and $t'=-0.3$ can be seen in Fig.~\ref{pGapChih03}(a). $T_{\bm{k}}^*$'s for three different $\bm{k}$ points are presented: the X point $\bm{k}=(\pi,0)$, those located at the intersection of the Fermi surface and the first Brillouin zone boundary $\bm{k}=(\pi,\,\tfrac{3}{32}\pi)$ and at the arrow tip of C in Fig.~\ref{FSh03}(b) $\bm{k}=(\tfrac{27}{32}\pi,\,\tfrac{3}{32}\pi)$. $\bm{k}=(\tfrac{27}{32}\pi,\,\tfrac{3}{32}\pi)$ is the last $\bm{k}$ point, at which the pseudogap structure of the spectral function is lost with increasing $p$. Since the center of this peak remains at $\omega=0$, the double peak structure is stable and  $T_{\bm{k}}^*$ is continuously reduced to zero at $p^*=0.2$. For the X point, the double peak located at the Fermi level at $p=0.17$ in Fig.~\ref{spec_h03}(a) is reduced to the broad asymmetric peak, whose center of the gravity slightly below the Fermi level, at $p=0.19$ in Fig.~\ref{spec_h03}(b) and is developed into the sharp single peak at $p=0.225$ in Fig.~\ref{spec_h03}(c), which makes $T_{\bm{k}}^*$ discontinuously drops to zero similar to $T_{\bm{k}}^*$ of $\bm{k}=(\pi,\,0)$ with $t'=-0.2$. However, unlike $\bm{k}=(\pi,\,0)$ with $t'=-0.2$ case, the peak position approaches to $\omega=0$ with decreasing $T$ for $p \ge 0.17$ as shown in Fig.~\ref{eps0T1}(f) and thus the intensity of $A_{\bm{k}}(\omega)$ at $\omega=0$ increases with decreasing $T$ as indicated in Fig.~\ref{pGapChih03}(b). The situation of $T_{\bm{k}}^*$ for $\bm{k}=(\pi,\,\tfrac{3}{32}\pi)$ is similar to the X point. Indeed, abrupt drop of $T_{\bm{k}}^*$ is found in the angle resolve photoemission spectra at the intersection of the Brillouin zone boundary and the hole-like Fermi surface of Bi2212 experiments \cite{SDChen2019}.  

The same as $T^*$ of $U=8$ and $t'=-0.2$, the upper bound of the PG phase $T^*$ as the maximum $T_{\bm{k}}^*$ among the three $\bm{k}$ points approximately coincide to $|\Delta|$ in the PG phase within $p<p^*$ and thus $T^*$ can be regarded as the crossover temperature from the quantum critical to renormalized classical regimes within $p<p^*$. Note that the above mentioned discontinuous change of $T_{\bm{k}}^*$ only occurs at $\bm{k}$ where the peak position of $A_{\bm{k}}(\omega)$ departs from the Fermi level and the $p$ dependence of $T_{\bm{k}}^*$ corresponding to a peak at Fermi level is continuous. Hence, the change of $T^*$ near $p^*$ is considered to be steep but continuous. 

\section{Absence of the vertex corrections in LDFA\label{vertex}}
The expectation value of the current $j_{\mu}$ induced by an external electromagnetic field $A_{\nu}$ within linear response can be described by the electromagnetic response kernel $K_{\mu,\nu}$ as
\begin{align}
  j_\mu(\bm{q},i\omega_n)=\sum_{\nu} K_{\mu,\nu}(\bm{q},i\omega_n)A_{\nu}(\bm{q},i\omega_n)
\end{align}
and the optical conductivity in the real frequency $\Omega$ can be obtained as
\begin{align}
  \sigma_{\mu,\mu}(\Omega)=-\left.\frac{K_{\mu,\mu}(\bm{q}=\bm{0},i\Omega_m)}{\Omega_m}\right|_{i\Omega_m\to \Omega+i0}.\label{optcnd}
\end{align}

$K_{\mu,\mu}(\bm{q}=\bm{0},i\Omega_m)$ can be expressed as \cite{GDMahan1990,AGeorges1996}
\begin{align}
  &K_{\mu,\mu}(\bm{q}=\bm{0},i\Omega_m)=\nonumber\\
  &~~-2\frac{T}{N}\sum_{\bm{k},n}\frac{\partial^2\varepsilon_{\bm{k}}}{\partial k_\mu^2}G_{\bm{k}}(i\omega_n)\nonumber\\
  &~~-2\frac{T}{N}\sum_{\bm{k},n}\left(v_{\bm{k}}^{\mu}\right)^2 G_{\bm{k}}(i\omega_n)G_{\bm{k}}(i\omega_n+i\Omega_m) \nonumber\\
  &~~+2\frac{T^2}{N^2}\sum_{\bm{k},n}\sum_{\bm{k}',n'}v_{\bm{k}}^{\mu}G_{\bm{k}}(i\omega_n)G_{\bm{k}}(i\omega_n+i\Omega_m)\nonumber\\
  &~~~~~~~~~~~~\times \Gamma^{({\rm ch})}_{\bm{k},\bm{k}';\bm{q}=\bm{0}}(i\omega_n,i\omega_{n'};i\Omega_m)\nonumber\\
  &~~~~~~~~~~~~\times v_{\bm{k}'}^{\mu}G_{\bm{k}'}(i\omega_{n'})G_{\bm{k}'}(i\omega_{n'}+i\Omega_m),\label{karnel}
\end{align}
where $v_{\bm{k}}^{\mu}\equiv\partial\varepsilon_{\bm{k}}/\partial k_{\mu}$ and $\Gamma^{({\rm ch})}_{\bm{k},\bm{k}';\bm{q}}(i\omega_n,i\omega_{n'};i\Omega_m)$ is the charge component of reducible vertex, which can be expressed those with the spin indexes $\Gamma_{\bm{k},\bm{k}';\bm{q}}^{\sigma,\sigma'}(i\omega_n,i\omega_{n'};i\Omega_m)$ as
\begin{align}
  \Gamma^{({\rm ch})}_{\bm{k},\bm{k}';\bm{q}}(i\omega_n,i\omega_{n'};i\Omega_m)\equiv\frac{1}{2}\sum_{\sigma,\sigma'}\Gamma_{\bm{k},\bm{k}';\bm{q}}^{\sigma,\sigma'}(i\omega_n,i\omega_{n'};i\Omega_m).
\end{align}
The first term in Eq.~(\ref{karnel}) is the so called diamagnetic term and the second term is the contribution from the particle-hole bubble diagram and together $\sigma_{\rm dc}$ in Eq.~(\ref{sigma_dc}) is obtained from Eq.~(\ref{optcnd}) in the limit of $\Omega\to 0$.

The third term, i.e., the vertex correction term, vanishes exactly within LDFA and this can be shown as follows.
In LDFA, the reducible vertex is approximately obtained with two steps \cite{SBrener2008}: (i) the reducible vertex of the ladder diagram of the particle-hole channel of the dual fermion $\Gamma^{d,({\rm ch})}$ is calculated from the charge component of reducible vertex of the effective IAM $\gamma^{({\rm ch})}$ as an effective interaction between the dual fermions and then (ii) $\Gamma^{({\rm ch})}$ is obtained from the one-to-one relation between the electron and dual fermion fields.
The reducible vertex of the dual fermion $\Gamma^{d,({\rm ch})}$ is obtained from the Bethe--Salpeter equation:
\begin{align}
&\Gamma^{d,({\rm ch})}_{\bm{q}}(i\omega_n,i\omega_{n'};i\Omega_m)\nonumber=\gamma^{({\rm ch})}(i\omega_n,i\omega_{n'};i\Omega_m)\nonumber\\
  &~~~~~~~+T\sum_{n''}\gamma^{({\rm ch})}(i\omega_n,i\omega_{n''};i\Omega_m)\chi^{d,0}_{\bm{q}}(i\omega_{n''},i\Omega_m)\nonumber\\
  &~~~~~~~~~~~~\times\Gamma^{d,({\rm ch})}_{\bm{q}}(i\omega_{n''},i\omega_{n'};i\Omega_m),
\end{align}
where $\chi^{d,0}$ is defined by the Green's function of the dual fermion $G^d_{\bm{k}}(i\omega_n)$ as
\begin{align}
\chi^{d,0}_{\bm{q}}(i\omega_{n},i\Omega_{m})\equiv-\frac{1}{N}\sum_{\bm{k}}G^d_{\bm{k}}(i\omega_n)G^d_{\bm{k}+\bm{q}}(i\omega_n+i\Omega_m).
\end{align}
Since the reducible vertex of the effective IAM $\gamma^{({\rm ch})}$ is purely local, $\Gamma^{d,({\rm ch})}$ does not depend on $\bm{k}$ or $\bm{k}'$ but only the transfer wave vector $\bm{q}$.
The reducible vertex of electrons and that of the dual fermions are related by the transformation as
\begin{align}
&\Gamma^{({\rm ch})}_{\bm{k},\bm{k}';\bm{q}}(i\omega_n,i\omega_{n'};i\Omega_m)=\nonumber\\
&~~~~~~R_{\bm{k}}(i\omega_{n})R_{\bm{k}+\bm{q}}(i\omega_{n}+i\Omega_m)\Gamma^{d,({\rm ch})}_{\bm{q}}(i\omega_{n},i\omega_{n'};i\Omega_m)\nonumber\\
&~~~~~~~~~~~~\times R_{\bm{k}'}(i\omega_{n'})R_{\bm{k}'+\bm{q}}(i\omega_{n'}+i\Omega_m),\label{GammaTrans}
\end{align}
where $R_{\bm{k}}(i\omega_n)$ is defined as
\begin{align}
&R_{\bm{k}}(i\omega_n)\equiv \nonumber\\
&~~~~~~G^d_{\bm{k}}(i\omega_n)g^{-1}(i\omega_n)\left(\Delta(i\omega_n)-\varepsilon_{\bm{k}}\right)^{-1}G^{-1}_{\bm{k}}(i\omega_n).
\end{align}
Here, $g(i\omega_n)$ and $\Delta(i\omega_n)$ are the Green's function and the hybridization function of the effective IAM, respectively. Utilizing Eq.~(\ref{GammaTrans}), the third term in Eq.~(\ref{karnel}) can be rewritten as 
\begin{align}
&2T^2\sum_{n,n'}X^{\mu}(i\omega_n,i\Omega_m)\nonumber\\
&~~~~~~~~~~~\times\Gamma^{d,({\rm ch})}_{\bm{q}=\bm{0}}(i\omega_{n},i\omega_{n'};i\Omega_m)X^{\mu}(i\omega_{n'},i\Omega_m),
\end{align}
where
\begin{align}
&X^{\mu}(i\omega_n,i\Omega_m)\equiv \frac{1}{N}\sum_{\bm{k}}v^{\mu}_{\bm{k}}G_{\bm{k}}(i\omega_{n})G_{\bm{k}}(i\omega_{n}+i\Omega_m)\nonumber\\
  &~~~~~~~~~~~~~~~~~~~~~~~\times R_{\bm{k}}(i\omega_{n})R_{\bm{k}}(i\omega_{n}+i\Omega_m).\label{Vtrans}
\end{align}
While $v^{\mu}_{\bm{k}}$ is an odd parity function of $\bm{k}$, $R_{\bm{k}}(i\omega_{n})$ and $G_{\bm{k}}(i\omega_{n})$ are even parity functions. This makes the summation over $\bm{k}$ on the right hand side of Eq.~(\ref{Vtrans}) amounts to zero and thus $X^{\mu}(i\omega_n,i\Omega_m)=0$. Hence, the third term in Eq.~(\ref{karnel}) vanishes and, therefore, no vertex correction in LDFA. The lack of the vertex corrections in LDFA also means that LDFA does not satisfy the Ward identity \cite{HHafermann2014}.
\section{Magnetic correlation length and the critical exponents $\nu$ and $\gamma$\label{corr0}}
\begin{figure*}
\includegraphics[width=17cm]{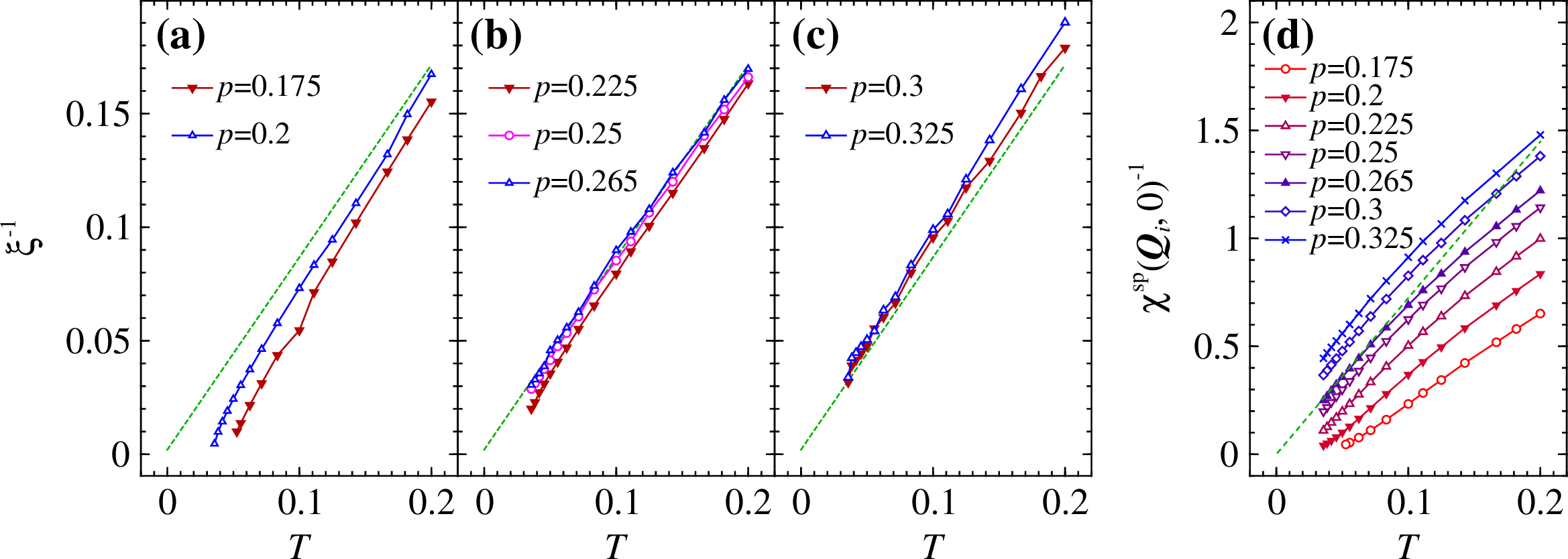}
\caption{Temperature dependence of the inverse magnetic correlation length $\xi^{-1}$ (a)-(c) and inverse static spin susceptibility $\chi^{\rm sp}(\bm{Q}_i,0)^{-1}$ (d) for $U=8$ and $t'=-0.2$ with various values of $p$. In each panel, the green dotted line denotes the least squares fit to a $T$-linear function for the data points within $T<0.1$ at the antiferromagnetic QCP $p_{\rm QCP}=0.265$.\label{corr0chi0}}
\end{figure*}
Since ${\rm Im}\chi^{\rm sp}(\bm{q},\omega)$ near $p_{\rm QCP}$ is well approximated by Eq.~(\ref{chi_Q}), using the Kramers-Kronig relation, $\chi^{\rm sp}(\bm{q},\omega)$ near $p_{\rm QCP}$ is expected to be described as
\begin{align}\label{chiq0}
  \chi^{\rm sp}(\bm{q},\omega)=\frac{2}{\pi}\sum_{i=1}^4\frac{2C}{-i\omega+\sqrt{(2T)^2+[(\bm{q}-\bm{Q}_i)/W]^2}}.
\end{align}
The magnetic correlation length $\xi$ can be extracted from the width $W\,(=2\pi/\xi)$ of the magnetic peak at $\bm{Q}_1=(\delta,\pi)$ with the shape around the peak top as 
\begin{align}\label{peakQ1}
  \chi^{\rm sp}(\bm{q},\omega=0)\propto\frac{1}{\sqrt{(2T)^2+[(\bm{q}-\bm{Q}_1)/W]^2}}.
\end{align}
To do so, a least squares fit was made to the inverse square of Eq.~(\ref{peakQ1}) located near $\bm{Q}_1$ along the Y--M symmetry line $\bm{q}=(q_x,\pi)$, i.e., fitting to a quadratic function of $q_x$. In Fig.~\ref{corr0chi0}, the $T$ dependence of $\xi^{-1}$ (a)-(c) and $\chi^{\rm sp}(\bm{Q}_i,0)^{-1}$ (d) for $U=8$ and $t'=-0.2$ are indicated for various values of $p$. As seen in Fig.~\ref{corr0chi0}(b), $\xi^{-1}$ at $p_{\rm QCP}=0.265$ is well fitted by a line proportional to $T$, at least within the temperature range presented in the figure. Moreover, apart from a slight deviation from the $T$-linear dependence at low temperatures, $\xi^{-1}$ remains nearly unchanged within the range $p_{\rm QCP}\le p\le 0.225=p^*$, indicating the extended quantum critical behavior of the spin fluctuations within this range of $p$ in the SM phase. These properties of $\xi^{-1}$ are consistent with the discussions on the $T$ and $p$ dependence of $\omega_{\rm max}$ of the dynamical spin susceptibility in the SM phase in Sec.~\ref{chi}, where the extended quantum critical behavior caused by the pinning of $\tilde{\varepsilon}^*_{\bm{X}}$ is found within the range $p_{\rm QCP}\le p\le p^*$. On the other hand, as shown in Fig.~\ref{corr0chi0}(a), $\xi^{-1}$ decreases rapidly at low temperatures below $T^*$, indicating exponential growth of $\xi$ with decreasing $T$ in the PG phase.

As discussed in Sec.~\ref{epsX}, $\chi^{\rm sp}(\bm{Q}_i,0)$ is well described by Eq.~(\ref{chi_static}) near $p_{\rm QCP}$ at low temperatures and as seen in Fig.~\ref{corr0chi0}(d), $\chi^{\rm sp}(\bm{Q}_i,0)^{-1}$ is approximately proportional to $T$ for $T<0.1$ at $p_{\rm QCP}$. The critical exponents $\xi\propto T^{-\nu}$ and $\chi^{\rm sp}(\bm{Q}_i,0)\propto T^{-\gamma}$ at $p_{\rm QCP}$ estimated from the data presented in Fig.~\ref{corr0chi0}(b) and those within $T<0.1$ in Fig.~\ref{corr0chi0}(d) are $\nu = 0.99~(\pm 0.01)$ and $\gamma = 1.01~(\pm 0.01)$, respectively. These values are in agreement with the model functions in Eq.~(\ref{chi_Q}) and Eq.~(\ref{chiq0}), where $\gamma=\nu=1$.

%%%%%%%%%%%%%%%%%%%%%%%%%%%%%%%%%%%%%%%%%%%%%%%%%%%%%%%%%%%

\end{document}